\documentclass[12pt, english]{article}

\usepackage[]{geometry}
\usepackage{amsmath}
\usepackage{mathtools}
\usepackage{epsfig,graphicx}
\usepackage{multirow}
\usepackage{float}
\usepackage{longtable}
\usepackage{authblk}
\usepackage{caption, setspace}
\usepackage{subfig}
\usepackage{comment}
\usepackage{lineno} 
\usepackage{appendix}
\usepackage[sectionbib]{chapterbib}
\usepackage{natbib}
\setcitestyle{aysep={}}
\setcitestyle{citesep={, }}
\usepackage{lscape}
\usepackage{fancyhdr}
\usepackage{color} 
\usepackage{soul}

\author[1,*]{Murat Ersalman}
\author[2,3]{Mervi Kunnasranta}
\author[4,5]{Markus Ahola}
\author[5]{Anja M. Carlsson}
\author[5]{Sara Persson}
\author[5]{Britt-Marie Bäcklin}
\author[3]{Inari Helle}
\author[5]{Linnea Cervin}
\author[1,6,*]{Jarno Vanhatalo}

\affil[1]{\normalsize Department of Mathematics and Statistics,  Faculty of Science, University of Helsinki, 00014 Helsinki, Finland}
\affil[2]{\normalsize Department of Environmental and Biological Sciences, University of Eastern Finland, 80100 Joensuu, Finland}
\affil[3]{\normalsize Natural Resources Institute Finland, 00790 Helsinki, Finland}
\affil[4]{\normalsize Marine Environment Research Group, Sustainable Environment Unit, Turku University of Applied Sciences, 20520 Turku, Finland}
\affil[5]{\normalsize Department of Environmental Monitoring, Swedish Museum of Natural History, 104 05 Stockholm, Sweden}
\affil[6]{\normalsize Organismal and Evolutionary Biology Research Programme, Faculty of Biological and Environmental Sciences, University of Helsinki, 00014 Helsinki, Finland}
\affil[*]{\normalsize Corresponding authors: muratersalman@gmail.com; jarno.vanhatalo@helsinki.fi}

\title{\LARGE Integrated population model reveals human and environment driven changes in Baltic ringed seal (\textit{Pusa hispida botnica}) demography and behavior}

\begin{document}

{\let\newpage\relax\maketitle}

\begin{flushleft}

\setcounter{page}{1}

\section*{Abstract}

Integrated population models (IPMs) are a promising approach to test ecological theories and assess wildlife populations in dynamic and uncertain conditions. By combining multiple data sources into a unified model, they enable the parametrization of versatile, mechanistic models that can predict population dynamics in novel circumstances. Here, we present a Bayesian IPM for the ringed seal (\textit{Pusa hispida botnica}) population inhabiting the Bothnian Bay in the Baltic Sea. Despite the availability of long-term monitoring data, traditional assessment methods have faltered due to dynamic environmental conditions, varying reproductive rates, and the recently re-introduced hunting, thus limiting the quality of information available to managers. We fit our model to census and various demographic, reproductive, and harvest data from 1988 to 2023 to provide a comprehensive assessment of past population trends, and predict population response to alternative hunting scenarios. We estimated that 20,000 to 36,000 ringed seals inhabited the Bothnian Bay in 2024, increasing at a rate of 3\% to 6\% per year. Reproductive rates have increased since 1988, leading to a substantial increase in the growth rate up until 2015. However, the re-introduction of hunting has since reduced the growth rate, and even minor quota increases are likely to reduce it further. Our results also support the hypothesis that a greater proportion of the population hauls out under lower ice cover circumstances, leading to higher aerial survey results in such years. In general, our study demonstrates the value of IPMs for monitoring wildlife populations under changing environments, and supporting science-based management decisions.

\paragraph{Key words:} 
IPM; state-space model; Bayesian; pinniped; environmental change; haulout; long-term monitoring; wildlife management;

\section{Introduction}

The rapid pace of environmental change presents unprecedented challenges for monitoring and managing wildlife populations \citep{berkes2008, waltner2008, chapin2009}. Conventional approaches often rely on long-term historical trends, and are increasingly insufficient as populations are exposed to novel conditions that result in unfamiliar dynamics \citep{chapin2009, marolla2021}. Management actions in response to changing conditions are frequently delayed until a more complete understanding is achieved through research, yet the urgency of these problems seldom allow for lengthy deliberation \citep{chapin2009, dietze2018}.

\bigskip

There is, thus, a growing need for versatile, mechanistic population models capable of testing alternative hypotheses on development of natural populations and accommodating uncertainties in population assessments. The integration of such models with long term monitoring data can lead to efficient generation of explanatory and anticipatory predictions \citep{mouquet2015, maris2018, marolla2021}. The repeated application of short-term predictions, combined with informed management decisions and continuous monitoring, can facilitate testing and refining model assumptions and accelerate the pace of research -- a process that is at the heart of adaptive management \citep{holling1978, lahoz2014, dietze2018}.

\bigskip

The demand for mechanistic, data-driven population models was among the key motivations behind the development of integrated population models (IPMs), which combine multiple data sources in a single, unified model to simultaneously infer key demographic parameters and population processes \citep{besbeas2002, Buckland2004, schaub2011, zipkin2018}. This is in contrast to traditional approaches where independent empirical estimates for model parameters are typically incorporated into a population projection matrix such as a Leslie matrix \citep{caswell2001}. The main advantages of IPMs include their flexibility, their ability to separate process variability from observation error, and their ability to more precisely estimate a larger number of parameters by leveraging synergies between multiple data sources \citep{abadi2010, schaub2011, zipkin2018}. Additional information can also be incorporated under the Bayesian framework in the form of prior distributions, which can be determined based on previous empirical work on similar species or expert opinions \citep{Buckland2004}. Sufficiently developed IPMs could potentially function as "digital twins" of their target populations, where the data is continuously updated and the model recalibrated in a way that enables managers to swiftly respond to changing conditions \citep{de2023, trantas2023, lecarpentier2024}. Despite the great promise of IPMs as digital twins of wild animal populations, their potential has yet to be fully realized.

\bigskip

To demonstrate the value of IPMs in monitoring and managing animal populations in dynamic and uncertain conditions, we developed a Bayesian IPM for the Baltic ringed seal (\textit{Pusa hispida botnica}) population inhabiting the Bothnian Bay, the northernmost region of the Baltic Sea and home to over 75\% of all Baltic ringed seals \citep{Harkonen1998, Sundqvist2012, halkka2017}. We adopted a state-space formulation for our model \citep{Buckland2004}, and parametrized it using count data obtained from aerial transect surveys, demographic data from hunted and by-caught seal samples, hunting records from Finland and Sweden, and assessments of reproductive status in sampled females. The ringed seal population in the Bothnian Bay is a notable example of an ice-dependent pinniped population for which long term monitoring data is available. However, despite the abundance of data, uncertainties stemming from a multitude of changing conditions have precluded assessments of population size and growth for over a decade \citep{HELCOMindicator}.

\bigskip

The population size of Baltic ringed seals had plummeted from an estimated 100,000-450,000 in the year 1900 to about 5,000 by the late 1970s, driven largely by unsustainable bounty hunting practices and extremely low reproductive rates caused by organochlorine contamination \citep{helle1976, durant1986, Harkonen1998, Harding1999, Kokko1999, Harkonen2008}. Both seal hunting and the use of PCB and DDT were subsequently prohibited, and the Bothnian Bay population has shown signs of recovery since the late 1980s. Aerial transect surveys conducted in 1988-2012 suggested an annual growth rate of approximately 5\% \citep{Sundqvist2012}. However, more recent census estimates have shown unusually high variability and systematic deviations from historical trends, casting uncertainty over the current status of the population \citep{HELCOMindicator}. 

\bigskip

In the Bothnian Bay, aerial transect surveys are conducted annually around the third week of April, and ringed seals are counted when they are visible while molting on the sea ice (i.e. "hauled out") after having abandoned their subnivean lairs \citep{Harkonen1992}. Aerial survey results have traditionally been used as an index of population size and trend, based on the assumption that approximately the same proportion of the population hauls out on ice during the surveys each year \citep{Harkonen1992}. It has been speculated, however, that a substantially larger fraction of seals may haul out when aerial surveys coincide with low ice cover or early ice breakup, resulting in extremely large population counts that are not comparable to typical results \citep{HELCOMindicator}. Moreover, previous analyses of survey estimates have often assumed constant population growth, implying constant demographic rates \citep{Sundqvist2012}. However, reproductive rates of Baltic ringed seals have likely been improving since the use of PCB and DDT were banned \citep{helle1980, Kauhala2019, HELCOMindicator_reproduction}. Unsurprisingly, the population growth rate may have been increasing as well \citep{HELCOMindicator}. Thus, we hypothesized that the atypical aerial survey estimates observed over the past decade stem from a combination of improvements in reproductive rates and a higher visibility of seals on ice following mild winters.

\bigskip

The challenges of managing the ringed seal population in the Bothnian Bay are amplified by the recent re-introduction of seal hunting in Finland and Sweden in an attempt to mitigate the rising tension between fisheries and the growing seal population. The effectiveness of hunting as a strategy for alleviating seal-fishery conflicts hinges on achieving a compromise between conservation goals and the interests of coastal fisheries \citep{oksanen2014, cummings2019}. This requires reliable assessments of population size, growth rate and demography, an understanding of seal-human interactions, as well as the ability to predict population responses to alternative management decisions. However, due to the ongoing difficulties with monitoring ringed seals in the Bothnian Bay, the impact of hunting on ringed seal demography is unknown.

\bigskip

The growing conflict between seals and fisheries, and the dynamic and uncertain conditions brought on by climate change, improving reproductive rates and the re-introduction of hunting highlight the need for mechanistic models of ringed seal population dynamics, and an increasingly holistic analysis of the available data. Using a Bayesian IPM, we provide a solution to the challenges that have crippled ringed seal monitoring efforts in the Bothnian Bay throughout the last decade. In addition to the population size and growth rate, we estimate a large number of ecologically important parameters, and make short-term predictions regarding population response to changes in hunting quotas.

\section{Materials and Methods}

\subsection{Study population}

The Baltic ringed seal is a subspecies of ringed seal endemic to the Baltic Sea (\citealt{rice1998}; but see \citealt{palo2001}). Recognized sub-populations inhabit the Bothnian Bay, the Archipelago Sea, the Gulf of Riga and the Gulf of Finland \citep{Harkonen1998, halkka2017}. Our study focused on the Bothnian Bay population, which comprises  over 75\% of all Baltic ringed seals \citep{Sundqvist2012}. The Bothnian Bay is the northernmost region of the Baltic Sea, with a surface area of approximately 37,000 km$^2$ (Figure~\ref{fig:study_area}). Since the southern sub-populations in the Archipelago Sea, the Gulf of Finland and the Gulf of Riga are small \citep{Sundqvist2012}, and ringed seals show high breeding site fidelity \citep{Harkonen2008, Kelly2010}, we treated the dynamics of the Bothnian Bay population independently of the other sub-populations.

\bigskip

Ringed seals are among the smallest and most strongly ice-associated pinnipeds in the world \citep{smith1991}. Their annual life cycle has been classified into three main 'ecological seasons' (Figure \ref{fig:lifecycle}): the foraging period, the subnivean period, and the molting period \citep{born2004, Kelly2010, Oksanen2015}. 

\bigskip

During the foraging period between June and January, Baltic ringed seals spend over 90\% of their time in the water, hauling out on land primarily at night for resting \citep{Oksanen2015}.
This phase is marked by intense feeding activities, as ringed seals seek to accumulate blubber reserves in preparation for the approaching winter \citep{Oksanen2015, Kauhala2019}. Interactions with coastal fisheries are therefore most common between late spring and late fall. During this period, bycatch mortality is most typical for young ringed seals \citep{oksanen2015b, jounela2019}.

\bigskip

During the winter, ringed seals spend most of their time out of the water in subnivean lairs built on sea ice \citep{Kelly2010}.
Around February-March, a single pup is born inside the snow lair, and nursed for 5-8 weeks \citep{helle1979b, lydersen1993}. The snow lair provides shelter to pups against harsh weather conditions and predators. Therefore, pups born without the protection offered by a stable lair are unlikely to survive \citep{smith1975, Ferguson2005, Sundqvist2012}. 
Because whelping and nursing require sufficient ice and snow conditions, adult seals are often associated with stable pack ice during the breeding season, whereas juveniles are more commonly observed in productive foraging areas near the ice edge or in the open water \citep{crawford2012, Oksanen2015}.
Mating is thought to take place within one month of parturition \citep{mclaren1958, stirling1983}, although the implantation of the embryo is delayed until around July \citep{mclaren1958}. If females do not have enough blubber reserves, implantation may not occur  \citep{boyd1999}.

\bigskip

The molting period begins as the snow cover starts to melt and subnivean lairs collapse, typically in April. Molting ringed seals spend a significant portion \citep[50-80\%;][]{Harkonen1992, born2002, Kelly2010} of their time basking on the sea ice in order to increase blood flow to the skin and maintain elevated skin temperatures \citep{feltz1966, thometz2021}. 
Sub-adults and adults often continue molting until the ice melts, typically around late May in the Bothnian Bay \citep{helle1980survey, Harkonen2008}, whereas pups typically complete their natal hair molt in lairs \citep{smith1970, lydersen1993}.

\bigskip

The molting period also marks the beginning of the ringed seal hunting season in Sweden and Finland, which was re-introduced in 2015 and in 2016, respectively. Some small scale hunting took place earlier, mainly for research purposes \citep{nyman2003, routti2009}.
Annual hunting quotas were initially set to about 100 seals in each country, and have gradually been raised to 420 seals in Sweden and 375 seals in Finland as of 2022 (Section~\ref{sec:hunting_bag_data}). Sweden reduced quotas to 350 seals in 2023, whereas the Finnish quota was kept the same.

\bigskip

Hunting of ringed seals in Sweden is legal between May 1 and January 15, although the actual distribution of hunting activities is bimodal, with distinct spring and autumn seasons (Figure \ref{fig:lifecycle}). 
Sweden practices "protective" hunting, which largely restricts hunting to within 200m of fishing sites where seals have previously caused damage to gear or taken catch. Protective hunting is also allowed within the vicinity of fish farms, fish conservation areas and fish release sites. Hunting in Sweden is therefore likely to be opportunistic.

\bigskip

Hunting of ringed seals in Finland is legal between April 16 and December 31. Finland allows recreational hunting of ringed seals, and hunters are therefore more likely to actively seek out seals. Hunting in Finland almost exclusively takes place on sea ice, and the hunting season effectively ends once sea ice melts (Figure \ref{fig:lifecycle}).
The total harvest in Finland is influenced by the severity of winter. For example, following mild winters seals tend to congregate in smaller areas of stable ice closer to land, making them more accessible to hunters \citep{Harkonen1992, Harding1999, managementplan2007}.

\subsection{Data}

\subsubsection{Aerial surveys}

The basking population size of ringed seals during the peak molting season was estimated by the Swedish Museum of Natural History (SMNH) using aerial strip surveys conducted in the Bothnian Bay in mid to late April each year between 1988 and 2023. The 800 m wide strips were placed evenly over the study area, and the number of observed seals was recorded from an aircraft flying at 90 m altitude \citep{Harkonen1992, Harkonen1998}. 
The proportion of the total ice area covered by the surveys ranged from 13\% to 53\%, with the exception of 2016, when only 7\% of the total ice area was surveyed. 
The average density of seals over the surveyed transects was extrapolated over the whole ice covered area to estimate the total size of the basking population, which is thought to account for 50-80\% of the overall population \citep{Harkonen1992, born2002, Kelly2010}.
The precision of the estimates improves with increasing coverage, although the gain in precision begins to diminish beyond a coverage of about 13\% \citep{Harkonen1992}. 

\subsubsection{Harvest totals and quotas}\label{sec:hunting_bag_data}

Finnish hunting quotas and harvest totals from 2016-2022 were provided by the Finnish Wildlife Agency. Hunting records, quotas and sampling protocols from Sweden between 2015-2022 were provided by SMNH. 

\subsubsection{Demographic and reproductive data}\label{sec:demographic_data}

Demographic age and sex data on the ringed seal population, and data on the reproductive status of female seals, were obtained from the Natural Resources Institute Finland (Luke) and SMNH. 
These data were compiled from samples taken from hunted (2016-2021 for Finland and 2015-2021 for Sweden), and by-caught (years 1988-2021) seals sent to Luke or SMNH by hunters and fishermen. 
Age determination was done by counting the growth layer groups in the cementum of either canines or molars from the lower jaw \citep{stewart1996}.
We assumed that misidentifications of sex or age were negligible, and samples that were missing information on both age and sex were excluded from our analysis. The resulting number of samples per year ranged between 46-240 for Finland, and 28-80 for Sweden.
Among those, the number of samples missing either age or sex information ranged from 1-10 per year for Finland, and 1-9 per year for Sweden.

\bigskip

Female ringed seals were examined for the presence of a visible fetus and placental scar in the uterus horn, and a \textit{corpus albicans} (CA) within the ovary. Because the implantation of embryos does not occur until July \citep{mclaren1958, boyd1991}, we relied on evaluations of a visible fetus in samples obtained between August and January. Placental scars and CA may both fade with time \citep{boyd1984, HELCOMindicator_reproduction}. In our samples, the proportion of seals with placental scars in recent years was nearly 20\% lower in May than in April. In contrast, the proportion of seals with CA were similar between April and June. To minimize observation errors associated with fading in post-partum signs, we relied on placental scar data from April, and CA data from April through June. 

\bigskip

We used only samples of females that were 5+ years old for the post-partum signs, and 4+ years old for a visible fetus (see Section~\ref{sec:pop_model}). Samples which were not evaluated for any reproductive sign were excluded, resulting in yearly sample sizes ranging from 0-18 for visible embryos, 0-62 for placental scars and 0-83 for CA (\ref{app:post_pred_checks}).

\bigskip

From here on, we define the pregnancy rate as the proportion of 4+ year old females carrying a fetus during the fall. In order to account for possible pregnancy losses, postpartum signs of pregnancy were modeled separately. We define the birth rate as the proportion of 4+ year old females that remain pregnant long enough to have a placental scar at the time of parturition, as this is thought to be the closest measurement available to assess the percentage of adult females that produce a live pup. Our definition of birth may include some late-term abortions and stillbirths, which are potentially common in pinnipeds \citep{mckenzie2005, stenson2016} and treated as pup mortality in our analysis. 

\subsubsection{Sea ice extent data}

Weekly raster maps of ice concentration across the Baltic Sea between 1988-2023 were acquired from the Finnish Meteorological Institute. We used ice concentration data from north of 63 degrees latitude around the third week of April, when both aerial surveys and most of Finnish hunting take place \citep{Harkonen1992}. The ice cover was calculated by multiplying the total sea area with the proportion of raster cells containing an ice concentration greater than zero.

\subsection{Population Dynamics Model}

\subsubsection{Structure of the population dynamics model}\label{sec:pop_model}

We modeled the population dynamics of ringed seals using an age and sex structured model with demographic stochasticity in births and deaths. Because female ringed seals typically begin reproducing at the age of five years \citep{mclaren1958, Lydersen1987, Kauhala2019, Reimer2019}, and reproductive senescence is rare \citep{mclaren1958, ellis2018}, we included six age classes in our model, grouping 5+ year olds in a single age class. We refer to 0 year old individuals as pups, 1-4 year olds as sub-adults, and 5+ year olds as adults. Note that while 4 year old sub-adults can become pregnant in our model, they do not give birth until reaching adulthood at age 5.
As hunting quotas do not discriminate between males and females, and the sex ratio of the harvests do not necessarily reflect the sex ratio of the population, we included males in our population model but assumed that females were the limiting sex in reproduction. 

\bigskip

We assumed a post-breeding census, and formulated the dynamics of the population between each census in terms of three successive sub-processes: mortality, aging, and birth (Figure \ref{fig:model_structure}). Aging was assumed to occur immediately before parturition. We denote the state of the population at the census of year $t$ by the vector $\mathbf{n}_t$. The elements of $\mathbf{n}_t$, denoted by $n_{s,a,t}$, correspond to the number of seals with sex $s \in \{f,m\}$ and age $a \in \{1,\dots,5+\}$ in the population immediately after parturition. The states of the population after the mortality and aging sub-processes are denoted respectively by the intermediate state vectors $\mathbf{u}^{(1)}_t$, and $\mathbf{u}^{(2)}_t$. 

\subsubsection{Survival and mortality}\label{sec:survival_and_mortality}

In each year, the mortality sub-process was described as a multi-state transition model, where seals are assigned to one of five states: those that survived ($S$), were harvested in Finland ($H^\text{fi}$), were harvested in Sweden either during the spring ($H^\text{sw(1)}$) or during the fall ($H^\text{sw(2)}$), or died due to other causes ($D$). 
Due to a lack of data on the magnitude of bycatch, it was included with other sources of mortality. 

\bigskip

Assuming mortality depends on sex and age and is mutually independent among seals, this process can be modeled stochastically using the multinomial distribution
\begin{equation}\label{eq:state_transition_model}
    \mathbf{u}^{(1)}_{s,a,t} | \mathbf{n}_t
    \sim \text{Multinomial}(n_{s,a,t}, \boldsymbol{\rho}_{s,a,t})
\end{equation}
where $n_{s,a,t}$ is the total number of seals with sex $s$ and age $a$ at the census of year $t$,
\begin{equation}
    \boldsymbol{\rho}_{s,a,t} = 
    \{ \rho^S_{s,a,t}, \rho^D_{s,a,t}, \rho^{H^\text{sw(1)}}_{s,a,t}, \rho^{H^\text{sw(2)}}_{s,a,t}, \rho^{H^\text{fi}}_{s,a,t} \}
\end{equation}
is a vector of transition probabilities and 
\begin{equation}
    \mathbf{u}^{(1)}_{s,a,t} = 
    \{ u^{(1)}_{s,a,S,t}, u^{(1)}_{s,a,D,t}, u^{(1)}_{s,a,H^\text{sw(1)},t}, u^{(1)}_{s,a,H^\text{sw(2)},t}, u^{(1)}_{s,a,H^\text{fi},t} \}
\end{equation}
is a vector of the number of seals that transition to each state. 

\bigskip

We modeled the mortality rates in the absence of hunting, $\mu_{s,a}$, as constant throughout the years. 
We explicitly modeled only the mortality rates of female pups and adults. 
The mortality rates of sub-adult females were interpolated between the mortality rates of pups and adults, so that
\begin{equation}\label{eq:natural_mortality}
\text{log}(\mu_{f,a}) = \text{log}(\mu_{f,0}) - \bigg(\frac{a}{5}\bigg)^c[\text{log}(\mu_{f,0})-\text{log}(\mu_{f,5+})], \ \ \ \  1 \leq a \leq 4    
\end{equation}
where $\mu_{f,a}$ is the mortality rate of female seals of age $a$, and $c$ is a parameter that determines how quickly the mortality rate approaches that of adults as the seals age. 
Male survival rates were modeled in terms of deviations from female survival rates, so that \textit{a priori} mortality rates were expected to be identical between the sexes, while allowing the data to inform us on potential differences (\ref{app:priors}). 
The probability, $\phi_{s,a}$, that a seal survives all sources of mortality other than hunting throughout the year is then given by the exponential function
\begin{equation}\label{eq:phi_adult}
    \phi_{s,a} = e^{- \mu_{s,a}}.
\end{equation}

\bigskip

To be able to predict future population responses to alternative management decisions, we explicitly modeled the within-year dynamics of hunting as a function of hunting quotas. We assumed all Finnish hunting takes place in April and May when ringed seals are basking on sea ice, and that the harvest rate is proportional to the density of seals hauled out on ice. 
We estimated the density of seals along the Finnish coast, in each age and sex group, as $\omega_{s,a,t} n_{s,a,t}/C_t$, where $\omega_{s,a,t}$ denotes the proportion of seals expected to be hauled out on sea ice in year $t$ (see Section~\ref{sec:haul-out_intensity}), and $C_t$ is the ice extent across the Bothnian Bay during peak molting. 
We further assumed that the number of active hunters in Finland is proportional to the number of unused hunting licenses, implying a Holling Type I functional response in the absence of a quota \citep{holling1959}. 
The within-year dynamics of hunting in Finland can therefore be expressed as a system of two ordinary differential equations,
\begin{equation}\label{eq:dHi}
    \frac{dH_{s,a,t}^\text{fi}}{d \tau} = E^\text{fi}_{s,a,t} (Q^\text{fi}_t - H_t^\text{fi}(\tau)) \frac{ \omega_{s,a,t} \tilde{n}_{s,a,t}(\tau)} {C_t}
\end{equation}
\begin{equation}\label{eq:dni}
    \frac{d\tilde{n}_{s,a,t}}{d \tau} = - \mu_{s,a} \tilde{n}_{s,a,t}(\tau) - E^\text{fi}_{s,a,t} (Q^\text{fi}_t - H_t^\text{fi}(\tau)) \frac{\omega_{s,a,t} \tilde{n}_{s,a,t}(\tau)} {C_t},
\end{equation}
where $H_{s,a,t}^\text{fi}$ denotes the expected number of seals with sex $s$ and age $a$ that have been harvested in Finland, $Q^\text{fi}_t$ is the hunting quota for year $t$, $H_{t}^\text{fi}(\tau) = \sum_{s,a}H_{s,a,t}^\text{fi}(\tau)$ is the total harvest in Finland at a time $\tau$ after the onset of the hunting season, $\tilde{n}_{s,a,t}(\tau)$ is the number of seals that are alive at time $\tau$ and $E^\text{fi}_{s,a,t}$ is the per capita harvest rate directed towards each demographic group in year $t$.

\bigskip

We assumed that there was stochastic variation in per capita harvest rates across the years, without any temporal trends:
\begin{equation}\label{eq:hunting_effort}
    E^\text{fi}_{s,a,t} = \hat{E}^\text{fi}_{s,a} e^{\epsilon_t},
\end{equation}
Here, $\hat{E}_{s,a}$ is the median per capita harvest rate, and $\epsilon_t \sim \text{N}(0,\sigma_{E_\text{fi}}^2)$ is a noise term. 
Setting the initial harvest to zero, an approximate closed form solution to Equations \eqref{eq:dHi} and \eqref{eq:dni} can be obtained to estimate the expected composition of the harvests, and hence the probability, $\rho^{H^\text{fi}}_{s,a,t}$, that a seal with sex $s$ and age $a$ is harvested in Finland during year $t$ (\ref{app:hunting_probability}). 
Swedish hunting during both the spring and the fall were modeled similarly to Finnish hunting. However, unlike hunting in Finland which takes place almost entirely on sea ice, protective hunting in Sweden is likely to be opportunistic. We therefore assumed that the harvest rate in Sweden was proportional to the total number of ringed seals in the Bothnian Bay, rather than the density of ringed seals on sea ice (\ref{app:hunting_probability}). Spring hunting in Sweden was assumed to take place during May and June, and fall hunting during September and October (Figure~\ref{fig:lifecycle}). 

\bigskip

Once estimates for $\rho^{H^\text{sw(1)}}_{s,a,t}$, $\rho^{H^\text{sw(2)}}_{s,a,t}$ and $\rho^{H^\text{fi}}_{s,a,t}$ are obtained, the probability that a seal survives all sources of mortality is given by 
\begin{equation}
    \rho^{S}_\text{s,a,t} = (1 - \rho^{H^\text{sw(1)}}_{s,a,t} - \rho^{H^\text{sw(2)}}_{s,a,t} - \rho^{H^\text{fi}}_{s,a,t})\phi_{s,a},
\end{equation}

\bigskip

Note that while the expected harvests, as implied by Equations \eqref{eq:dHi} and \eqref{eq:dni}, are constrained by the hunting quotas, the realized harvests given by the stochastic model in Equation \eqref{eq:state_transition_model} may exceed these quotas. This formulation not only simplifies the model considerably but also accounts for the possibility of illegal or unrecovered hunting.

\subsubsection{Haul out probability}\label{sec:haul-out_intensity}

The proportion of seals that are visible on ice during the molting period depends on the frequency with which seals move into and out of the water. We assumed that during the molting period, seals with sex $s$ and age $a$ move onto the ice at a per-capita rate of $\zeta^1_{s,a,t}$, and move back into the water at a per-capita rate of $\zeta^0_{s,a,t}$. The proportion of seals hauled out on ice is then given by
\begin{equation}\label{eq:haul_out_eq}
    \omega_{s,a,t} = \frac{\zeta^1_{s,a,t}}{\zeta^1_{s,a,t} + \zeta^0_{s,a,t}}.
\end{equation}
Although our assumption of constant movement rates implies exponentially distributed haul out and foraging durations, Equation~\eqref{eq:haul_out_eq} remains valid for any distribution with a finite expectation \citep{janssen2006}.

\bigskip

Deteriorating ice conditions may force seals to spend longer periods of time in the water or on land, as suitable haul-out sites on ice become increasingly less accessible \citep{Harkonen1998, thometz2021}. We therefore assumed that the per-capita rate at which seals move onto the ice follows a sigmoid function of ice availability, obtaining a value of zero when no ice is present, and saturating at a maximum value when sea ice is abundant:
\begin{equation}
    \zeta^1_{s,a,t} = \frac{C_t^2}{f^2 + C_t^2}.
\end{equation}
Here, $f$ is a half-saturation constant. The value of $\omega_{s,a,t}$ in Equation \eqref{eq:haul_out_eq} is determined solely by the relative magnitudes of $\zeta^0_{s,a,t}$ and $\zeta^1_{s,a,t}$. Consequently, we have chosen to scale $\zeta^1_{s,a,t}$ such that it asymptotically approaches one for all sex and age classes. 

\bigskip

To account for the possibility that a larger fraction of seals haul out during low ice cover, we modeled the per-capita rate at which hauled-out seals move into the water as a logistic function that decreases as ice cover approaches zero. In other words, we assumed than on average, seals remain on the ice for longer periods of time when sea ice is scarce:
\begin{equation}\label{eq:haul_out_prob_3}
    \zeta^0_{s,a,t} = \frac{1-\hat{\omega}_{s,a}}{\hat{\omega}_{s,a}}
    \bigg[ d + \frac{1-d}{1+e^{-(\alpha_0 + \alpha_1 C_t)}} \bigg],
\end{equation}
Here, $\hat{\omega}_{s,a}$ is the expected proportion of seals hauled out when sea ice is abundant, $\alpha_0$ and $\alpha_1$ are parameters that determine the shape of the logistic curve, and $d$ is the ratio between the lower and upper asymptotes of $\zeta^0_{s,a,t}$. A non-zero lower asymptote for $\zeta^0_{s,a,t}$ accounts for the fact that the maximum amount of time seals can spend on ice may be constrained by the need to forage or cool off.

\bigskip

We assumed that on average sub-adult and adult seals haul out with the same probability (i.e. $\hat{\omega}_{s,a} = \hat{\omega}, \ \forall a > 0$), but note that during the time of aerial surveys sub-adults are typically thought to be underrepresented on ice compared to adults \citep{smith1970}. Hence, $\hat{\omega}$ should be regarded as the average haul out probability across sub-adults and adults. Pups largely complete molting by the time aerial surveys are conducted, and are significantly less likely to be on ice \citep{smith1970, lydersen1993}. Thus, we assumed that the baseline haul out probability of pups is a constant fraction $\delta \in [0,1]$ of that for sub-adults and adults (i.e. $\hat{\omega}_{s,0} = \delta \hat{\omega}$), with the effect of sea ice modeled in the same way as for sub-adults and adults using Equations \eqref{eq:haul_out_eq}-\eqref{eq:haul_out_prob_3}.

\bigskip

\subsubsection{Reproduction}

To account for the diminishing effects of organochlorine contamination on the reproductive rates of ringed seals, we modeled the probability that an adult female gives birth to a single pup in the absence of density dependent effects as a time-varying logistic function increasing from a historic low of $b_\text{min}$ to a theoretical maximum of $b_\text{max}$, so that
\begin{equation}
    b^0_t = b_\text{min} + \frac{b_\text{max} - b_\text{min}}{1+e^{-(\beta_0 + \beta_1 t)}}.
    \label{eq:logistic_preg_rate}
\end{equation}
Here, $b^0_t$ is the expected birth rate in year $t$ if population density were zero, and $\beta_0$ and $\beta_1$ are parameters that determine the shape of the logistic curve. 

\bigskip

Little is known about the mechanisms that regulate population density in marine mammals, although \citet{demaster1984} suggested that ringed seal populations in the Arctic may be predator-limited. Given the absence of large predators in the Baltic Sea, we assumed that Baltic ringed seals are resource-limited, with intra-specific competition primarily affecting fecundity.
We assumed that fecundity would be more sensitive to changes in population size at high population densities, and modeled the rate at which pregnancies fail (e.g. due to abortions) as an exponentially increasing function of population size. The density dependent birth rate, denoted by $b_t$, can then be modeled as
\begin{equation}
    b_t = b^0_t e^{- \theta_0 (e^{\theta_1 N_{t-1}}-1)},
\end{equation}
where $N_{t-1} = \sum_{s,a} n_{s,a,t-1}$ is the total population size during the previous year, $\theta_0$ is the average failure rate of pregnancies at zero population density and $\theta_1$ is the rate at which such failed pregnancies increase with population size (\ref{app:density_dependence}). 

\bigskip

Pup production was then modeled using the multinomial distribution
\begin{equation}
    \{ n_{f,0,t}, n_{m,0,t}, \bullet \} |\mathbf{u}^{(2)}_{t-1} \sim 
    \text{Multinomial}\left( u^{(2)}_{f, 5+, t-1}, \left\{ \frac{b_t}{2}, \frac{b_t}{2}, 1 - b_t \right\} \right)
\end{equation}
where the sex ratio at birth is assumed to be at parity \citep{mclaren1958}. Here, $u^{(2)}_{f, 5+,t-1}$ denotes the number of adult females immediately before pupping, and $\bullet$ is a dummy variable for the number of mature females that did not give birth, which occurs with probability $1 - b_t$.

\subsection{Observation Models}\label{sec:obs_models}

\subsubsection{Aerial survey estimates}

We modeled the estimated basking population size resulting from the aerial strip surveys as a negative binomial distribution with expectation equal to the true basking population size:
\begin{equation}
    y_t^\text{survey}|\mathbf{\tilde{n}}_t \sim \text{Negative-Binomial}(\boldsymbol{\omega}_t^\mathbf{T} \mathbf{\tilde{n}}_{t}, r).
\end{equation}
Here, $\mathbf{\tilde{n}}_t$ is the expected population state at the time of aerial surveys (Section~\ref{sec:survival_and_mortality}), $\boldsymbol{\omega}_t$ is a vector of haul out probabilities (Section~\ref{sec:haul-out_intensity}) and $r$ is a variance parameter. 
The negative binomial distribution is commonly used to model biological count data, and can be thought of as a Poisson distribution with a rate parameter that varies stochastically as a result of, e.g., changes in observability or demographic stochasticity on $\mathbf{\tilde{n}}_t$ that lead to over-dispersion in the counts \citep{Linden2011}.

\subsubsection{Harvest totals and samples}\label{sec:Hunting_bag_and_hunting_samples}

The total harvests for both Finland and Sweden are known exactly in principle, but we nonetheless assumed a small ($\approx 5\%$ CV) variation around them because, in addition to having computational benefits, it accounts for the possibility of unrecovered or illegally hunted seals, as well as rare instances of misreporting:
\begin{equation}
    y_t^\text{ht}|\mathbf{u}^{(1)}_t \sim N(H_t, (0.05 H_t)^2).
\end{equation}
Here, $y_t^\text{ht}$ denotes the observed harvest totals and $H_t$ denotes the actual harvest totals in either Finland or Sweden during year $t$. We used separate observation models for spring and fall hunting in Sweden, except in 2015 and 2016 when only the total harvest was known.

\bigskip

Assuming that sampling of seals from the harvests is independent, the age and sex composition of the sampled seals from Finland and Sweden can be modeled using the multinomial distribution
\begin{equation}\label{eq:hunting_obs_model}
    \mathbf{y}_t^\text{hs}|\mathbf{u}^{(1)}_t \sim \text{Multinomial}\left(\sum_{s,a} y_{s,a,t}^\text{hs}, \frac{\mathbf{v} \odot \mathbf{u}^{(1)}_{H,t}}{ 
    \mathbf{v}^\mathbf{T} \mathbf{u}^{(1)}_{H,t} }\right).
\end{equation}
where $\mathbf{y}_t^\text{hs}$ is the demographic composition of the recovered samples in year $t$, $\mathbf{u}^{(1)}_{H,t}$ is the composition of the harvests, $v_{s,a}$ is the relative probability that a seal with sex $s$ and age $a$ is sampled from the harvests, and $\odot$ denotes the element wise, or Hadamard product. We assumed that seals are sampled randomly from the harvests in Finland, and in Sweden during the fall (i.e. $v_{s,a}=1, \ \forall s,a$). However, sampling during the spring has not been random in Sweden since larger seals are specifically requested to assess reproductive health. The observation model to estimate $\mathbf{v}$ for spring hunting in Sweden is described in \ref{app:sampling_bias_sw}.
Observation models for samples that were missing either sex or age information are described in \ref{app:obs_model_incomplete_info}.

\subsubsection{By-catch samples}\label{sec:By-catch_samples}

We assumed that bycaught seals were sampled randomly and independently. Because we included bycatch with other sources of mortality, we conditioned the observation model for bycaught samples on the number of seals that have died due to causes other than hunting, and modeled the composition of bycaught samples using the multinomial distribution
\begin{equation}
    \mathbf{y}_t^\text{bs}|\mathbf{u}^{(1)}_t \sim \text{Multinomial}\left(\sum_{s,a} y_{s,a,t}^\text{bs}, \frac{\boldsymbol{\psi} \odot \mathbf{u}^{(1)}_{D,t}}{\boldsymbol{\psi}^\mathbf{T} \mathbf{u}^{(1)}_{D,t}} \right)
    \label{eq:bycatch_obs_model}
\end{equation}
where $\mathbf{y}_t^\text{bs}$ is the demographic composition of the bycaught samples in year $t$, and $\boldsymbol{\psi}$ is a 12-simplex of weights accounting for possible deviations between the composition of bycatch and sources of mortality other than hunting (\ref{app:priors}). 

\bigskip

Observation models for samples that were missing either sex or age information are presented in \ref{app:obs_model_incomplete_info}.

\subsubsection{Observations of reproductive success}\label{sec:reproduction_obs}

Given reports of significant late-term pregnancy losses in other pinniped populations and the increased energetic demands of late-stage gestation \citep{pitcher1998, mckenzie2005, stenson2016}, we assumed the rate of pregnancy losses increased linearly from zero at conception to a maximum at parturition. However, we note that there is little evidence of late-term pregnancy losses in Baltic ringed seals, and it has been suggested that the influence of nutritional factors on pinniped reproduction is greatest at the point of implantation \citep{boyd1991}. Hence, for sensitivity analysis we also considered a model where the rate of pregnancy losses was assumed to decrease linearly (\ref{app:density_dependence}).

\bigskip

Assuming the reproductive status of sampled females is representative of the population as a whole, and there are no errors in the detection of fetuses, the number of observed pregnancies during the fall of year $t$ can be modeled using the binomial distribution
\begin{equation}
    y_t^\text{pregnant}|p_t \sim \text{Binomial}(y_t^\text{pregnant} + y_t^\text{not pregnant}, p_t).
\end{equation}
where, 
\begin{equation}
    p_t = b_{t+1} e^{(1 - \tau_p^2) \theta_0 e^{\theta_1 N_t}}
\end{equation}
is the pregnancy rate of the population, and $\tau_p \approx 0.5$ is the average time, in years, between mating and the sampling of pregnant females (\ref{app:density_dependence}). 

\bigskip

In principle, all females that recently gave birth will possess both a placental scar and a CA. However, since both post-partum signs can fade with time, they might not always be identified \citep{boyd1984, HELCOMindicator_reproduction}. Moreover, the presence of a CA does not necessarily indicate recent birth. CA may also be present in females that had an infertile estrous cycle, and in some cases it may even be a remnant of the previous reproductive season \citep{boyd1984}. 

\bigskip

We denote by $z^{ij}_{k,t} \in \{0,1\}$ whether a sampled seal $k$ had a placental scar evaluation of $i$ and a CA evaluation of $j$; for example,  $z^{10}_{k,t}=1$ if a placental scar, but not a CA, was observed in a seal. The outcome of the reproductive assessment for each adult female can then be modeled using a categorical distribution with probability vector $\boldsymbol{\gamma_t}$:
\begin{equation}
    \{z_{k,t}^{11}, z_{k,t}^{10}, z_{k,t}^{01}, z_{k,t}^{00}\}|
    \boldsymbol{\gamma}_t \sim \text{Categorical}(
    \{\gamma_t^{11},\gamma_t^{10},\gamma_t^{01},\gamma_t^{00} \})
    \label{reproductive_category_obs_model}
\end{equation}
The probabilities $\gamma^{ij}_t$ for each assessment outcome will depend on the birth rate $b_t$, the probability $\kappa$ that a seal that has not given birth has a CA, and the detection probabilities for placental scars, $\pi_s$, and CA, $\pi_{c}$, respectively. 
Hence, for a randomly sampled seal both a placental scar and a CA are observed ($z^{11}_{k,t}=1$) with probability $\gamma_t^{11}=b_t\pi_{s}\pi_{c}$,
corresponding to sampling a seal that has recently given birth and successfully detecting both the placental scar and the CA.
If a placental scar is observed without a corresponding CA ($z^{10}_{k,t}=1$), it must be that a CA was present but not seen. Such an outcome will occur with probability $\gamma_t^{10}=b_t\pi_{s}(1-\pi_{c})$.
Finally, if a CA is observed without a corresponding placental scar ($z^{01}_{k,t}=1$), it means that either the female recently gave birth but the placental scar was not detected (i.e. because it faded), or the female did not recently give birth but nonetheless had a CA which was detected. Such an assessment will occur with probability $\gamma_t^{01}=b_t(1-\pi_{s})\pi_{c} + (1-b_t) \kappa \pi_{c}$.
The probability that neither a placental scar nor a CA is observed is then $\gamma_t^{00}=1-\gamma_t^{11}-\gamma_t^{10}-\gamma_t^{01}$.

\bigskip

Samples obtained in May and June, as well as those obtained before 2007, were only evaluated for the presence of CA. For these samples we replaced the categorical distribution in Equation~\eqref{reproductive_category_obs_model} with the Bernoulli observation model:
\begin{equation}
    z_{k,t}^{\bullet 1}|\boldsymbol{\gamma}_t \sim \text{Bernoulli}(
    \gamma_t^{11}+\gamma_t^{01})
    \label{CA_obs_model}
\end{equation}

\bigskip

An extension of the observation model described in this section to handle incomplete data is presented in \ref{app:obs_model_incomplete_info}.

\subsection{Posterior inference and model assessment}

\subsubsection{Posterior inference}\label{sec:posterior-inference}

Following standard practice in IPMs, we assumed the different data sources were independent and constructed the joint likelihood as a product of the individual data likelihoods presented in Section~\ref{sec:obs_models} \citep{abadi2010}.
Prior distributions for model parameters are given in \ref{app:priors}.
Posterior inference was conducted using Markov chain Monte Carlo (MCMC) methods implemented using Stan version 2.26.23 \citep{stan2018}, a probabilistic programming language that uses gradient-based methods to draw samples from the posterior distributions of the parameters. 
Because Stan relies on gradient information, it does not permit discrete valued parameters. 
As a result, all discrete valued parameters in our model, along with their corresponding probability distributions, were approximated using continuous numbers and distributions (\ref{app:continuous_approximations}). 

\bigskip

We simulated four Markov Chains, and drew 4,000 posterior samples with each chain. The first 2,000 samples drawn from each chain were treated as the warm-up period and were discarded. The resulting Effective Sample Size \citep{stan2018} was greater than 1,500 for all model parameters. Convergence checks were performed using the R-hat diagnostic \citep{stan2018} as well as visual inspection of trace plots. All R-hat values were less than 1.01, and the trace plots for each parameter appeared to be well-mixed, indicating that the model had converged. Model fit was assessed using posterior predictive checks (\ref{app:post_pred_checks}). A visual comparison of the observed distribution of data with the distribution of simulated data showed no signs of model misfit.

\subsubsection{Sensitivity analyses}\label{sec:sensitivity_analyses}

To examine how different data sources influenced posterior estimates of individual parameters, we refitted our model using various subsets of the available data \citep{gelman2020}. We focused on aerial survey data, reproduction data, harvest totals, and dead samples. For each of the four data types, we refitted the model using only data from odd years. For aerial survey data, we also ran models with only pre- and post-2012 data, corresponding respectively to periods with low and high variability in the observed counts. For reproduction data, we ran an additional model with only post-2007 data to determine if long-term reproductive trends could be estimated without direct observations of reproductive signs. In total, we fit our model to eight different datasets: the full dataset and seven reduced versions. For all models, we created prior-posterior comparison plots of key parameters and calculated prior-posterior overlaps (\ref{app:post_pred_checks}: Figures \ref{fig:survival_comp}-\ref{fig:popsize_comp}). For models using the reduced datasets, we also computed overlaps of posteriors with those from the full model (\ref{app:post_pred_checks}: Table \ref{table:post_post_overlap}).

\bigskip

Additionally, we computed Pareto-k statistics for each data type using Pareto smoothed importance sampling, implemented via the R package 'loo' \citep{loo}, in order to assess their relative influence on overall model fit \citep{vehtari2017}. The results of these analyses are presented in \ref{app:post_pred_checks}.

\bigskip

We also ran prior sensitivity analyses for the baseline haul out probability parameter $\hat{\omega}$, which plays a key role in estimating total population size, and was assigned an informative prior in our model due to a lack of direct data (Section \ref{sec:haul-out_intensity}). We refitted our model with two alternative priors, each shifted by $0.1$ to the left and right from the original, and compared the resulting estimates for key model parameters.

\bigskip

All models were fitted and their sampling convergence assessed as described in Section~\ref{sec:posterior-inference}. In four of the sensitivity analyses, one of the chains showed poor mixing and was therefore removed from consideration.

\subsubsection{Demographic analysis}\label{sec:life_table}

Since asymptotic demographic analysis is of interest to practical management of Baltic ringed seals, we applied analytical methods for matrix population models to our demographic parameter estimates to calculate key asymptotic quantities, such as the intrinsic population growth rate, stable stage structure, and critical harvest levels that would lead to population decline \citep{caswell2001, ersalman2024}. All demographic analyses were performed in R Statistical Software version 4.2.0 \citep{Rteam2021}.

\subsubsection{Predictive simulations}

Using posterior samples of the model parameters, we projected the population dynamics of Baltic ringed seals in the Bothnian Bay over the next 15 years. Since our model lacked an explicit link between demographic parameters and sea ice conditions, we chose not to include climate change projections in future simulations to avoid drawing misleading conclusions. Instead, we focused on comparing the effects of alternative hunting quota scenarios. We considered four management scenarios: Scenario 1 maintains current hunting quotas; Scenarios 2 and 3 implement annual increases of 15 and 35 licenses, respectively, for both Finland and Sweden; and Scenario 4 introduces a one-time reduction of 270 licenses in each country, equivalent to a decrease of approximately 75\%. These scenarios were applied starting from 2024. For each simulated year, the ice extent in April was randomly sampled from historically observed results. We ran 8,000 simulations of each scenario, corresponding to each posterior sample of the model parameters, using R Statistical Software version 4.2.0 \citep{Rteam2021}.

\section{Results}

Posterior median estimates from our model indicate that the total number of ringed seals in the Bothnian Bay has increased from 4,700 (95\%CI: 3,500 - 6,600) in 1988 to 26,700 (20,200 - 36,300) in 2023 (Figure \ref{fig:pop_size}). 
Driven by an estimated rise in ringed seal birth rates from 0.32 (0.13 - 0.53) to 0.74 (0.66 - 0.81) (Figure \ref{fig:birth_rate}), we estimated that the population growth rate increased from 1.7\% (-1.7\% - 4.5\%) in 1988 to a peak of 6.8\% (5.6\% - 8.1\%) in 2015 before declining to as low as 4.0\% (2.5\% - 5.5\%) in 2021 due to the re-introduction of seal hunting (Figure \ref{fig:growth_rate}).
The dominant eigenvalue of the population projection matrix suggests that under ideal conditions, the Bothnian Bay population can achieve a maximum growth rate of about 7.0\% (5.9\% - 8.8\%) (Figure~\ref{fig:growth_rate}), with a corresponding net reproduction rate of 3.3 (2.3 - 4.6) female pups produced per lifetime (Section~\ref{sec:life_table}). 

\bigskip

We estimated that pregnancy rates may have been as low as 0.27 (0.02 - 0.56) during the 1970s, and could reach a maximum of 0.83 (0.75 - 0.93) in a healthy Bothnian Bay population. Reproductive rates were primarily informed by direct data on reproductive signs, but the aerial survey data also provided some indication of an increasing trend in reproduction over time (\ref{app:post_pred_checks}: Figure~\ref{fig:reproduction_comp}). The current pregnancy rate was estimated to be 0.82 (0.74 - 0.89) (\ref{app:post_pred_checks}: Figure~\ref{fig:reproduction_pp}), although the birth rate in the spring was 0.74 (0.66 - 0.81). It is therefore possible that about 9\% (1\% - 21\%) of pregnancies are aborted between the fall and spring. However, the pregnancy estimates were somewhat sensitive to assumptions about the importance of late-term abortions. When late-term abortions were assumed to be relatively rare, corresponding to the model with decreasing rate of pregnancy losses (\ref{app:density_dependence}), the estimated current pregnancy rate was 0.78 (0.72 – 0.84), with 4\% (0 – 8\%) of pregnancies failing between fall and winter. The choice of the model for the pregnancy losses was insignificant for other results.

\bigskip

The reproductive status of ringed seals is commonly assessed using placental scars and CA. We estimated that by late April, approximately 7\% (1\% - 15\%) of placental scars from births may have already faded sufficiently to become undetectable. In contrast, we estimated that most CA were easily visible between April and June, and were correctly identified 97\% (91\% - 100\%) of the time. In addition to seals that had recently given birth, our estimates indicate that between April and June, CA can also be found in 9\% (0\% - 26\%) of seals that had not given birth.

\bigskip

Our model was not informative on the carrying capacity of the population, indicating that density dependent effects were not detectable at the present population size. Hence, the posterior distribution of the carrying capacity corresponded to our prior assumptions based on historical estimates of the population size (\ref{app:priors}). Assuming density dependence primarily affects fecundity, birth rates may eventually decline to as low as 0.24 (0.17 - 0.34) (Figure~\ref{fig:birth_rate}). It is worth noting that this latter result is independent of the estimated carrying capacity \citep{caswell2001}.

\bigskip

Improvements in reproductive rates have implied a substantial change in the age structure of the population (Figure~\ref{fig:age_str}). Our estimates suggest that the proportion of pups in the population may have increased from 11\% (5\% - 15\%) in 1988 to 18\% (16\% - 21\%) in 2023. 
During the same period, we estimated a decline in the proportion of adults from 67\% (54\% - 84\%) to 50\% (43\% - 55\%). Assuming density dependence primarily affects fecundity, the age structure at carrying capacity can be expected to consist of 9\% (6\% - 11\%) pups, 18\% (12\% - 29\%) sub-adults and 74\% (61\% - 82\%) adults.

\bigskip

We did not find evidence for a skewed sex ratio which, driven by our prior assumptions, was reflected in similar mortality rates between the sexes in the absence of hunting (\ref{app:priors}).  
The probability that pups survive all mortality sources other than hunting during their first year was estimated to be 0.63 (0.42 - 0.87), which increased to 0.86 (0.77 - 0.95) at age one and to 0.95 (0.91 - 0.97) by maturity (Figure~\ref{fig:survival_probs}). However, the large negative posterior correlation between juvenile and adult survival probabilities suggested that the model had limited capacity to estimate age-specific mortality rates (\ref{app:post_pred_checks}: Figure \ref{fig:survival_corr}). Refitting the model with different subsets of the data confirmed that survival probabilities were primarily informed by pre-2012 aerial survey and the reproduction data, both of which lack information on age-specific mortality (\ref{app:post_pred_checks}: Figure \ref{fig:survival_comp}).

\bigskip

We found significant juvenile bias in bycatch from small-scale fisheries, with pups nearly 8 (4 - 16) times more likely to be bycaught than one year olds (Figure~\ref{fig:bycatch_bias}). Male pups were also about 40\% (1\% - 88\%) more likely to be bycaught than female pups. We estimated that Swedish hunting during the spring was representative of population demography (Figure~\ref{fig:sw_bias_spring}). In contrast, the fall hunt was heavily aimed at adults of both sexes, with a slight bias towards males (Figure~\ref{fig:sw_bias_fall}). We estimated that pups were the most vulnerable age group to hunting in Finland (Figure~\ref{fig:fi_bias}). Sub-adults were also generally underrepresented in the Finnish harvest.

\bigskip

We estimated that about 60\% (48\% - 74\%) of sub-adult and adult seals and 35\% (12\% - 60\%) of pups may be visible on ice during aerial surveys that coincide with extensive ice cover. Historical ice conditions in the Bothnian Bay suggest that approximately 51\% (39\% - 65\%) of the total population may be visible during aerial surveys carried out in typical ice conditions (Figure~\ref{fig:haulout_yr}). We found that as ice cover diminishes, the fraction of hauled-out seals may gradually decline to as low as 44\% (30\% - 62\%). However, this proportion was found to increase dramatically to as much as 91\% (70\% - 99\%) when the ice cover declined below a critical threshold of about 13,000 km$^2$ (Figure~\ref{fig:haulout}). Although our mechanistic haul out probability model was relatively complex, all parameters appeared to be well informed, largely by the aerial survey data. The exception was the baseline haul out probability $\hat{\omega}_{s,a}$, which was given an informative prior (\ref{app:priors}) and was only weakly informed by data. Nonetheless, posterior estimates were generally robust to prior choice. Only estimates of total population size, and consequently population growth rate under harvesting, showed moderate sensitivity, with estimates varying by less than 10\% under sensible prior choices (\ref{app:post_pred_checks}: Table \ref{table:post_post_overlap}).

\bigskip

An analysis of the impact of hunting on population growth revealed that at present, hunting as few as 1,200 seals per year may be sufficient to cause population decline (Figure~\ref{fig:critical_hunting}), although the uncertainty around this estimate was high (1,200 - 2,600). Achieving a growth rate of 7\%, the benchmark set by the Baltic Marine Environment Protection Commission (HELCOM) for designating good status, might prove challenging with any level of hunting activity (Figure~\ref{fig:future_preds}), especially if more than 380 seals are hunted annually. Posterior predictive simulations suggested that maintaining hunting quotas at their current levels of 350 and 375 seals per year in Sweden and Finland, respectively, may allow for a slight improvement in the population growth rate over the next 10-15 years (Figure~\ref{fig:future_preds}). In contrast, increasing quotas by 15 licenses per year in each of Sweden and Finland may cause a slight decline in the growth rate, and an increase of 35 licenses per year in each country may significantly slow down population growth, and may even lead to population decline within the next 15 years. 

\section{Discussion}

So far, different sources of long-term monitoring data on Baltic ringed seals have been analyzed separately to assess population growth, reproductive health and other relevant metrics of population status. A major limitation of this approach is its inability to leverage all available information, as the synergies between different data sources are not exploited \citep{schaub2011, zipkin2018}. This limitation has become more apparent than ever during the last decade, when unexpected trends and large fluctuations in population counts from aerial surveys made it impossible to obtain reliable estimates of population growth from survey data alone. Growth in closed populations, however, is ultimately a consequence of births and deaths. Thus, data related to reproduction and hunting also contain information on the growth rate. By incorporating all available information into a mechanistic IPM, we were able to address recent challenges in ringed seal monitoring in the Bothnian Bay. Our model showed a good fit to all available data (\ref{app:post_pred_checks}), and was able to estimate a large number of ecologically important parameters. Moreover, fitting our model in Stan \citep{stan2018}, while requiring some approximations, proved highly efficient, achieving convergence in under an hour on a personal laptop computer despite the large number of parameters.

\bigskip

Earlier analyses of ringed seal population trends in the Bothnian Bay suggested that the population had been growing at a constant rate of approximately 5\% per year since 1988 \citep{Sundqvist2012}, although more recent analyses showed that an increasing trend was likely \citep{HELCOMindicator}. We estimated that improvements in reproductive rates may have led to a substantial increase in the growth rate between 1988 and the re-introduction of hunting in 2015, which accounted for much of the systematic deviation between recent population size estimates and earlier expectations based on historical trends. 

\bigskip

On the other hand, a small number of unusually large population estimates obtained in recent years were explained surprisingly well by an increase in the fraction of seals hauling out during aerial surveys conducted in poor ice conditions, supporting earlier speculations \citep{HELCOMindicator}. 
It is not clear why seals would haul out in significantly greater numbers under such circumstances. Sea ice conditions during the surveys likely correlate with air temperature, wind speed, and snow depth, which are known to affect ringed seal haul out probability \citep{Harkonen1992, hamilton2018}. Snow depth may additionally influence the proportion of seals that are hiding in lairs \citep{Kelly2010}. However, as surveys are conducted in late April, when most lairs have collapsed and seals are at the height of molting, and only on clear, calm days, the impact of these variables on population size estimates is likely limited. Another possibility is that the break up of sea ice causes adult seals to relinquish territorial behavior typical of winter, allowing younger sub-adults to haul out in greater numbers \citep{Harkonen1992, Harkonen1998, Kelly2010, Oksanen2015}. This may additionally result in large congregations of seals, potentially contributing to the high variability in survey estimates \citep{Harkonen1998, Linden2011}.

\bigskip

Since the sub-adult age classes account for about a third of the total population, it seems unlikely that an increase in their visibility alone can account for a nearly two-fold increase in population size estimates. Likewise, immigration from southern sub-populations, which are collectively believed to be nearly an order of magnitude smaller than the Bothnian Bay population \citep{Sundqvist2012}, also appears insufficient to explain the observed increase in counts. The increase in the proportion of visible seals during low ice cover could, in part, be attributed to an increase in the cost of foraging. During the molting period, ringed seals are likely to derive greater benefit from maintaining elevated skin temperatures than from consuming resources \citep{thometz2021}. Diminishing sea ice will most likely result in increased distances between optimal haul-out sites and productive foraging areas, and the trade-offs associated with foraging could compel seals to decrease the frequency with which they venture out to sea. Moreover, seals may anticipate that the molting period could be interrupted prematurely, causing them to spend longer periods of time out of the water to ensure that they attain adequate levels of solar exposure before the ice melts completely.

\bigskip

While our findings demonstrate a clear effect of ice cover on aerial survey estimates, this conclusion was largely driven by two outliers that corresponded to unusually mild winters. Establishing haul-out behavior as the causal mechanism demands further investigations, including additional data and model comparisons to evaluate alternative hypotheses. Regardless of the mechanism, the number of seals on ice must evidently be zero when no sea ice is present. 
Extreme inter-annual variability in population counts from aerial surveys may therefore be expected following mild winters (Figure~\ref{fig:haulout_ice}), which could pose additional challenges for future ringed seal monitoring efforts in the Bothnian Bay. 
As ice conditions in the Baltic Sea continue to deteriorate \citep{meier2004, haapala2015}, understanding the factors that affect ringed seal haul-out behavior will become increasingly important for reliable monitoring. Our study demonstrates that aerial surveys can be a valuable source of information to this end, and survey estimates that appear unreliable at first may provide important insights on ringed seal behavior. We therefore emphasize the need for continued monitoring efforts, as well as the need for detailed mechanistic models that go beyond simple analyses of trends. Future IPMs could also incorporate individual level data collected from telemetry devices with dry/wet sensors, enabling direct estimation of haul out probabilities \citep{london2024}.

\bigskip

Statements regarding the total population size of ringed seals require an estimate of the proportion of seals that are hauled out during the surveys. Haul-out behavior is likely to vary by age \citep{smith1970, lydersen1993, crawford2012, Oksanen2015}, but little is known to quantify age-specific differences. Most studies have focused on sub-adult and adult ringed seals, which suggested that 50\% to 84\% of seals may be visible during surveys \citep{smith1970, Harkonen1992, born2002, Kelly2010}. However, pups undergo their first molt and shed their lanugo hair before the molting season of older seals. Thus they spend most of their time in the water by the time they are 5 weeks old \citep{smith1970, lydersen1993}. Failing to account for their lower visibility during the surveys may lead to an underestimation of the total population size. We found that following typical winters, roughly half of the population may be visible on ice during surveys, which corresponds to the lower end of previous estimates, and is similar to what has been proposed by \citet{smith1970}.
The number of ringed seals in the Bothnian Bay may therefore be larger than previously expected \citep{Sundqvist2012}. However, our results were influenced by prior assumptions about haul out probabilities in abundant ice conditions. Detailed studies on the molting behavior of Baltic ringed seals will be crucial for obtaining reliable estimates of total abundance.

\bigskip

Little is known about the survival probabilities of ringed seal pups, but previous estimates range from as low as 0.34 in lake Saimaa before active conservation efforts, \citep{kokko1998} to 0.6-0.7 in the Arctic \citep{smith1970, Kelly1988}. A survival probability as high as 0.8 has been suggested for unharvested Arctic populations \citep{smith1970}. For Baltic ringed seals, a pup survival probability of 0.65 has previously been assumed \citep{Sundqvist2012}, which is very similar to our posterior median estimate. However, the uncertainty around our estimate was high due to weak parameter identifiability (\ref{app:post_pred_checks}: Figure \ref{fig:survival_corr}). Baltic ringed seal pups are safe from large predators that are common in the Arctic such as polar bears (\textit{Ursus maritimus}) and Arctic foxes (\textit{Vulpes lagopus}), which might lead to comparably higher survival probabilities \citep{stirling2004}. On the other hand, higher bycatch mortality and greater chance of early snow and ice melt in the Baltic Sea could compensate for the limited presence of predators. Snow and ice conditions in the Baltic Sea are expected to decline in the future, which will most likely lead to increased pup mortality \citep{smith1975, meier2004, Ferguson2005, Kelly2010, Sundqvist2012, Reimer2019}.

\bigskip

Although previous estimates of sub-adult survival probabilities are rare, our estimate aligns closely with the 0.81-0.93 range reported in earlier studies \citep{Sundqvist2012, koivuniemi2019}. Our posterior median estimate of 0.95 for adult survival probability corresponds to the upper end of previous estimates for ringed seals, which range from 0.88-0.96 \citep{durant1986, Kelly1988, kokko1998, Sundqvist2012, koivuniemi2019, Reimer2019}, and is consistent with previous assumptions for Baltic ringed seals \citep{Sundqvist2012}. Similarly high adult survival probabilities have also been estimated for Saimaa ringed seals \citep{koivuniemi2019} and British grey seals (\textit{Halichoerus grypus})  \citep{thomas2019}. However, the high posterior correlation between juvenile and adult survival probabilities suggests that these parameters were not individually identifiable (\ref{app:post_pred_checks}: Figure \ref{fig:survival_corr}). Direct data on survival, such as from mark-recapture studies, will be needed to obtain precise age-specific estimates. Such data can additionally support the estimation of overall abundance and, in turn, haul out probabilities.

\bigskip

We estimated a historical minimum pregnancy rate that was strongly in agreement with some empirical estimates from the late 1970s \citep{helle1980}. However, the uncertainty around our estimate was high due to limited data on female reproductive status during the initial years of our study period. Our estimates indicate that the current pregnancy rate may be nearly the same as the maximum attainable by a healthy Bothnian Bay population. Thus, the population appears to have almost fully recovered from the effects of organochlorine contamination. 
Assessments of the reproductive health of Baltic ringed seals is currently based on a tentative pregnancy rate threshold of 0.9 that was established by HELCOM based on studies of seal populations elsewhere \citep{HELCOMindicator_reproduction}. Our results suggest that this threshold may be too high even for a healthy Bothnian Bay population, possibly because the Bothnian Bay is a relatively unproductive environment \citep{Kauhala2019}. A threshold between 0.80-0.85 may be a more realistic target. 
As the population continues to grow, rising intra-specific competition may cause pregnancy rates to decrease. Body conditions of Baltic ringed seals may already be declining due to poor nutritional status \citep{Kauhala2019}, although it is unclear whether population density is a contributing factor. Our model did not provide information on density-dependent effects on reproductive rates, either because the current population size is substantially below carrying capacity, or because other demographic parameters, such as pup survival, were more strongly affected. Long-term predictions of ringed seal population dynamics will require detailed consideration of density-dependent mechanisms, which would greatly benefit from additional demographic data such as long-term mark-recapture studies \citep[e.g.][]{hiby2007, koivuniemi2019}. Understanding seasonal patterns of reproductive failure and their links to population density and nutritional stress would also be valuable for both assessing current reproductive health and improving future population projections.

\bigskip

The most reliable assessments of reproductive rates will be based on the presence of embryos in female seals sampled in the fall. Unfortunately, the current sample size for this measure remains relatively small, with substantial inter-annual variability in the observed pregnancy rate. Assessments based on placental scars may underestimate the pregnancy rate, as our results indicate that some scars that were present during parturition may have already faded by the end of April. Many of these scars are likely to be from early births in February. Additionally, a small fraction of fetuses may be aborted in early pregnancy, and placental scar from these pregnancies may have also faded by spring. Accuracy of pregnancy rate estimates from placental scars can be improved either by using the earliest available samples in the spring, or by adjusting the estimates to account for the effect of fading. Assessments based only on CA may be more difficult to interpret since a CA could also be present in seals that had an infertile estrous cycle \citep{boyd1984}. However, we found that CA may yield very accurate estimates of the pregnancy rate when the pregnancy rate is high. An additional challenge when estimating pregnancy rates is accounting for potential sampling biases. Most assessments, including the present study, assume that reproductive signs in harvested or bycaught females is representative of the entire female population. This assumption may introduce bias if reproductive status affects a female’s likelihood of being harvested or bycaught. For instance, preferential harvesting of pups can inflate reproductive rate estimates if their mothers are harvested with them. Similarly, if postpartum females forage differently due to reduced energy reserves or molt at different locations, their susceptibility to hunting or bycatch may differ. Detailed studies on seal and hunter behavior will be essential for addressing potential sampling biases.

\bigskip

Our results suggest that hunting has had a substantial impact on the population growth rate since its re-introduction, possibly reducing it by more than three percentage points. An unexploited Bothnian Bay population could achieve a maximum growth rate of about 7\%, satisfying HELCOM’s current threshold for good status. However, this threshold is unlikely to be met in the presence of even a small amount of hunting, and a growth rate of 4\% to 5\% may be a more likely outcome from the current hunting practices (Figure~\ref{fig:future_preds}).

\bigskip

It is possible that the quotas in 2022 were close to a critical threshold beyond which population growth could no longer be maintained. However, this threshold will increase in the near future provided that the population continues to grow (Figure \ref{fig:critical_hunting}). Current annual hunting quotas are established at 350 and 375 seals in Sweden and Finland, respectively. Our simulation study suggested that while increasing quotas by fewer than 35 licenses per year in each of Sweden and Finland is unlikely to cause population decline within the next 10-15 years, increases in excess of 15 per year in each country may further decrease the population growth rate (Figure \ref{fig:future_preds}). Our model assumed that there is sufficient demand for seal hunting such that harvest rates will increase proportionally to hunting quotas. While this assumption will not hold for large quotas, it remains appropriate within a precautionary framework.

\bigskip

It should be emphasized that, due to the short time frame of available demographic data and the opportunistic nature of sampling, our model did not take into account likely effects of climate change and other environmental variables, such as prey availability, on demographic parameters like pup mortality and reproductive rates \citep{meier2004, Ferguson2005, Sundqvist2012, Reimer2019}.  Thus, our simulation results should be viewed as a comparison of the potential impact of different hunting quota decisions, rather than as forecasts of future population dynamics. As climate change is expected to lead to further declines in snow and ice cover in the Baltic Sea \citep{meier2004, haapala2015}, understanding its effects on ringed seal demography is essential for making longer term predictions. Future IPMs could explicitly link ringed seal survival and reproduction to environmental variables such as snow and ice conditions to predict population responses to various climate change scenarios \citep[e.g.][]{Sundqvist2012, Reimer2019}. However, estimating parameters of such models will likely require direct data on survival probabilities, such as from mark-recapture studies.

\bigskip

Our estimates suggest that the demographic composition of Swedish harvests may differ significantly between the spring and the fall. We found that hunting was largely aimed at adults during the fall, possibly because of the greater damage they cause to fishing gear and catch. Studies of grey seals have shown that some adults, typically males, may specialize in raiding fishing gear \citep{konigson2013}. In contrast, the adult bias was completely absent during the spring hunt. This is not particularly surprising considering that adult seals are molting on ice in large numbers during the spring, and seldom venture out at sea \citep{smith1970, Harkonen1992, born2002, Kelly2010}. Younger seals may be much more likely to be foraging in open water during most of the molting period, and thus have more frequent interactions with fishermen \citep{smith1970, lydersen1993, crawford2012, Oksanen2015}. In fact, we estimated that adult seals were generally more likely to be hunted than sub-adults during the spring in Finland, where hunting primarily takes place on sea ice. Pups were an exception and were overrepresented in Finnish harvests despite their relatively low presence on sea ice, although some late born pups are often still with their mothers during the Finnish hunting season. Since hunting in Finland is recreational, hunters may specifically seek out inexperienced pups for their meat and fur \citep{kingsley1998}.

\bigskip

A population's response to harvesting is strongly influenced by the age and sex composition of the harvest, with the removal of adult females typically having the greatest impact in long-lived mammals like ringed seals \citep{Kokko1999, heppell2000, Reimer2019}. Hence, it is important to ensure that recovered seals are either sampled randomly from the harvests or that any sampling biases are accounted for. In Sweden, hunter-reported data on the sex and body length of each seal, along with records of whether or not each harvested seal was recovered as a sample, allowed us to quantify size- and sex-related sampling biases. Although such biases are unlikely in Finland, similar data could help ensure accurate and unbiased estimates of harvest compositions.

\bigskip

We were not able to estimate the overall magnitude of bycatch mortality due to a lack of data. Such data could be obtained through interviews with fishermen \citep{vanhatalo2014, tubbs2024}.
However, consistent with findings from many other seal populations, we found that pups were significantly more at risk of getting bycaught in small fisheries than older seals \citep{sipila1990, backlin2011, niemi2013, vanhatalo2014, jounela2019, jounela2024}. As pups tend to favor shallow waters for foraging, and are generally curious and inexperienced, it is probable that they frequently engage with coastal fisheries and become caught in fishing gear \citep{Kelly2010, niemi2013, cronin2014}. Their relatively small body sizes may also make them more likely to enter smaller fishing gear, and make escaping more difficult once they are caught \citep{cronin2014}. The juxtaposition between the substantial juvenile bias in bycatch and the pronounced adult bias in protective hunting by fishermen may reflect the fact that young, inexperienced seals are inclined towards exploration, whereas experienced adults prioritize exploitation \citep{sjoberg2000}. On the other hand, larger seals may be more likely to detach from fishing nets, and fishermen often report the difficulty of lifting out larger seals from the water \citep{cronin2014}. Older seals may therefore be underrepresented in the samples compared to the actual bycatch. 

\section{Conclusions}

Amid the dynamic and uncertain conditions brought on by climate change, varying reproductive rates, and the re-introduction of seal hunting, our Bayesian IPM succeeded in providing a comprehensive assessment of the Baltic ringed seal population in the Bothnian Bay -- a task that has not been possible for over a decade. In addition to estimating population size and growth, we inferred key demographic parameters including survival probabilities, reproductive rates and the relative vulnerabilities of different demographic groups to hunting and bycatch. Moreover, our model revealed possible changes in ringed seal haul-out behavior in response to changing sea ice patterns. The mechanistic nature of our model additionally enabled near-term predictions of population response to changes in hunting quotas, informing management decisions. 

\bigskip

Wildlife populations across the globe are increasingly exposed to novel conditions that fundamentally alter their dynamics. The recent difficulties in monitoring and managing ringed seals in the Bothnian Bay may foreshadow similar challenges elsewhere.
Our study demonstrates the value of mechanistic IPMs for monitoring populations in uncertain and rapidly changing environments, testing ecological hypotheses regarding mechanisms of change, and supporting science-based management decisions. IPMs such as ours could be developed into digital twins of their target populations \citep{de2023, trantas2023}. Embedding these models and Bayesian inference software within a user-friendly interface could substantially streamline research and monitoring efforts, as well as hasten management responses. We believe that the modeling approach presented here will pave the way towards the embrace of IPMs as digital twins of wildlife populations, which could prove to be a critical component of the adaptive management toolkit in a rapidly changing world. Given the accelerating pace of anthropogenic change, we anticipate increasing adoption of IPMs in studies of wild animal populations.

\section*{Acknowledgements}

The present work is a spin-off from Murat Ersalman’s Master's thesis \citep{ersalman2024}. The work received funding from the Helsinki Institute of Life Science (HiLIFE) (ME), Research Council of Finland (grant 317255) and Jane \& Aatos Erkko Foundation (JV). In addition, JV acknowledges funding from the European Union (ERC Consolidator Grant BEFPREDICT, 101087409). The monitoring of ringed seals in the Bothnian Bay was conducted by the Swedish Museum of Natural History and funded by the Swedish Environmental Protection Agency and the Swedish Agency for Marine and Water Management. The Finnish monitoring of seals was funded by the Ministry of Agriculture and Forestry. We thank Yessenia Rojas, Jannikke Räikkönen, Petri Timonen and Charlotta Moraeus for help in age determinations, and Eva Kisdi for support in building the mathematical models. Finally, we are grateful to the two anonymous reviewers for their valuable insights and constructive feedback, which have greatly strengthened this manuscript.

\section*{Code and data availability}

Code and data are available on Zenodo at: https://doi.org/10.5281/zenodo.14243458

\section*{Conflicts of interest}

The authors declare no conflict of interest.

\bibliographystyle{inter_research}
\bibliography{ringed_seal_citations}

\begin{singlespace}
\begin{longtable}{ |p{2cm} p{14cm}|  }
\caption{\setstretch{1.0} Notation for commonly referenced parameters.}\label{table:params} \\
    \hline
    Symbol & Description \\
    \hline
    $n_{s,a,t}$ & Number of seals of sex $s$ and age $a$ at the start of year $t$ \\
    $N_{t}$ & Total population size at the start of year $t$ \\
    $\rho^{X}_{s,a,t}$ & Probability that a seal of sex $s$ and age $a$ transitions to mortality state $X$ \\
    $u^{(1)}_{s,a,X,t}$ & Number of seals of sex $s$ and age $a$ that transition to mortality state $X$ \\
    $u^{(2)}_{s,a,t}$ & Number of seals of sex $s$ and age $a$ at the end of year $t$ \\
    $H^{i}_{t}$ & Harvest totals of country $i$ in year $t$ \\
    $\mu_{s,a}$ & Mortality rate of seals of sex $s$ and age $a$ \\
    $\phi_{s,a}$ & Survival probability of seals of sex $s$ and age $a$ in the absence of hunting \\
    $\omega_{s,a,t}$ & Haul out probability of seals of sex $s$ and age $a$ in year $t$ \\
    $\hat{\omega}_{s,a}$ & Haul out probability of seals of sex $s$ and age $a$ when sea ice is abundant \\
    $C_{t}$ & Ice extent in the Bothnian Bay in late April \\
    $b_{t}$ & Probability that an adult female gives birth at the start of year $t$ \\
    \hline
\end{longtable}
\end{singlespace}

\setcounter{figure}{0}

\begin{figure}[H]
\begin{center}
 \includegraphics[width=0.9\linewidth]{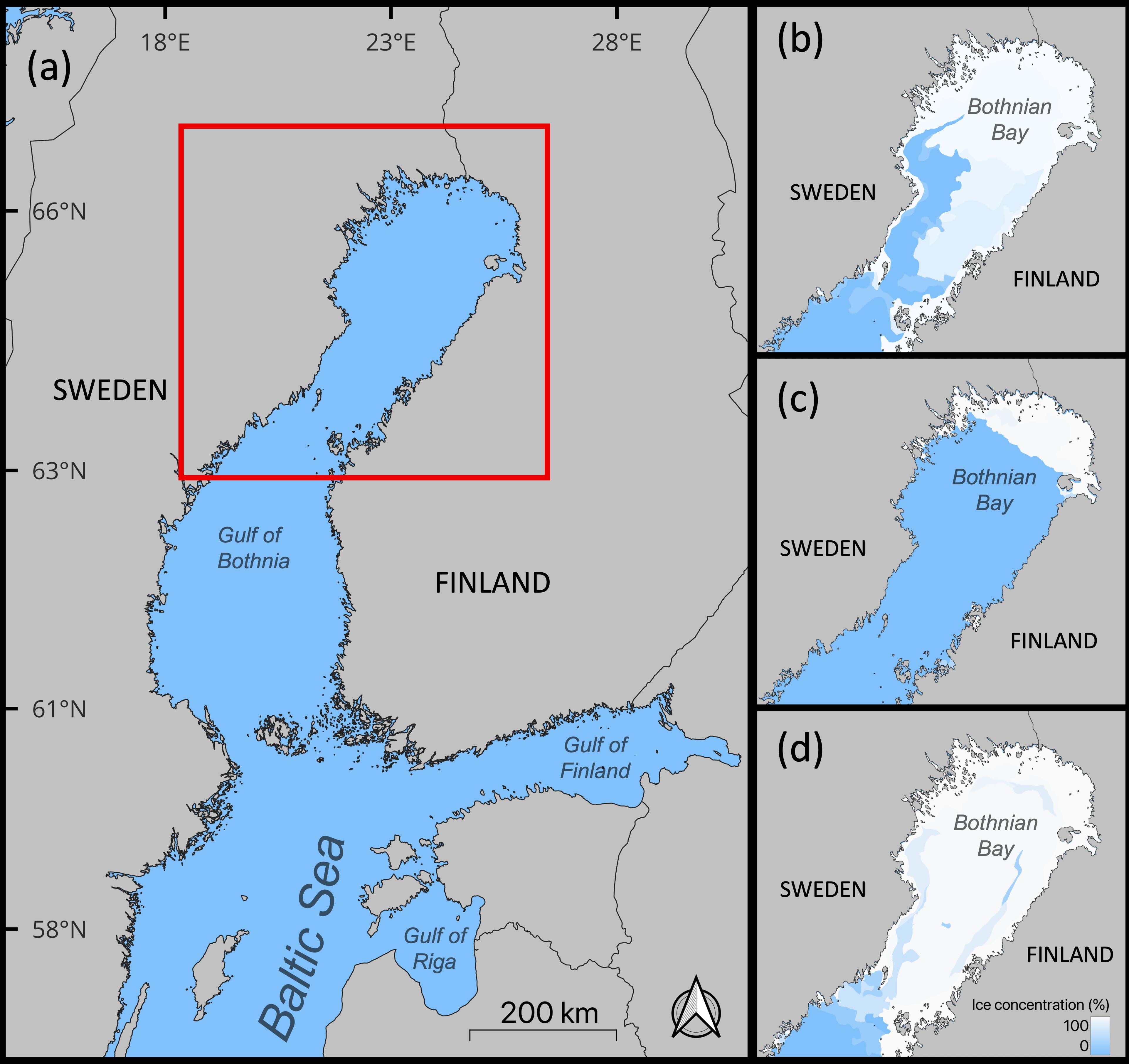}
 \caption{\setstretch{1.0} A map of the study area (a) and the extent of ice cover in 2004 (b), 2015 (c) and 2018 (d) around the third week of April. These three years respectively illustrate the median, minimum and maximum ice extents observed during the study period.}
 \label{fig:study_area}
\end{center}
\end{figure}

\begin{figure}[p!]
\begin{center}
 \includegraphics[width=\linewidth]{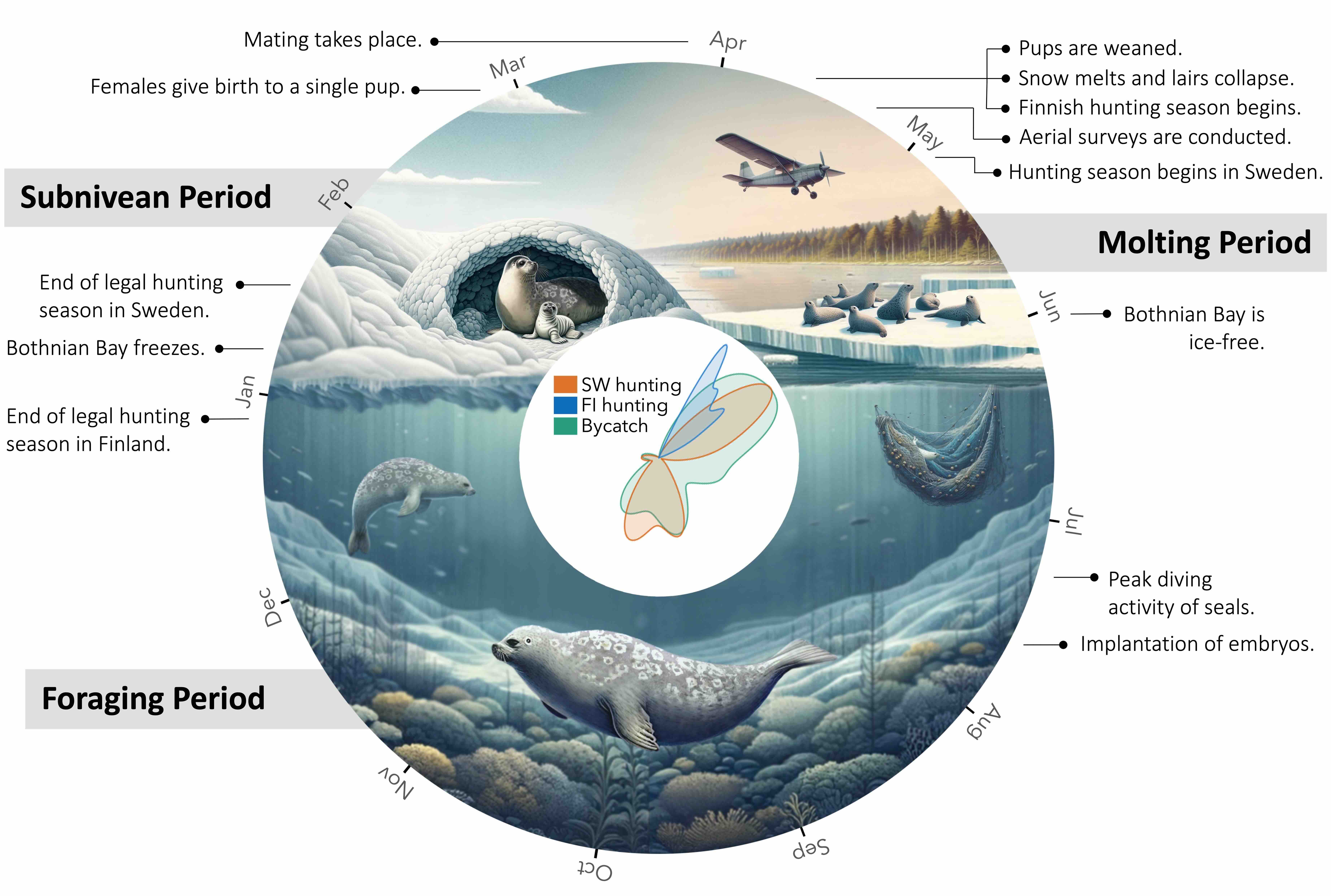}
 \caption{\setstretch{1.0} Illustration of the Baltic ringed seal annual life-cycle. The density plots at the center show, in polar coordinates,  the temporal distribution of hunting and bycatch based on sampled seals. (The illustration was created with the aid of DALL-E 3.)}
 \label{fig:lifecycle}
\end{center}
\end{figure}

\begin{figure}[p!]
\begin{center}
 \includegraphics[width=\linewidth]{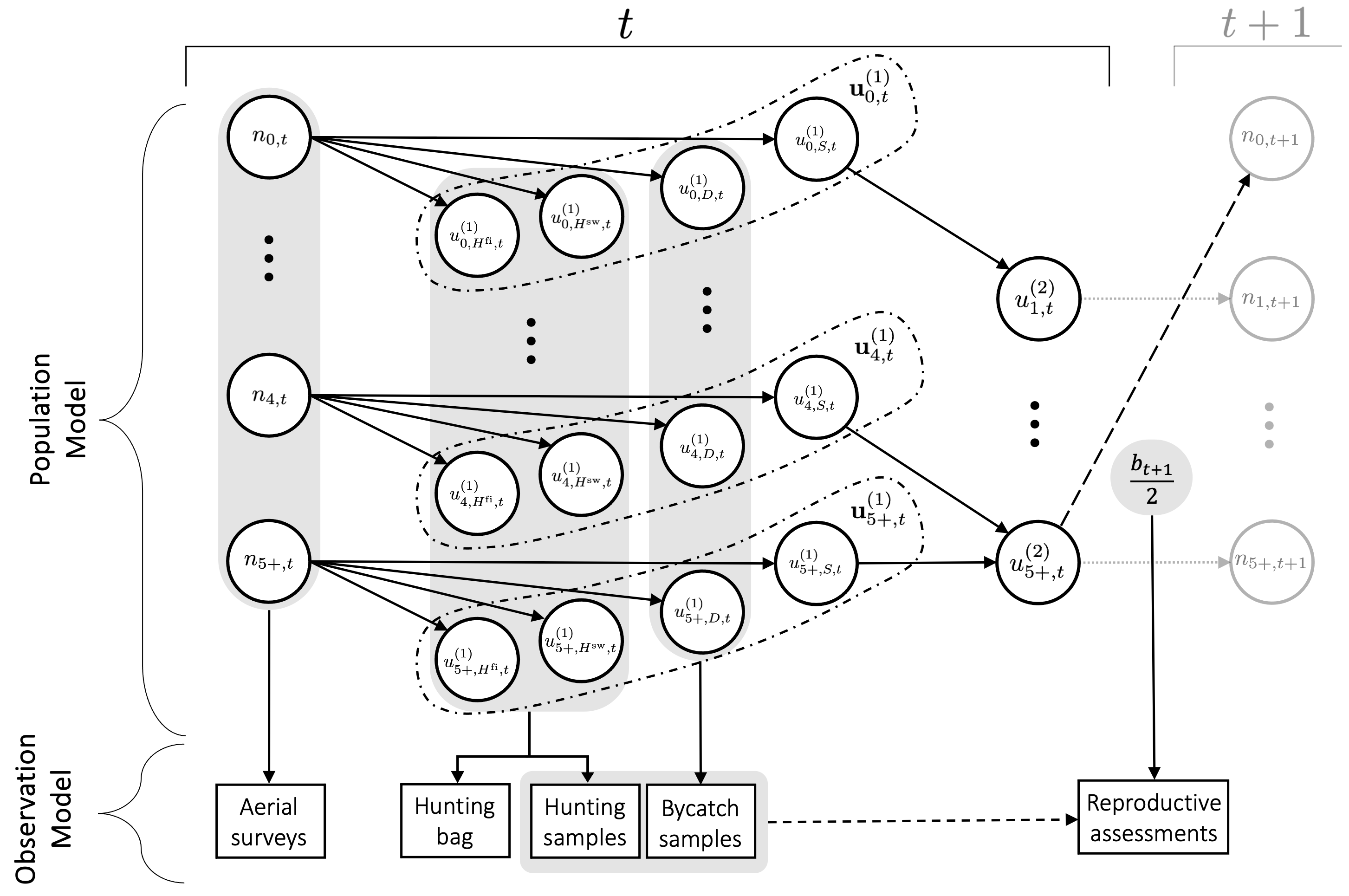}
 \caption{\setstretch{1.0} Schematic diagram of the model structure for female seals. Swedish hunting is further split into spring and fall hunting.}
 \label{fig:model_structure}
\end{center}
\end{figure}

\begin{figure}[p!]
\centering
\captionsetup[subfigure]{position=top,justification=raggedright,singlelinecheck=false}
\subfloat[][]{
    \includegraphics[width=0.49\linewidth]{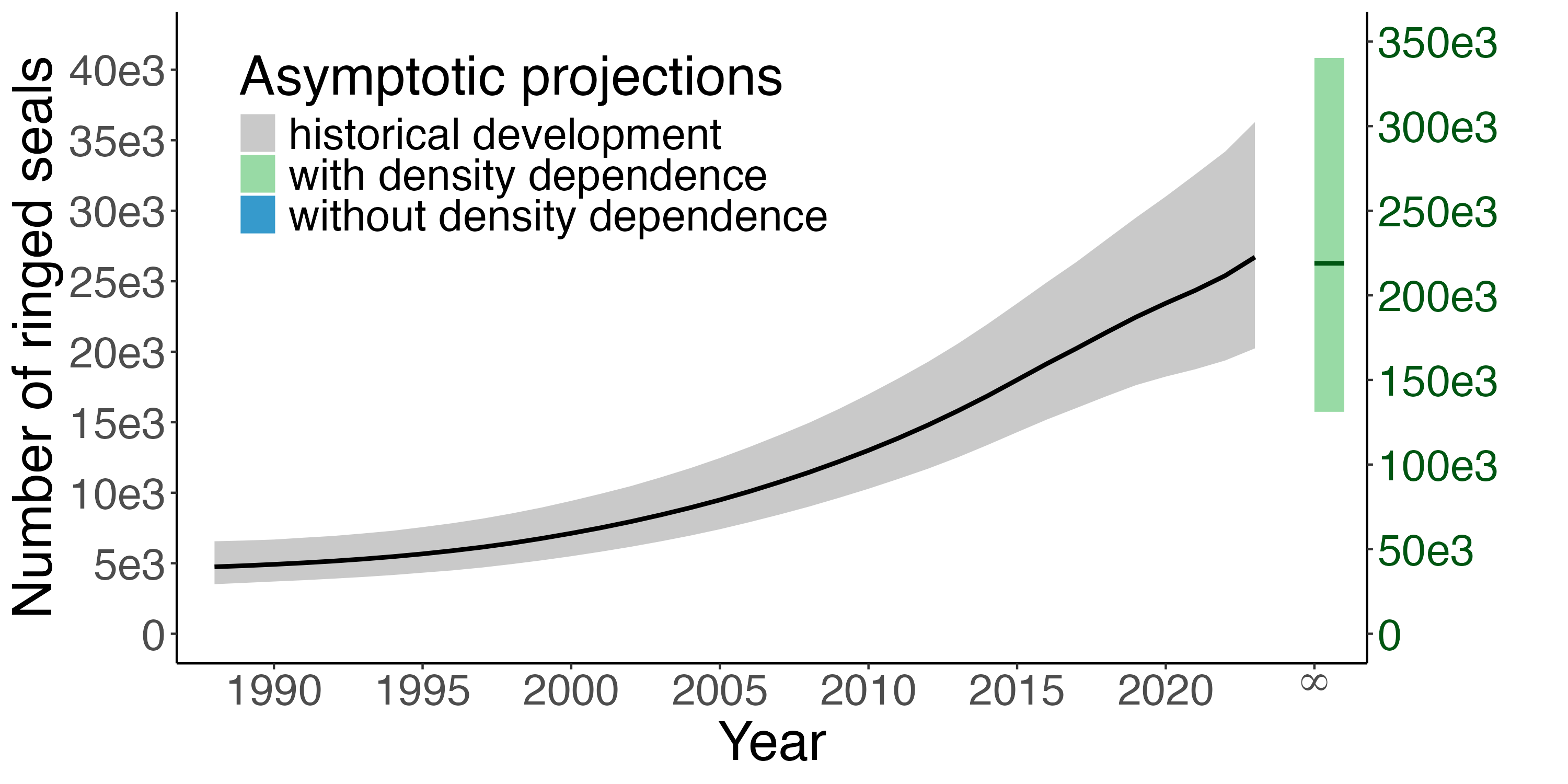}\label{fig:pop_size}}
\subfloat[][]{
    \includegraphics[width=0.49\linewidth]{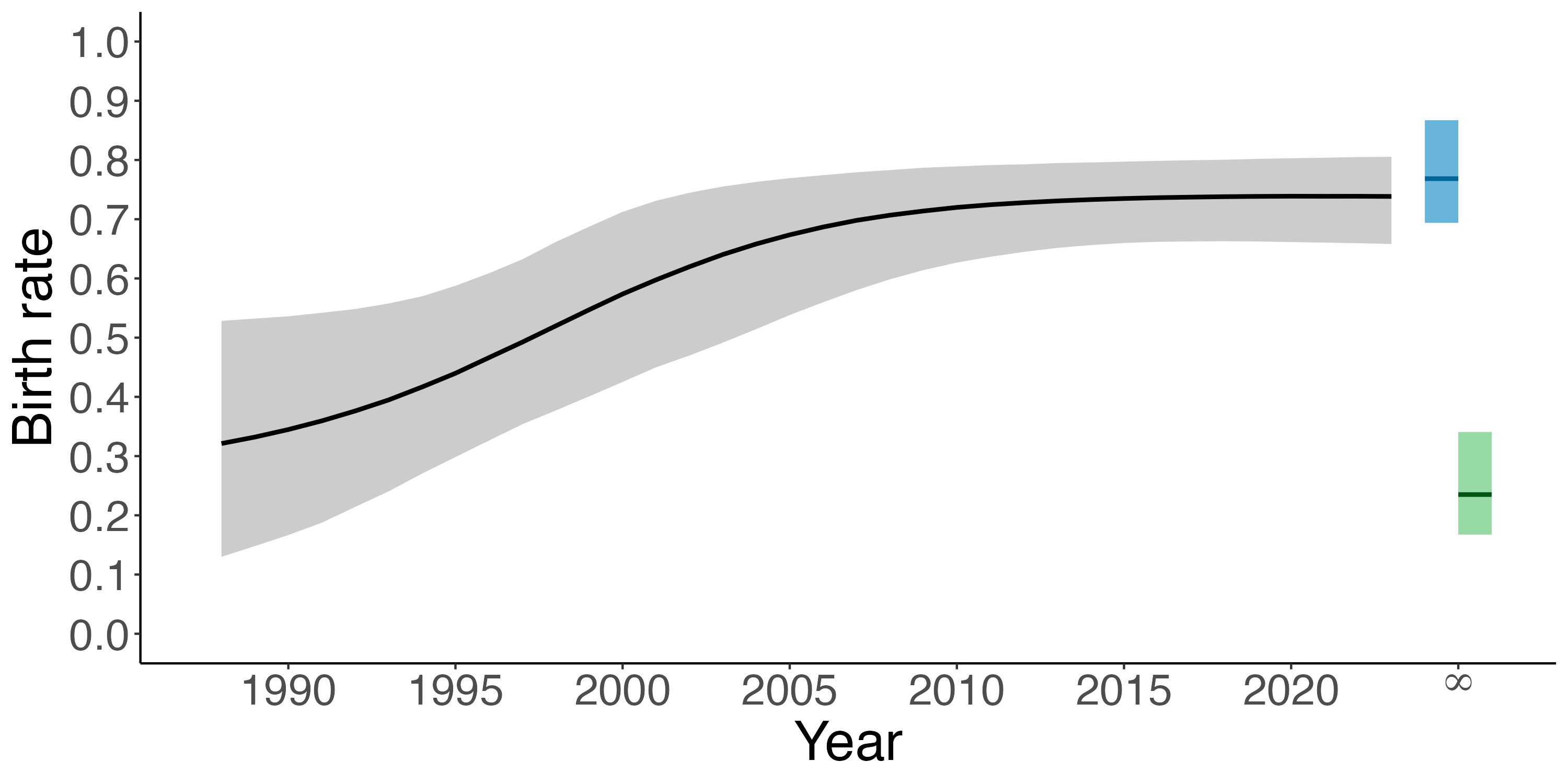}\label{fig:birth_rate}} \\
\subfloat[][]{
    \includegraphics[width=0.49\linewidth]{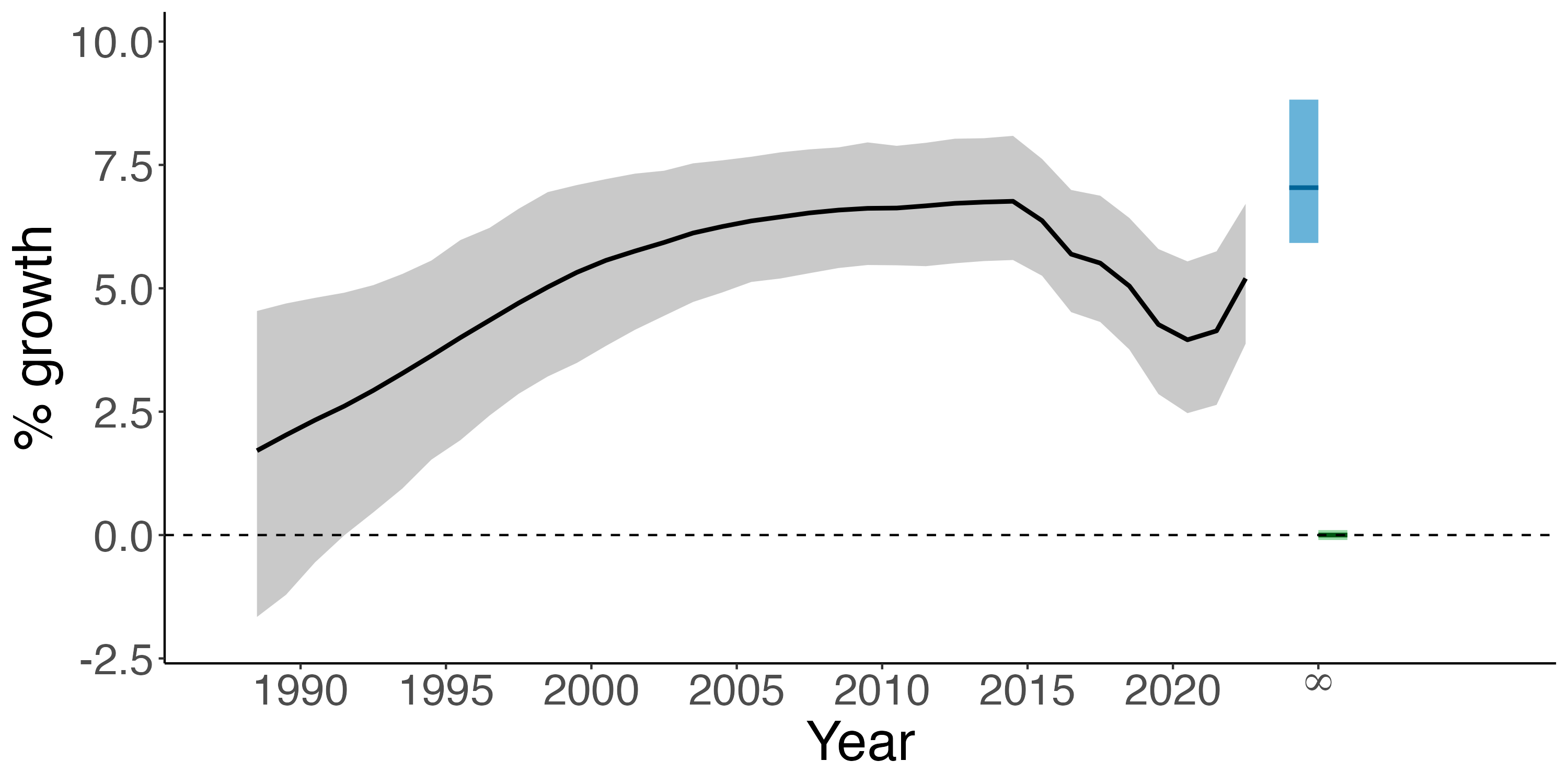}\label{fig:growth_rate}}
\subfloat[][]{
    \includegraphics[width=0.49\linewidth]{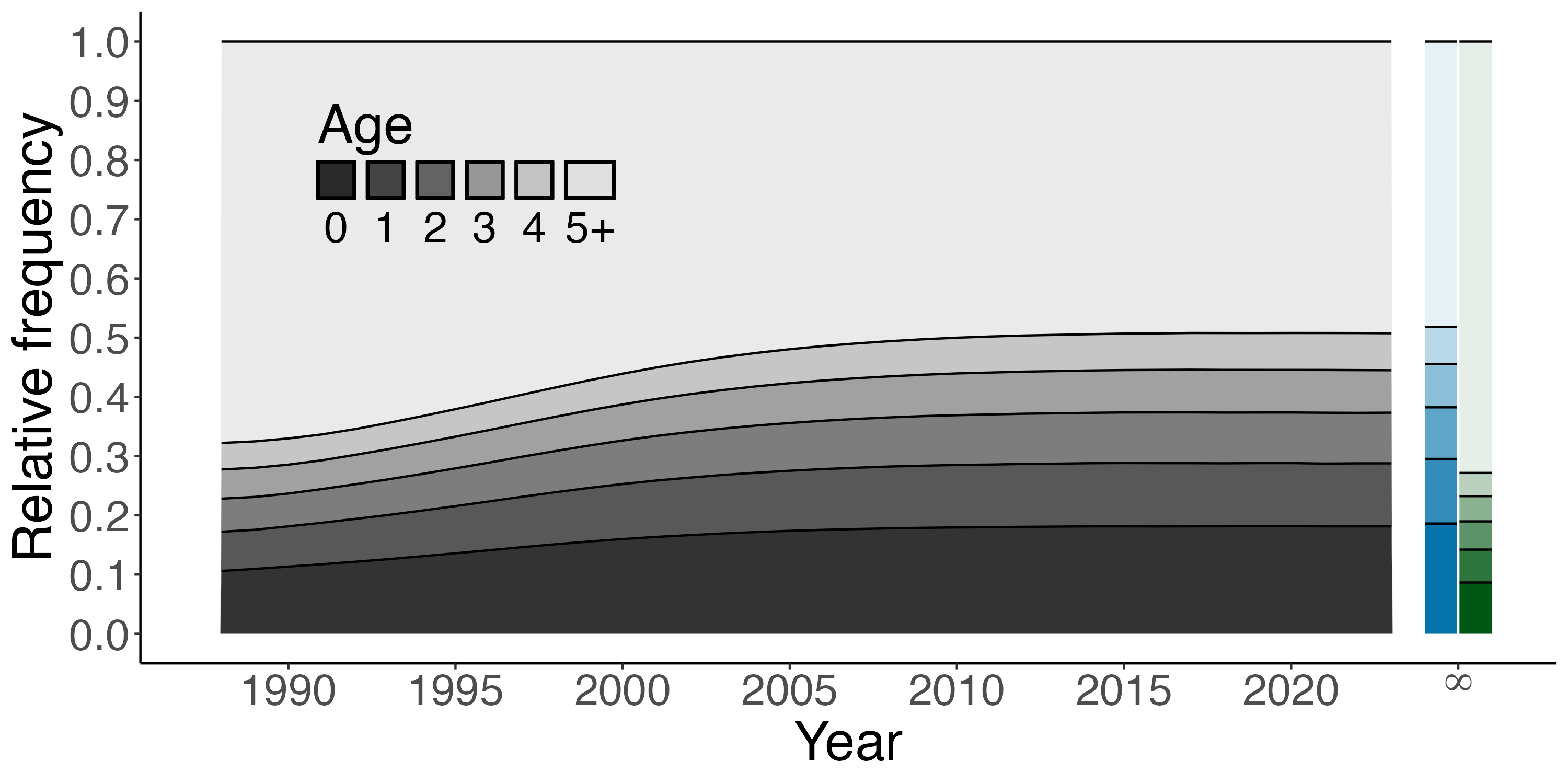}\label{fig:age_str}}
\caption{\setstretch{1.0} Estimated \protect\subref{fig:pop_size} population size, \protect\subref{fig:birth_rate} birth rate, \protect\subref{fig:growth_rate} population growth rate and \protect\subref{fig:age_str} age structure for ringed seals in the Bothnian Bay. The rightmost segments show the expected asymptotic values for an unexploited population with and without density dependent effects.}
\label{fig:demography}
\end{figure}

\begin{figure}[p!]
\centering
\captionsetup[subfigure]{position=top,justification=raggedright,singlelinecheck=false}
\subfloat[][]{
    \includegraphics[width=0.49\linewidth]{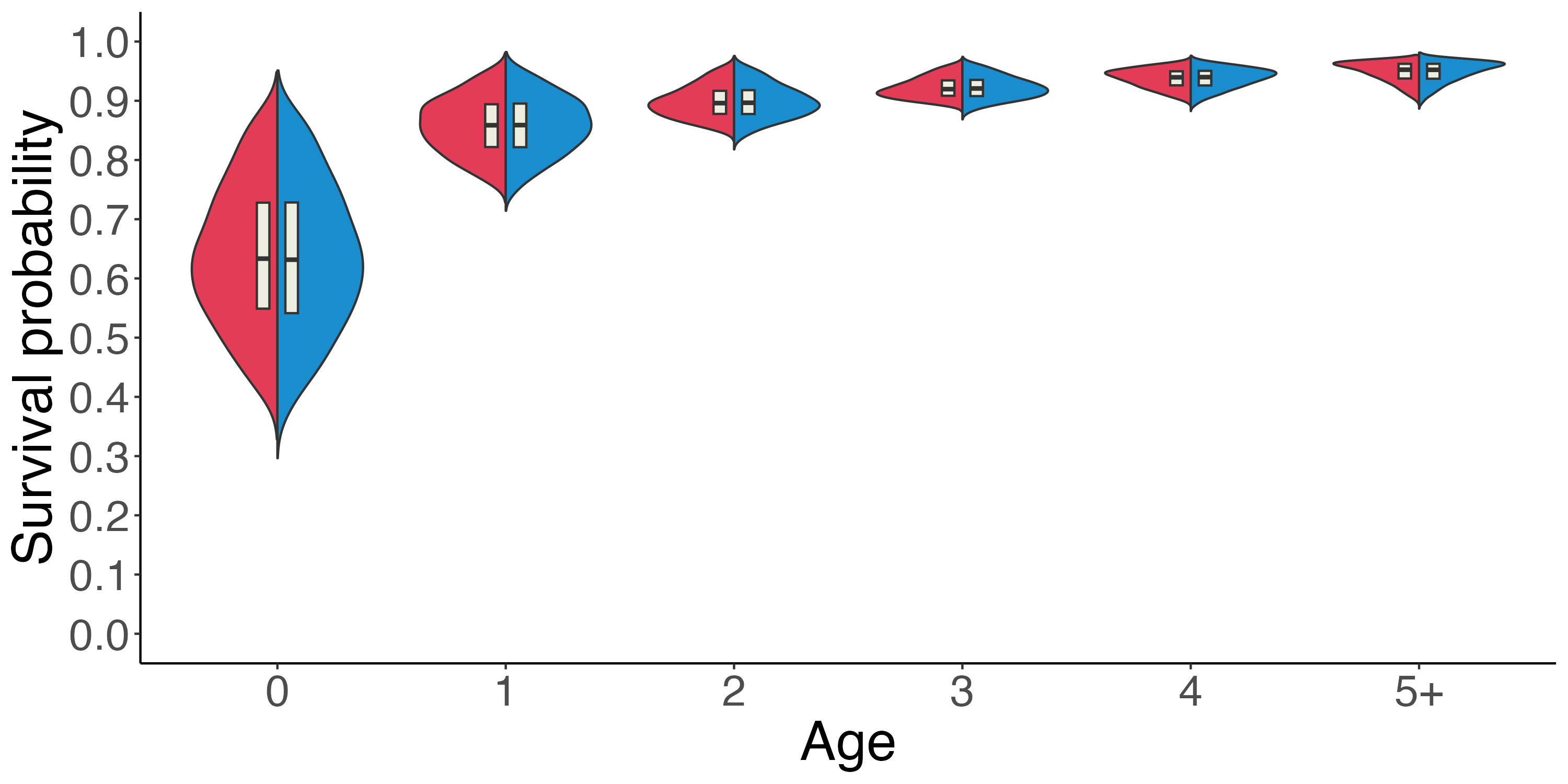}
    \label{fig:survival_probs}}
\subfloat[][]{
    \includegraphics[width=0.49\linewidth]{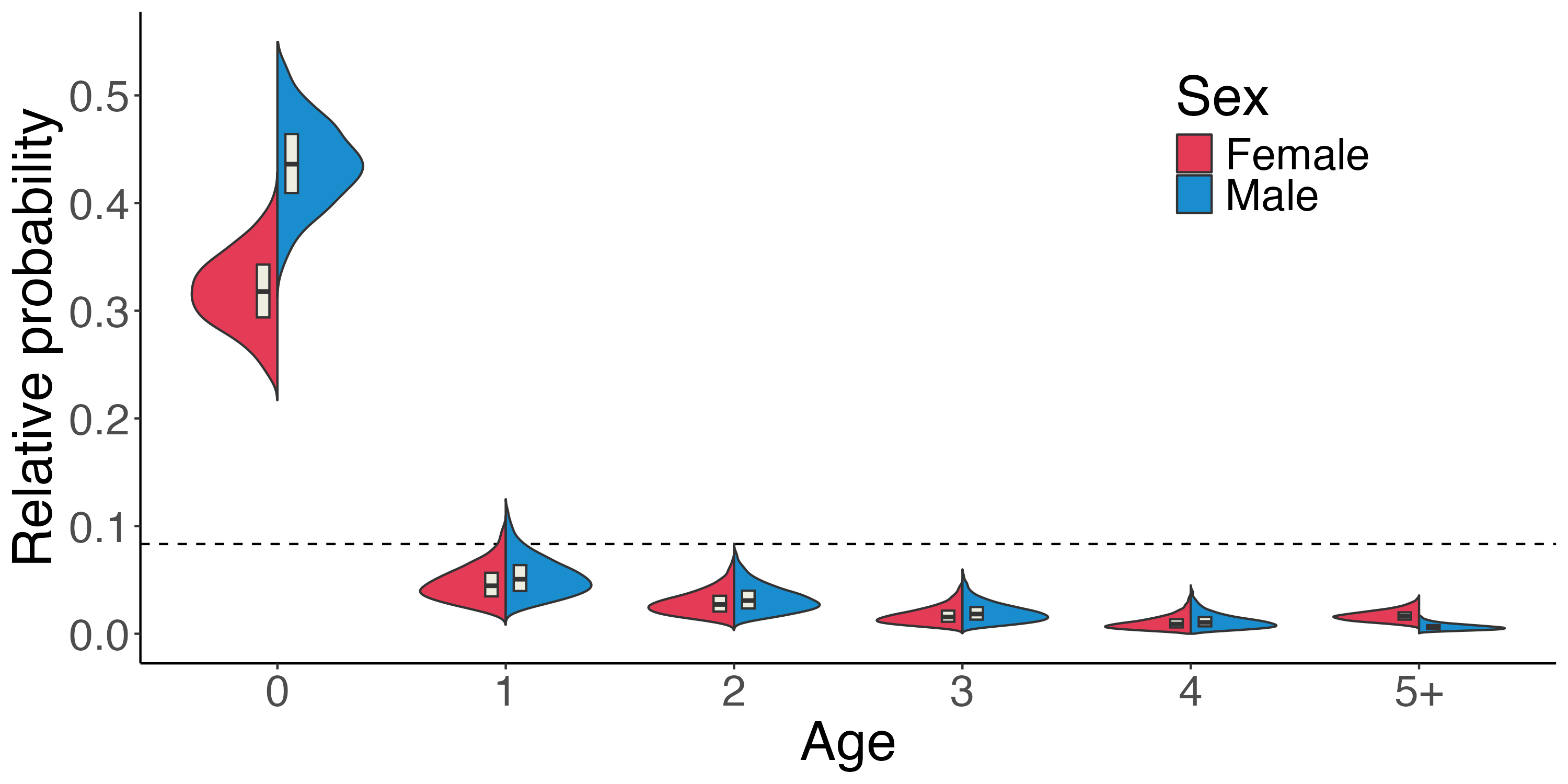}
    \label{fig:bycatch_bias}} \\
\subfloat[][]{
    \includegraphics[width=0.33\linewidth]{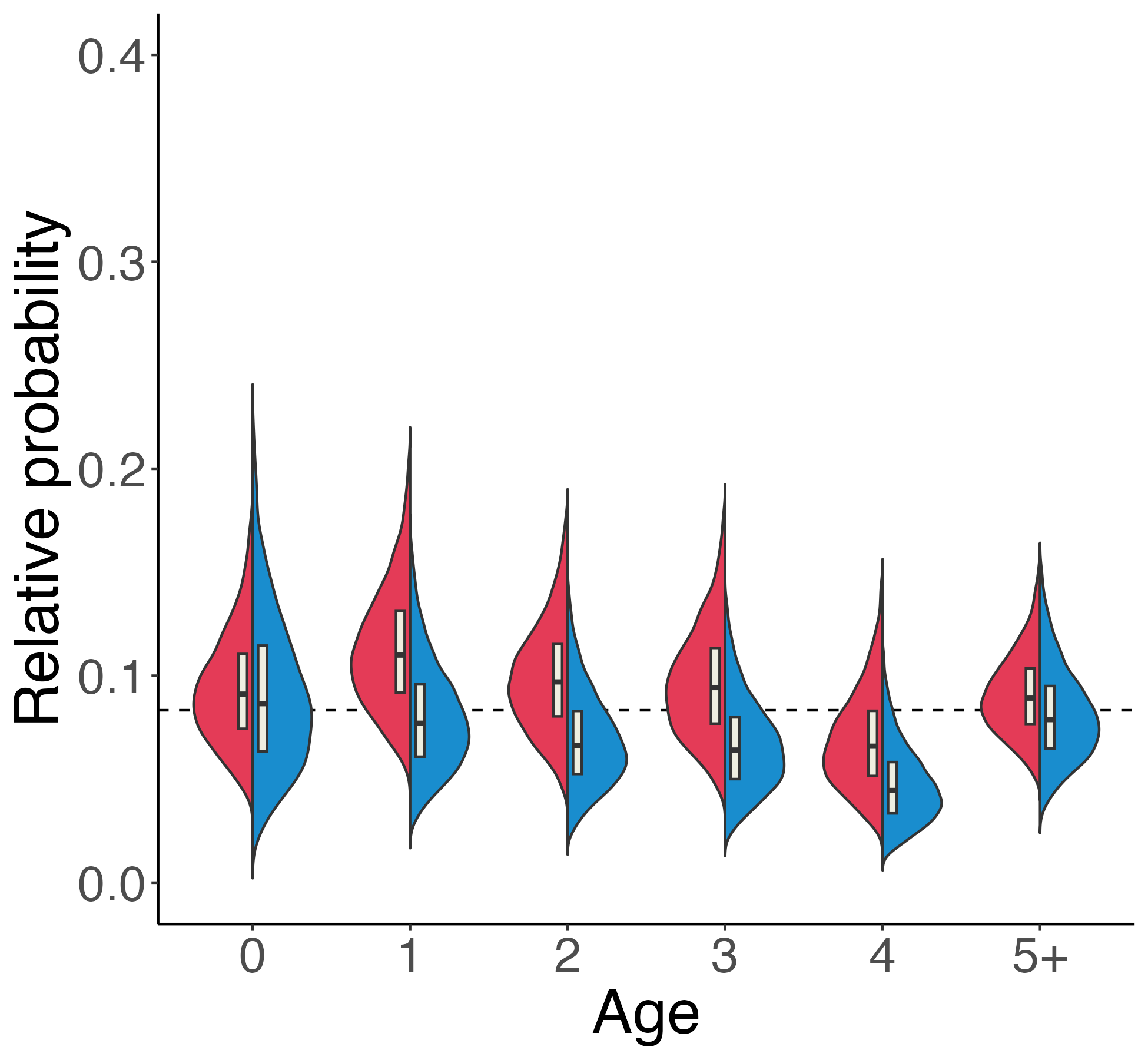}
    \label{fig:sw_bias_spring}}
\subfloat[][]{
    \includegraphics[width=0.33\linewidth]{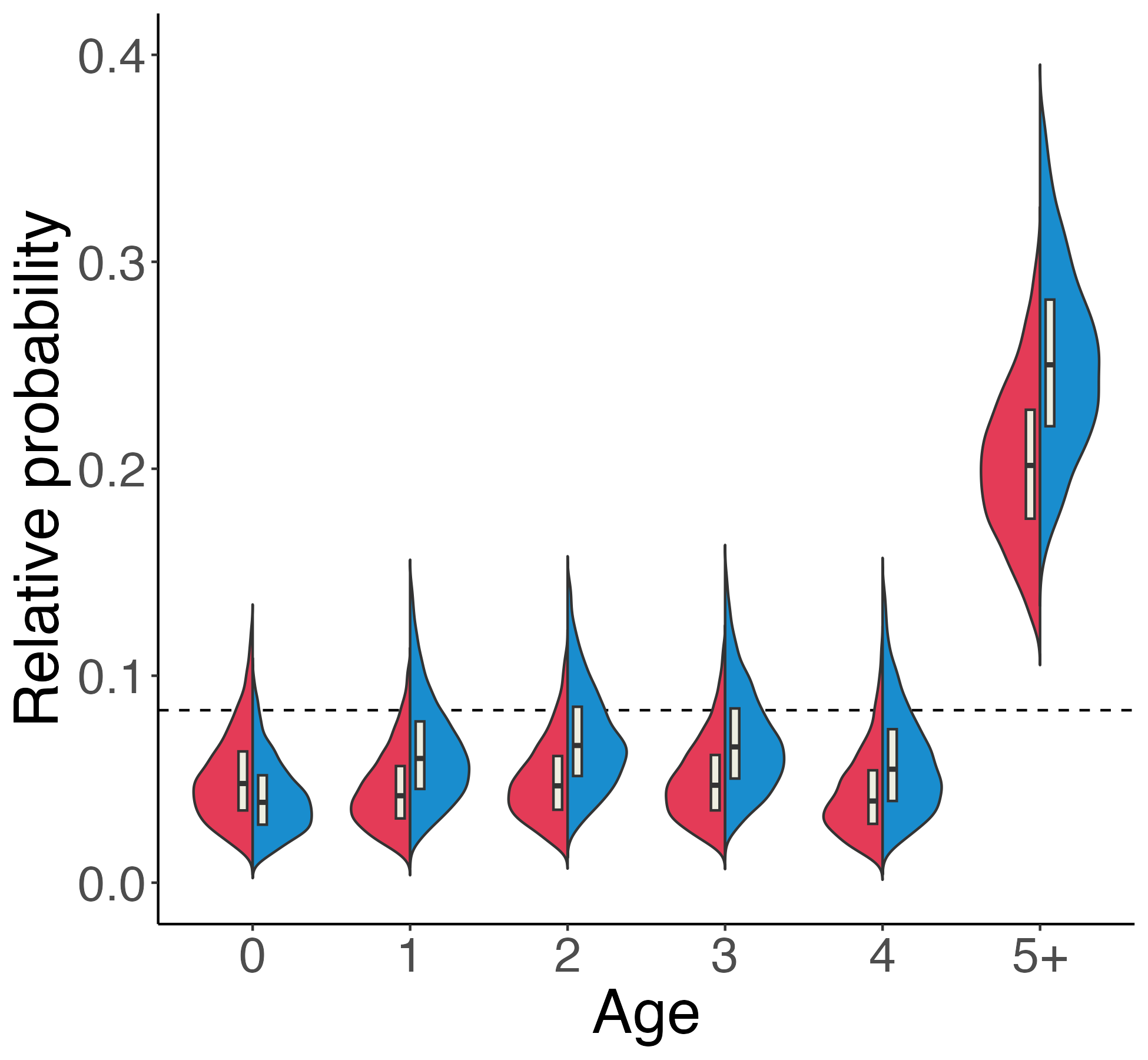}
    \label{fig:sw_bias_fall}}
\subfloat[][]{
    \includegraphics[width=0.33\linewidth]{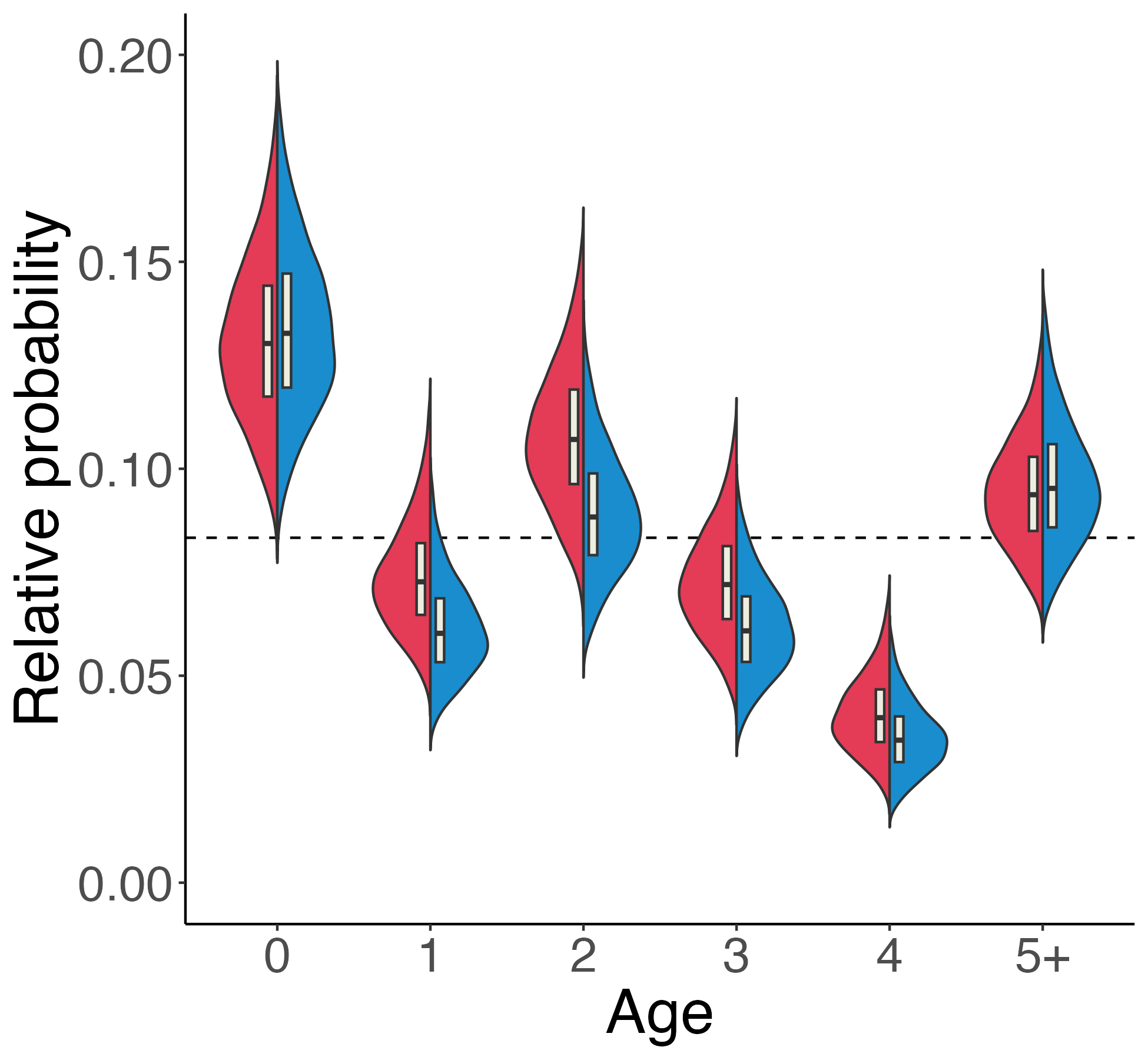}
    \label{fig:fi_bias}}
\caption{\setstretch{1.0} \protect\subref{fig:survival_probs} Probabilities of surviving natural mortality (including bycatch), and the relative probabilities that seals are \protect\subref{fig:bycatch_bias} bycaught, \protect\subref{fig:sw_bias_spring} hunted in Sweden during the spring, \protect\subref{fig:sw_bias_fall} hunted in Sweden during the fall, and \protect\subref{fig:fi_bias} hunted in Finland. The dashed lines in \protect\subref{fig:bycatch_bias}-\protect\subref{fig:fi_bias} show the relative mortality probabilities if all individuals were equally likely to be hunted or bycaught. Calculation of the relative mortality probabilities are presented in \ref{app:hunting_probability}.}
\label{fig:mortality}
\end{figure}

\begin{figure}[p!]
\centering
\captionsetup[subfigure]{position=top,justification=raggedright,singlelinecheck=false}
\subfloat[][]{
    \includegraphics[width=0.99\linewidth]{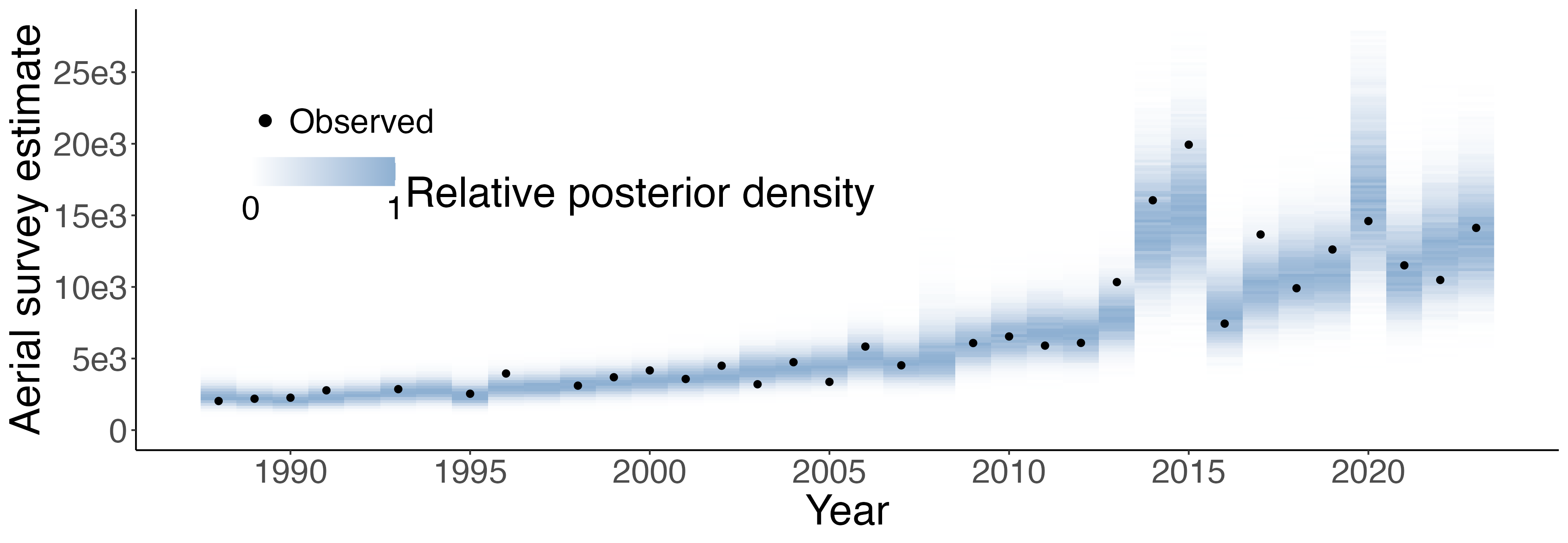}
    \label{fig:aerial_survey}} \\[-3.75ex]
\subfloat[][]{
    \includegraphics[width=0.99\linewidth]{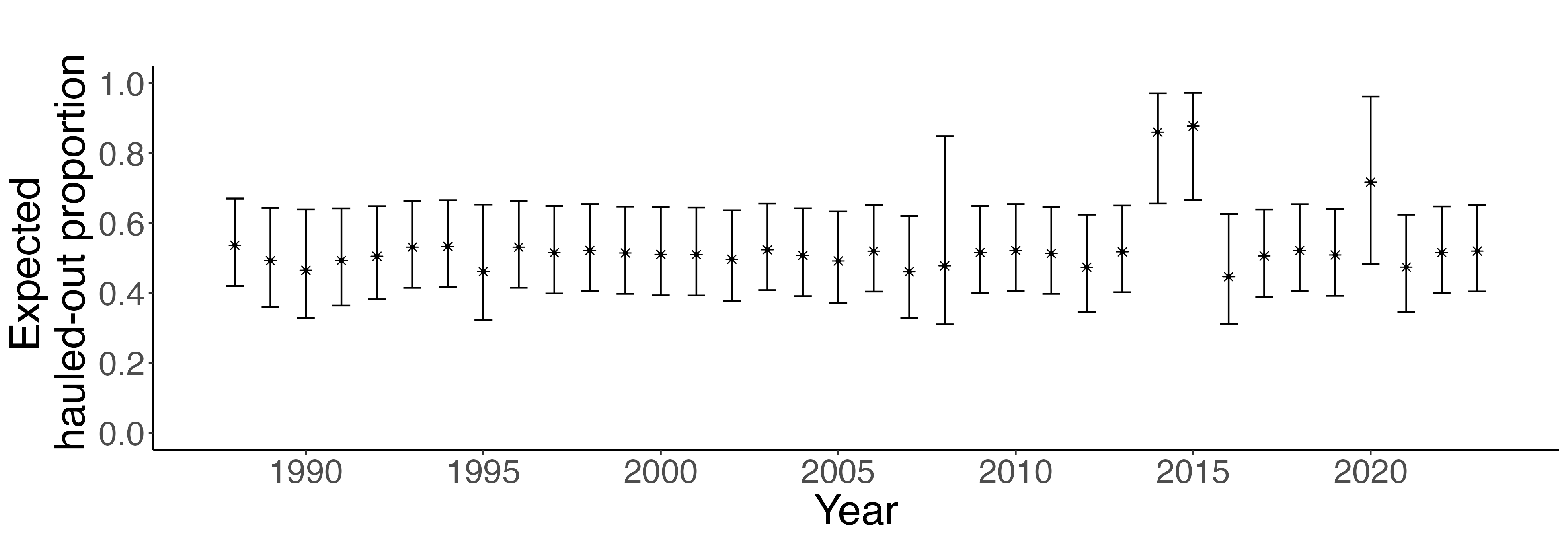}
    \label{fig:haulout_yr}} \\[-3.5ex]
\subfloat[][]{
    \includegraphics[width=\linewidth]{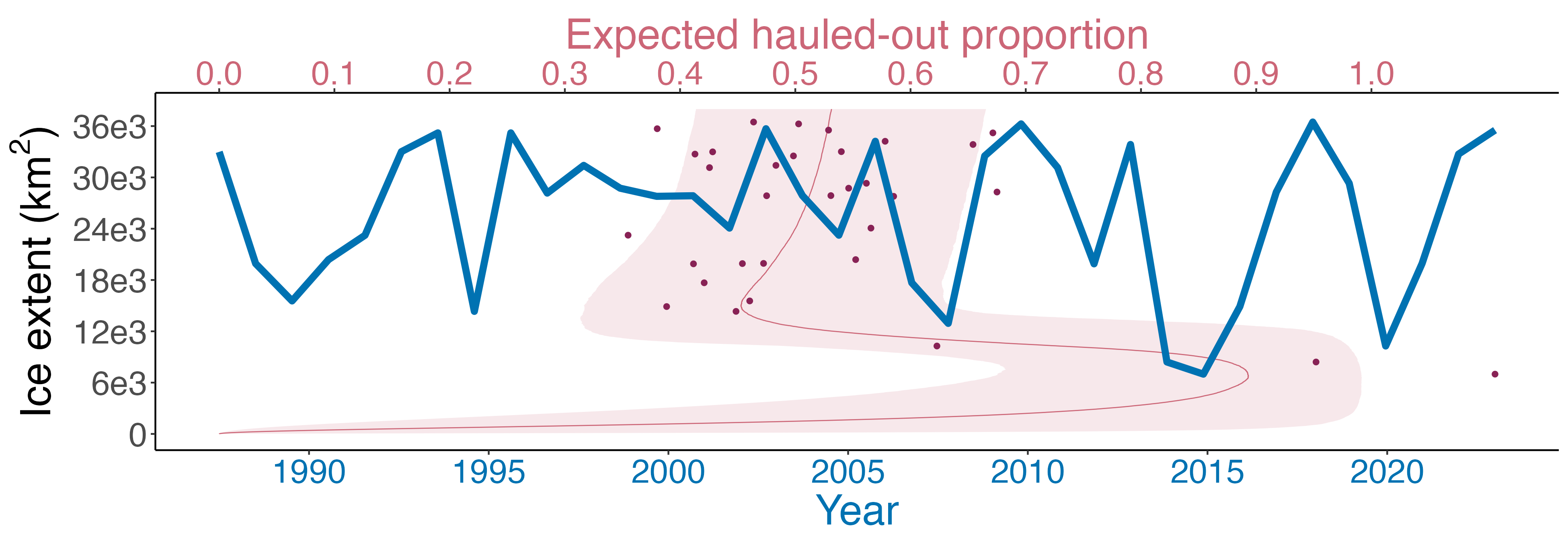}
    \label{fig:haulout_ice}}
\caption{\setstretch{1.0} \protect\subref{fig:aerial_survey} Posterior predictive distribution for aerial survey estimates. \protect\subref{fig:haulout_yr} Expected proportion of seals hauled-out in each year. \protect\subref{fig:haulout_ice} Annual variation in the ice covered area of the Bothnian Bay around the 3rd week of April (bottom and left axes) and the expected proportion of seals hauled-out during aerial surveys as a function of ice covered area (top and left axes). The points depict the median ratio of observed aerial survey estimates to the posterior estimates of total population size. Note that the overall proportion of seals hauled-out depends on the population age structure, and figures \protect\subref{fig:haulout_yr}-\protect\subref{fig:haulout_ice} are based on the most recent estimate of the age structure.}
\label{fig:haulout}
\end{figure}

\begin{figure}[p!]
\centering
\captionsetup[subfigure]{position=top,justification=raggedright,singlelinecheck=false}
\subfloat[][]{
    \includegraphics[width=0.49\linewidth]{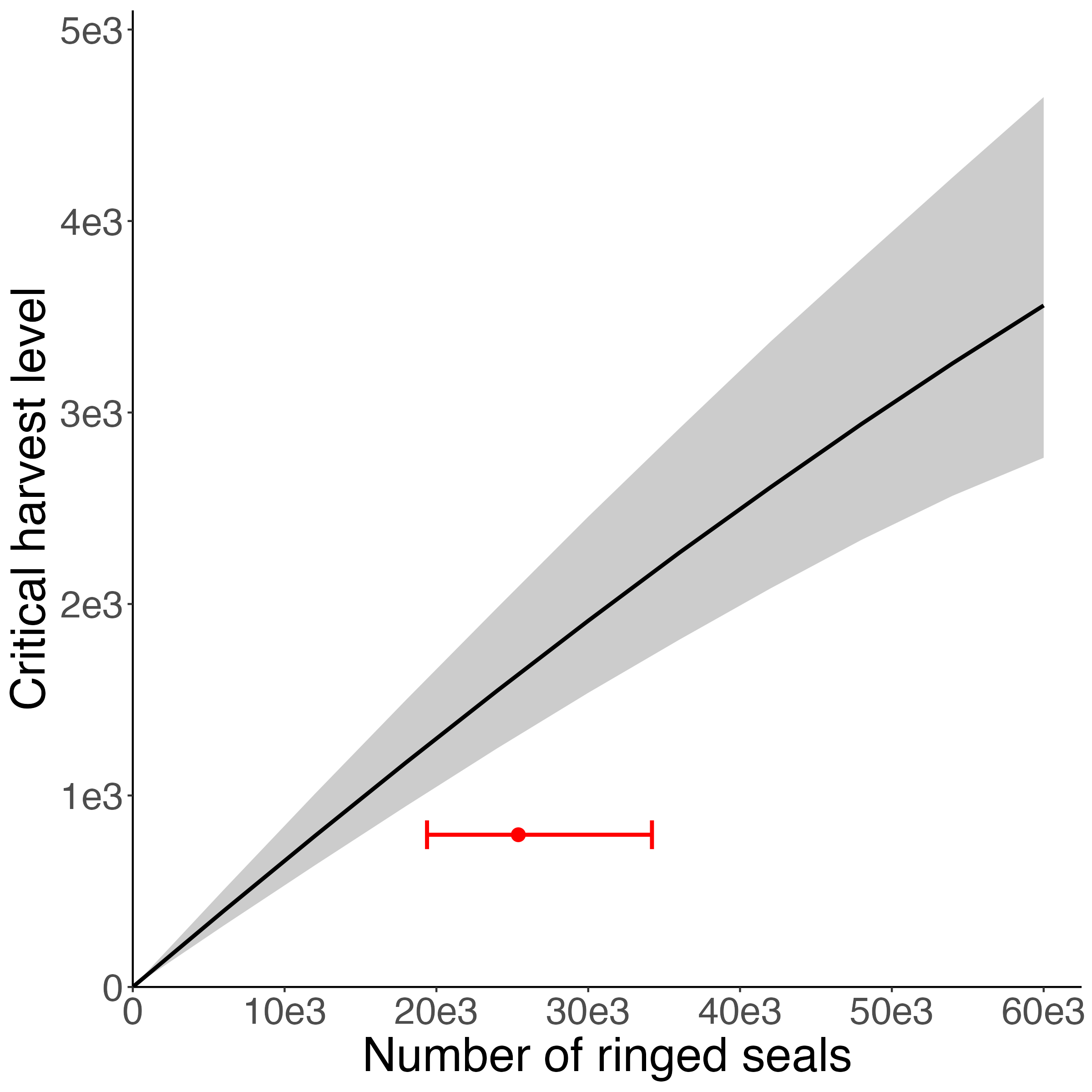}
    \label{fig:critical_hunting}}
\subfloat[][]{
    \includegraphics[width=0.49\linewidth]{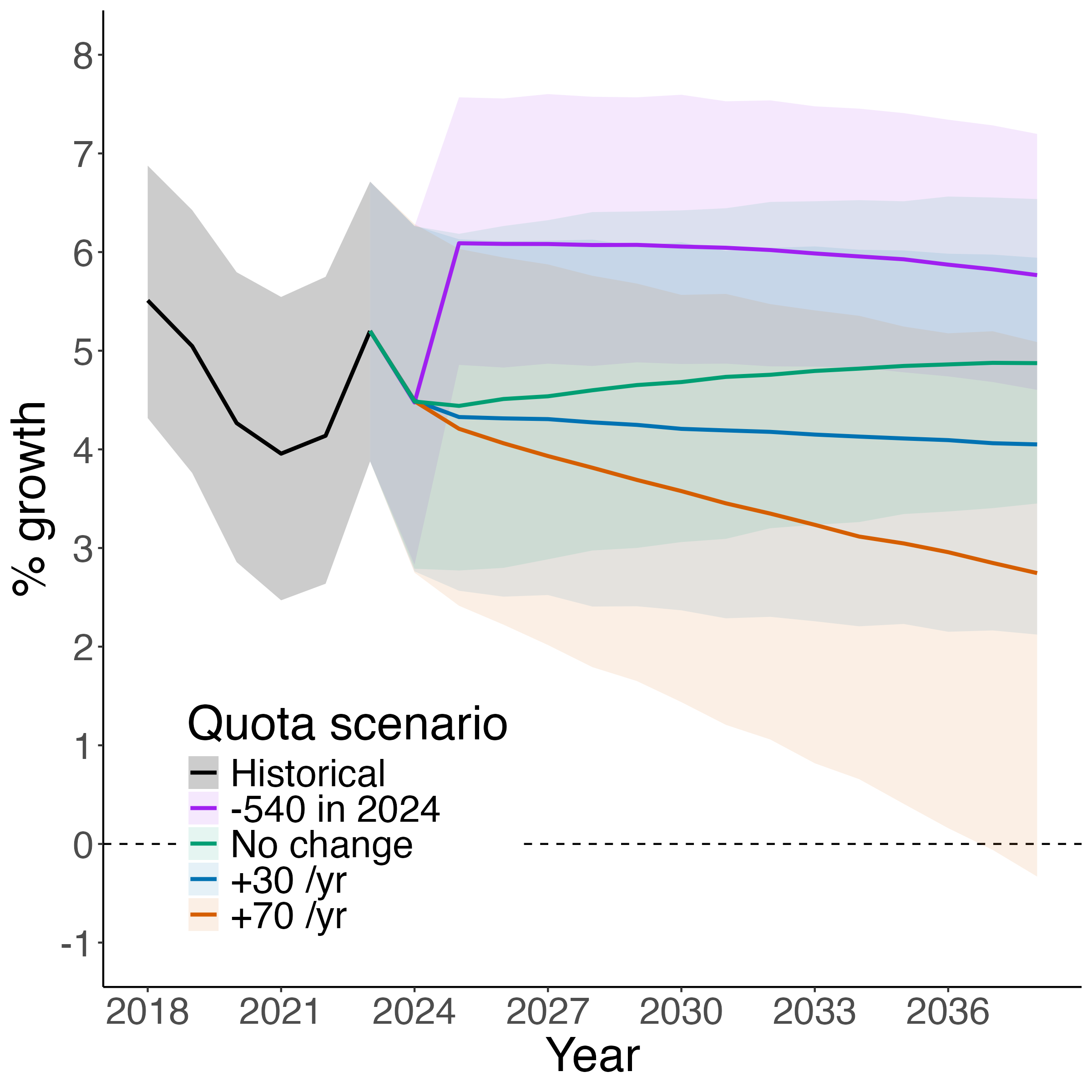}
    \label{fig:future_preds}}   
\caption{\setstretch{1.0} \protect\subref{fig:critical_hunting} Critical harvest level for which the population growth rate is expected to be zero, as a function of total population size. The critical harvest level is calculated by setting the net reproduction rate equal to one. The red error bar shows the estimated population size and hunting quota (Finland and Sweden combined) in 2022. \protect\subref{fig:future_preds} Posterior predictions for future population growth rates under four different hunting quota scenarios: (1) quotas maintained at their current level of 350 and 375 seals in Sweden and Finland respectively, (2) quotas increased by a total of 30 licenses per year (15 in both Sweden and Finland), (3) quotas increased by a total of 70 licenses per year (35 in both Sweden and Finland) and (4) quotas reduced by a total of 540 in 2024 (270 in both Sweden and Finland) and kept constant thereafter.}
\label{fig:hunting}
\end{figure}

\end{flushleft}

\newpage

\begin{appendices}
\nolinenumbers

\renewcommand{\thesection}{S\arabic{section}}
\renewcommand{\appendixname}{Appendix}

\renewcommand{\thefigure}{S\arabic{figure}}
\renewcommand{\thetable}{S\arabic{table}}
\renewcommand{\theequation}{S\arabic{equation}}
\renewcommand{\thesection}{Supplement \arabic{section}}
\renewcommand{\thesubsection}{Section S\arabic{subsection}:}

\makeatletter
\@addtoreset{figure}{section}
\@addtoreset{figure}{subsection}
\@addtoreset{table}{section}
\@addtoreset{table}{subsection}
\@addtoreset{equation}{section}
\@addtoreset{equation}{subsection}
\makeatother

\setcounter{page}{1}

\section{Model assessment}\label{app:post_pred_checks}

\begin{figure}[H]
\begin{center}
 \includegraphics[width=0.9\linewidth]{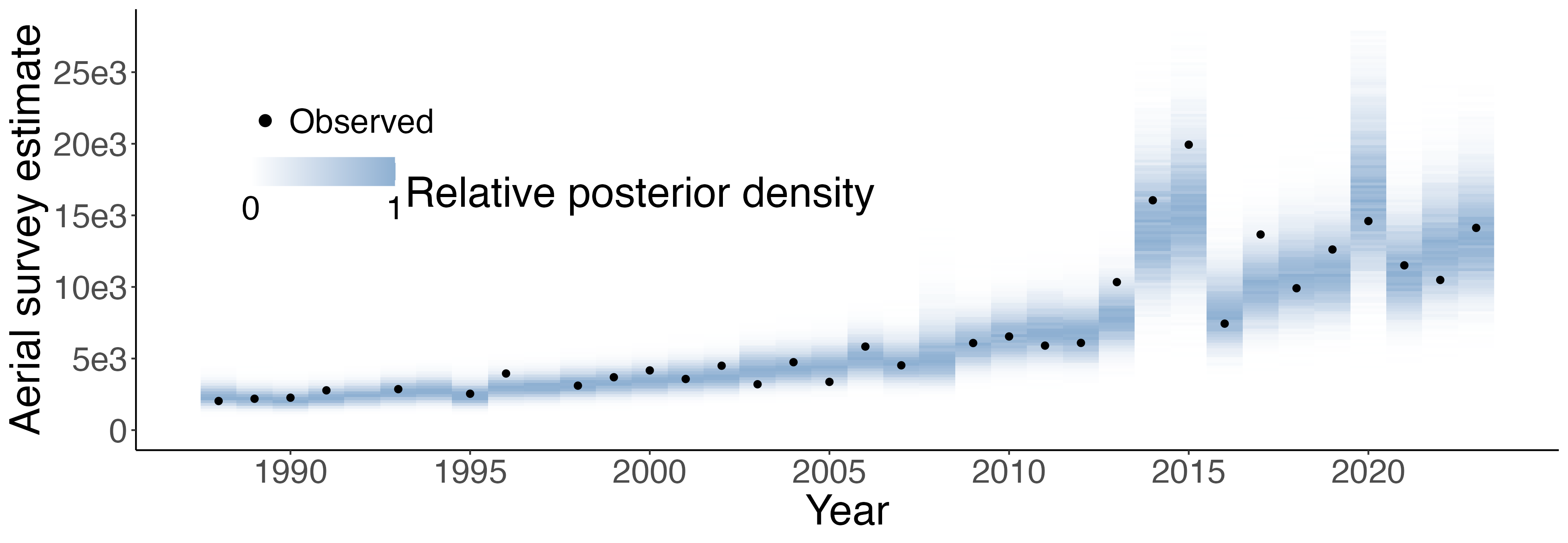}
 \caption{\setstretch{1.0} Posterior predictive check for aerial survey data.}
 \label{fig:survey_pp}
\end{center}
\end{figure}

\begin{figure}[H]
\centering
\captionsetup[subfigure]{position=top,justification=raggedright,singlelinecheck=false}
\subfloat[][]{
    \includegraphics[width=0.28\linewidth]{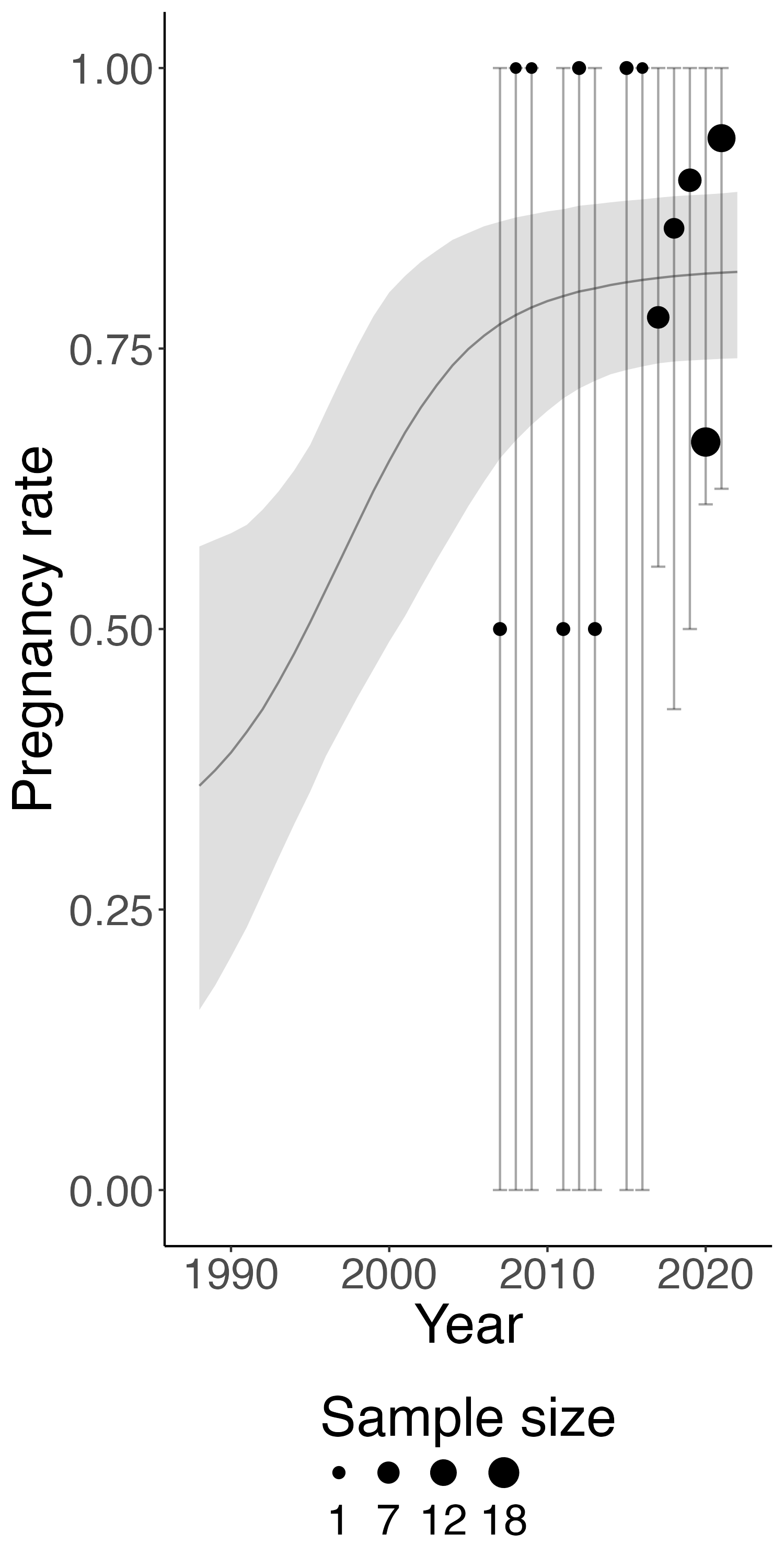}
    \label{fig:preg_rate}}
\subfloat[][]{
    \includegraphics[width=0.28\linewidth]{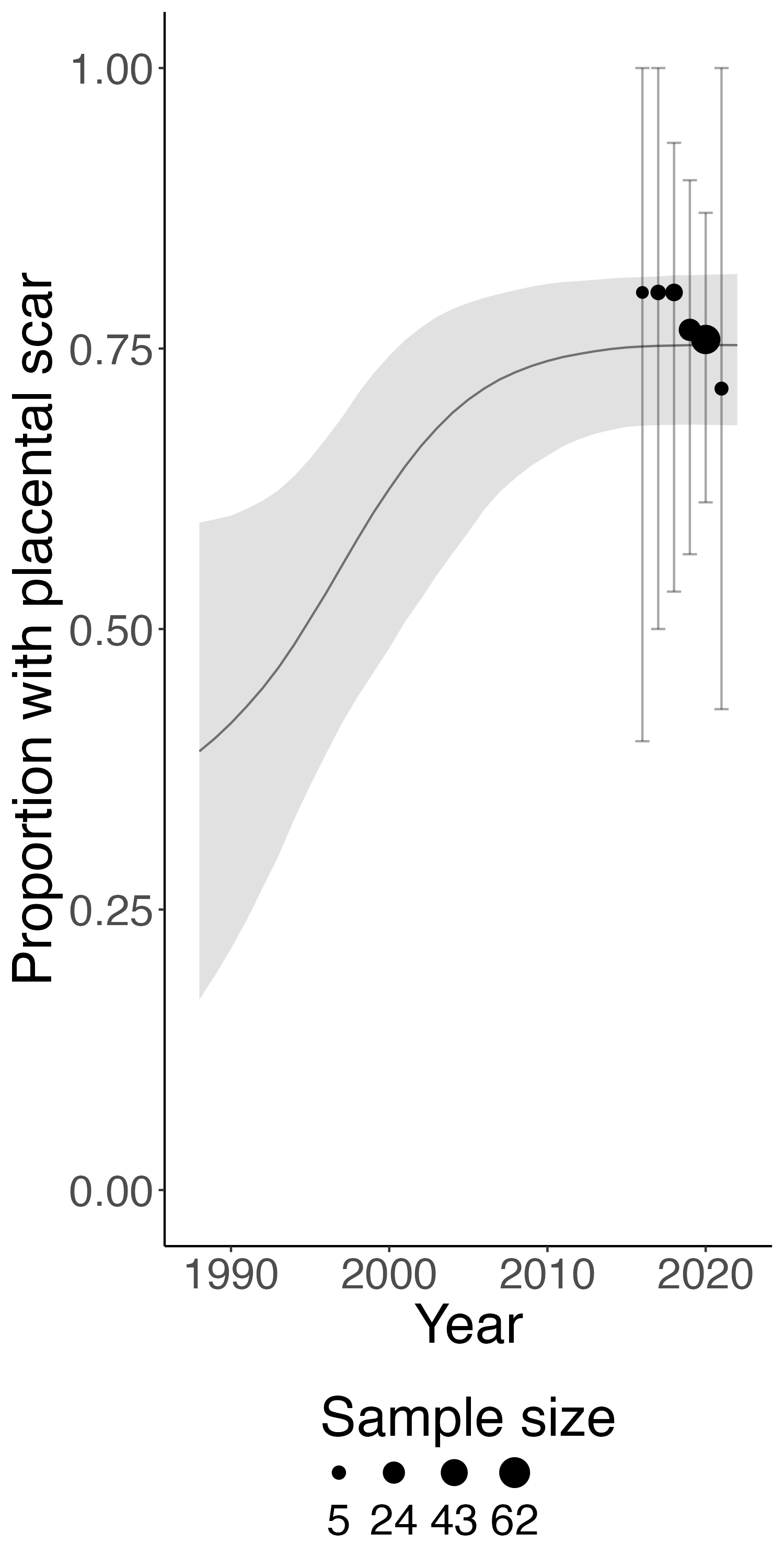}
    \label{fig:scar_rate}}
\subfloat[][]{
    \includegraphics[width=0.28\linewidth]{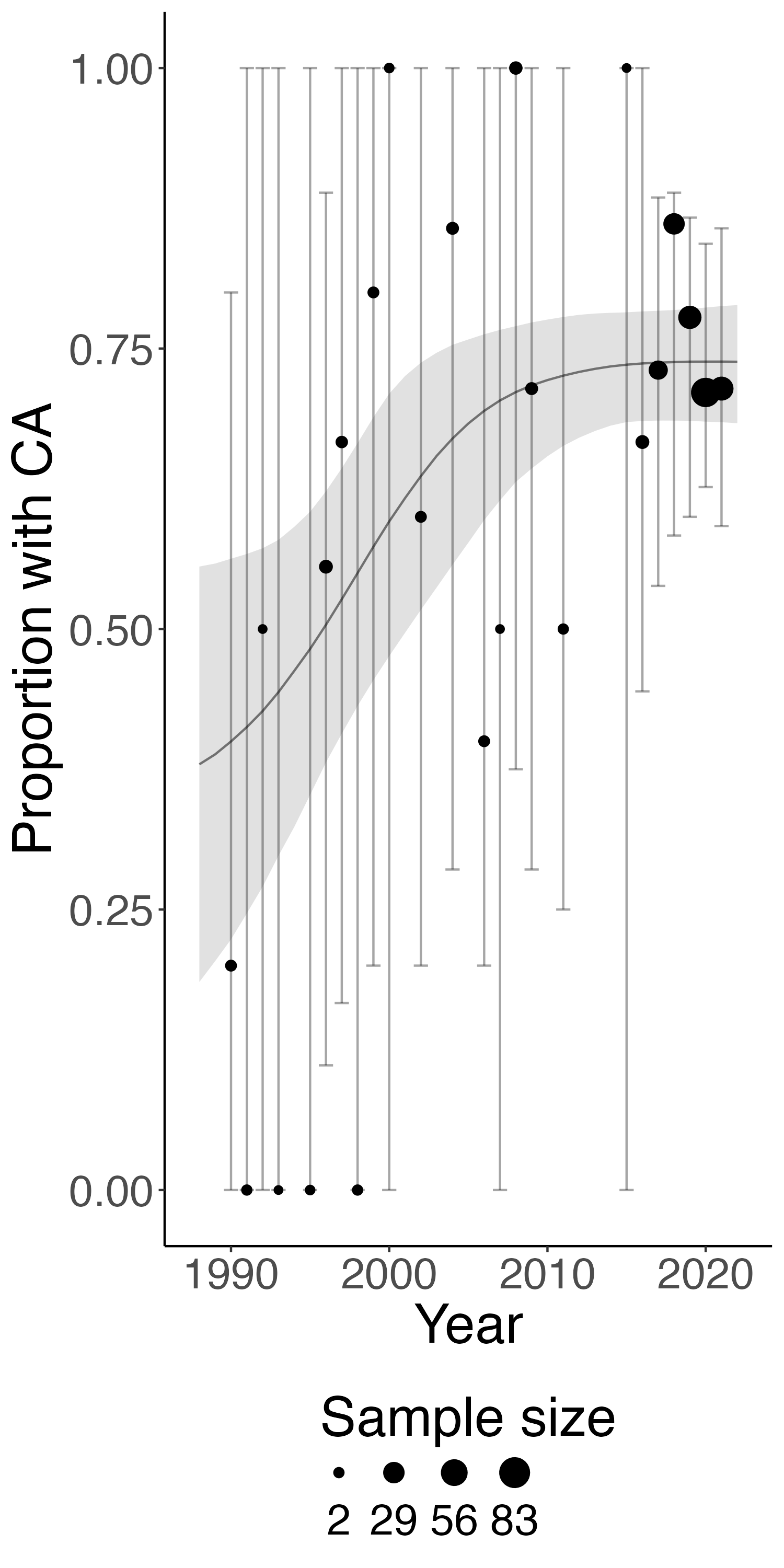}
    \label{fig:CA_rate}}
\caption{\setstretch{1.0} Posterior predictive check for the reproductive sign parameters. Posterior median estimates and central 95\% credible intervals for the (population level) proportion of adult females with a \protect\subref{fig:preg_rate} visible fetus, \protect\subref{fig:scar_rate} placental scar, and \protect\subref{fig:CA_rate} \textit{corpus albicans} are denoted by the black line and grey shading respectively. The observed sample proportions are denoted with dots, whose size reflects the sample size, and the 95\% predictive intervals for the sample proportions are shown by the error bars.}
\label{fig:reproduction_pp}
\end{figure}

\begin{figure}[H]
\centering
\captionsetup[subfigure]{position=top,justification=raggedright,singlelinecheck=false}
\subfloat[][]{
    \includegraphics[width=0.49\linewidth]{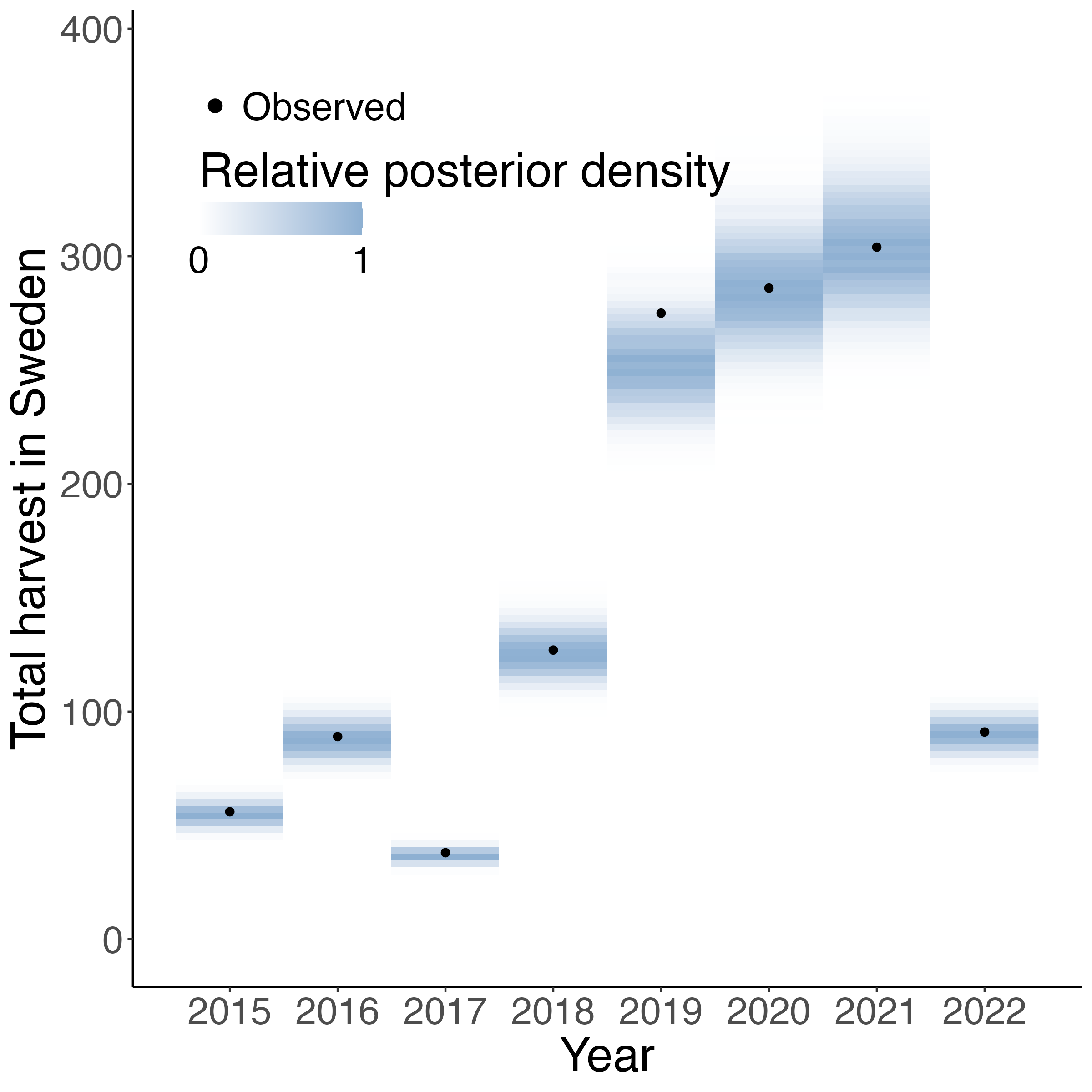}
    \label{fig:hbag_sw}}
\subfloat[][]{
    \includegraphics[width=0.49\linewidth]{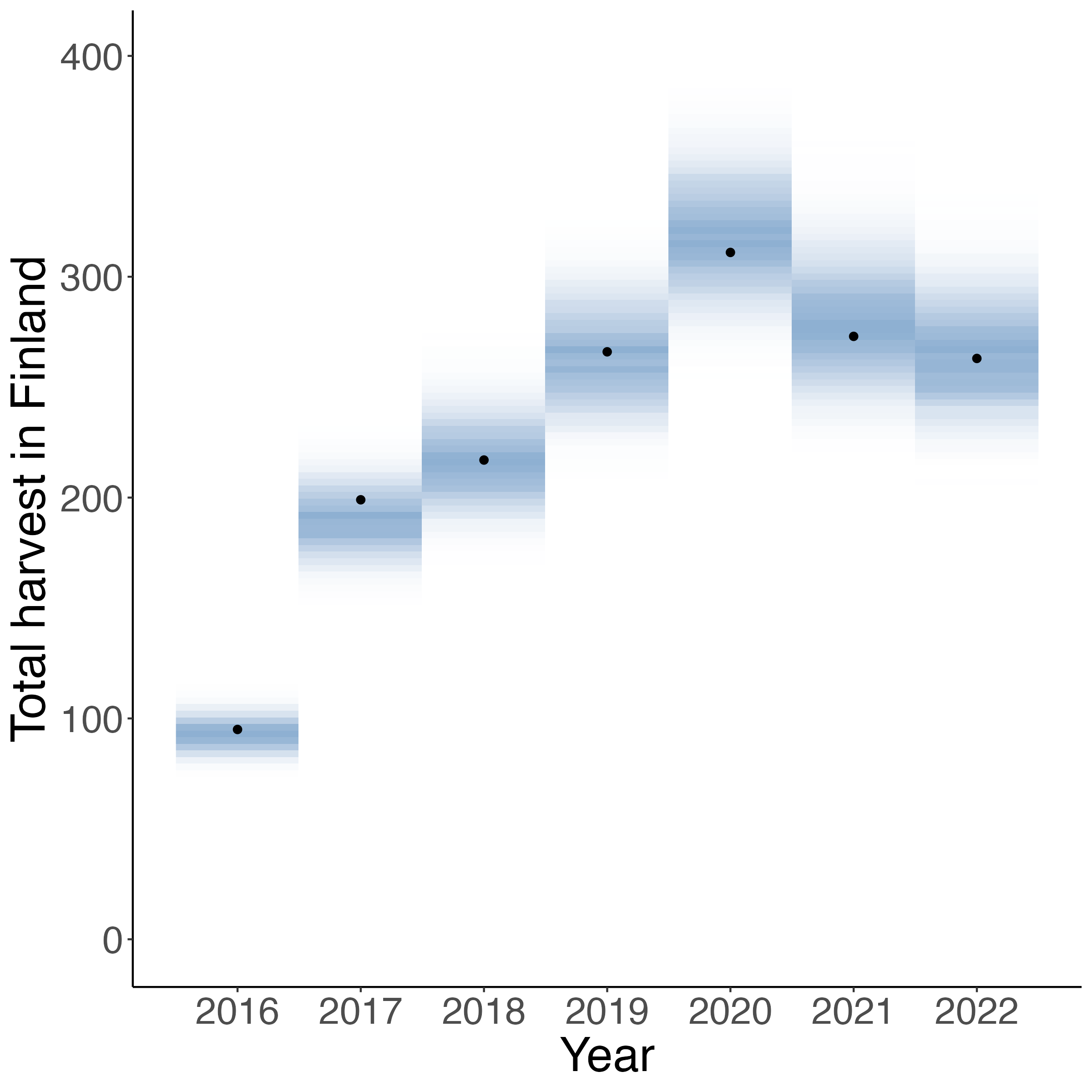}
    \label{fig:hbag_fi}}
\caption{\setstretch{1.0} Posterior predictive checks for the \protect\subref{fig:hbag_sw} Swedish and \protect\subref{fig:hbag_fi} Finnish ringed seal harvest totals.}
\label{fig:hbag_pp}
\end{figure}

\begin{figure}[H]
\centering
\captionsetup[subfigure]{position=top,justification=raggedright,singlelinecheck=false}
\subfloat[][]{
    \includegraphics[width=0.33\linewidth]{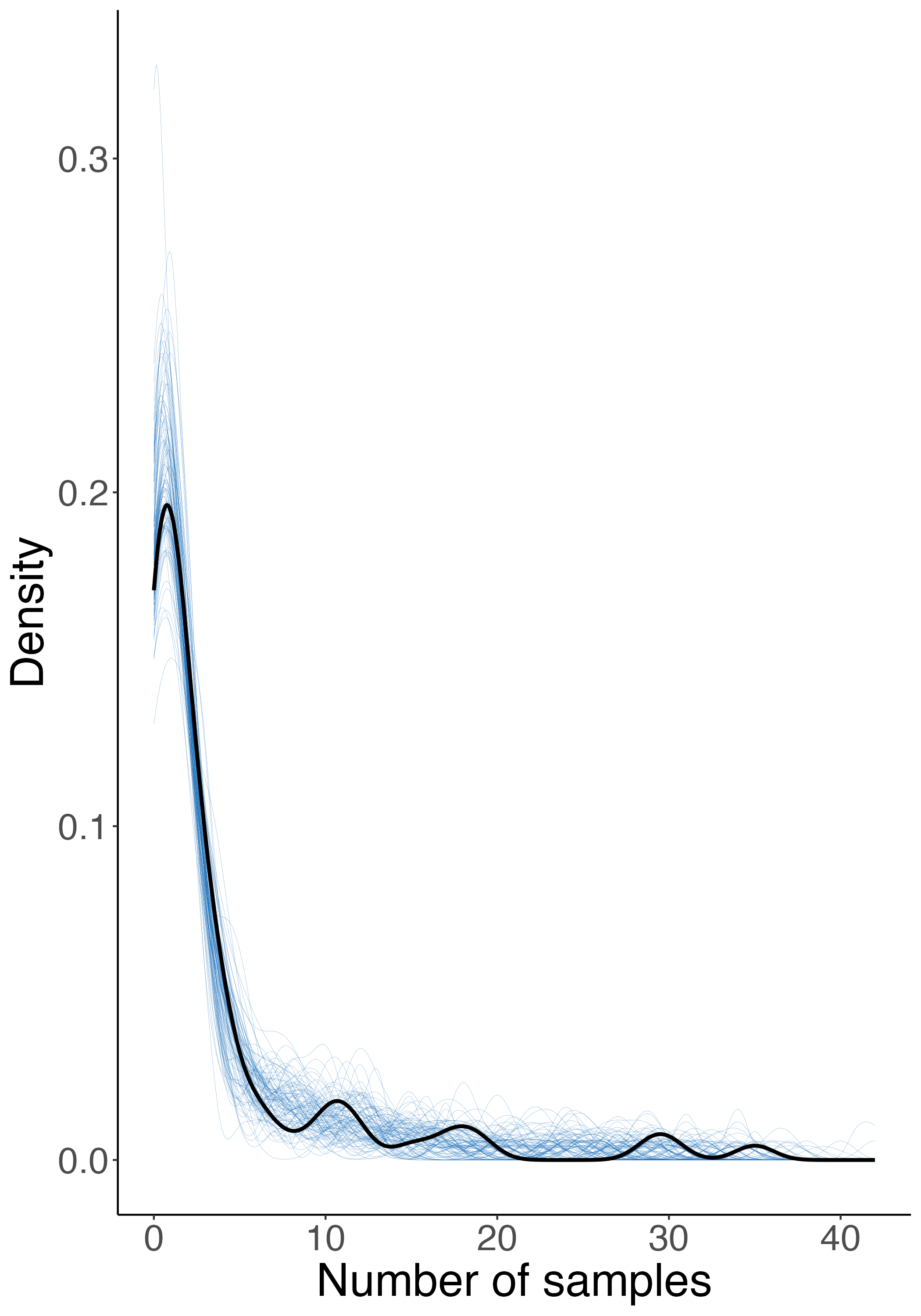}
    \label{fig:h_samples_sw}}
\subfloat[][]{
    \includegraphics[width=0.33\linewidth]{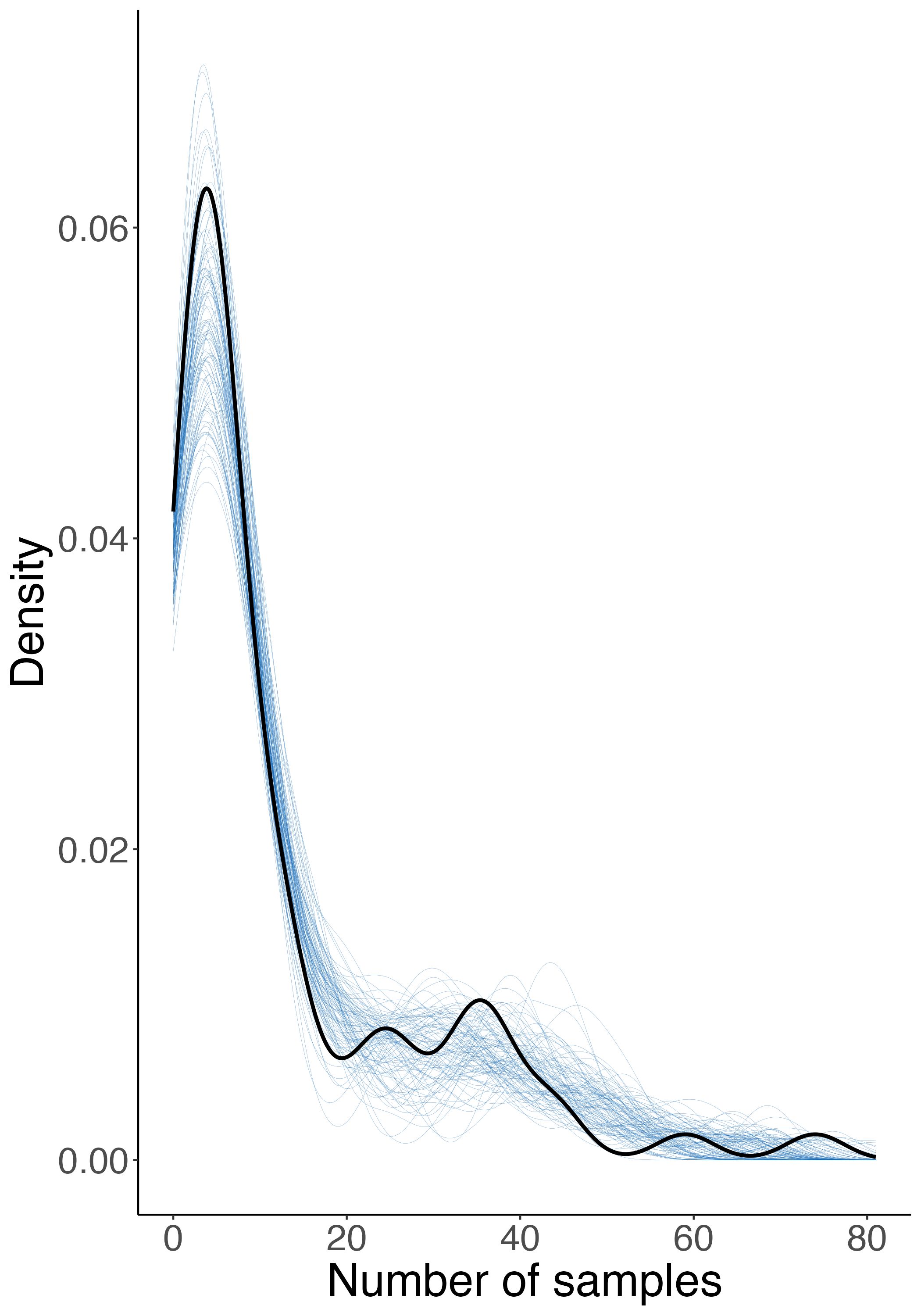}
    \label{fig:h_samples_fi}}
\subfloat[][]{
    \includegraphics[width=0.33\linewidth]{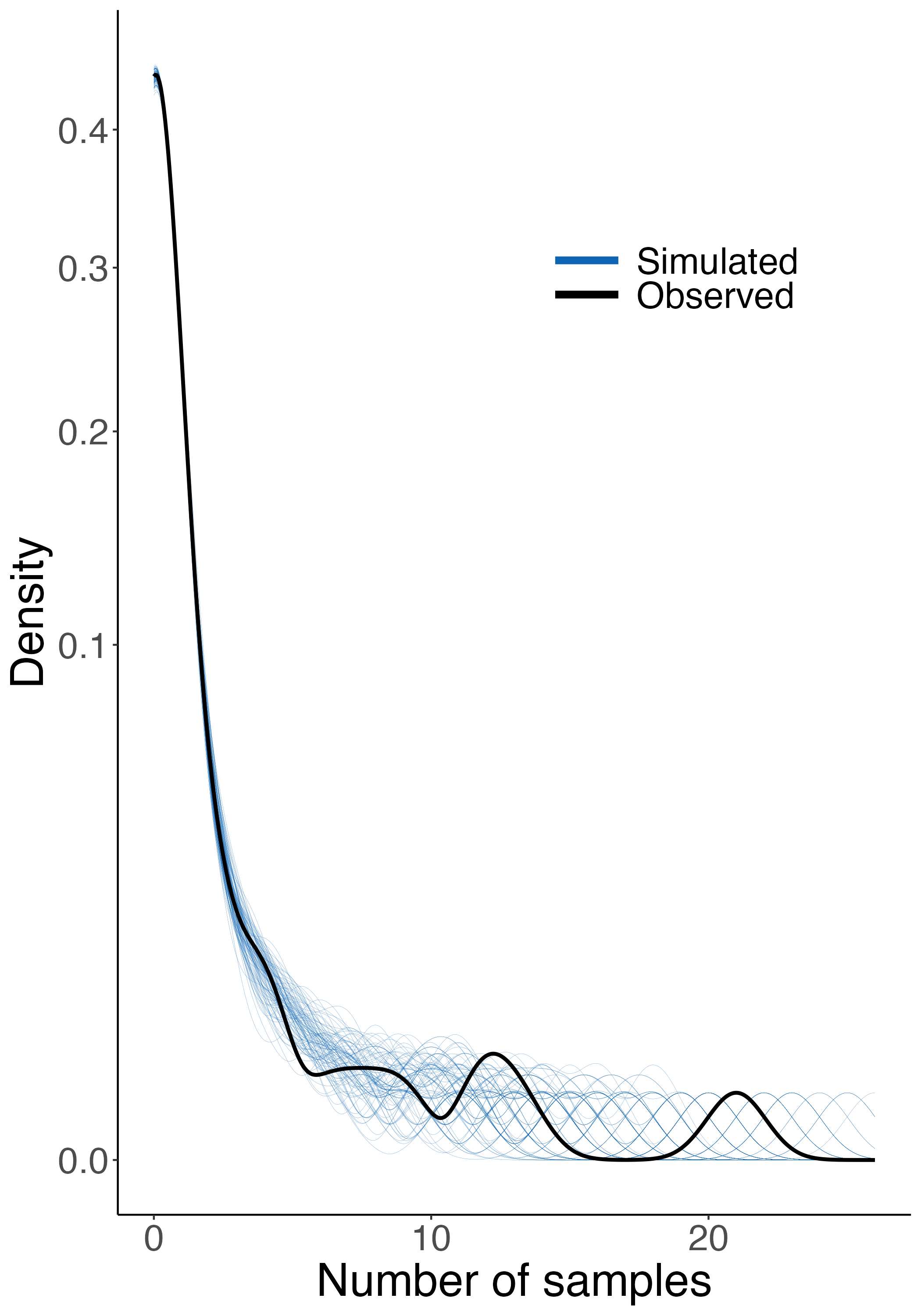}
    \label{fig:bycatch_samples}}
\caption{\setstretch{1.0} Posterior predictive checks for samples from \protect\subref{fig:h_samples_sw} Swedish harvests, \protect\subref{fig:h_samples_fi} Finnish harvests and \protect\subref{fig:bycatch_samples} bycatch. The density estimates are based on the frequency distribution of the number of sampled seals, pooled across all years and demographic classes. }
\label{fig:samples_pp}
\end{figure}

\begin{figure}[H]
\centering
\captionsetup[subfigure]{position=top,justification=raggedright,singlelinecheck=false}
\subfloat[][]{
    \includegraphics[width=0.49\linewidth]{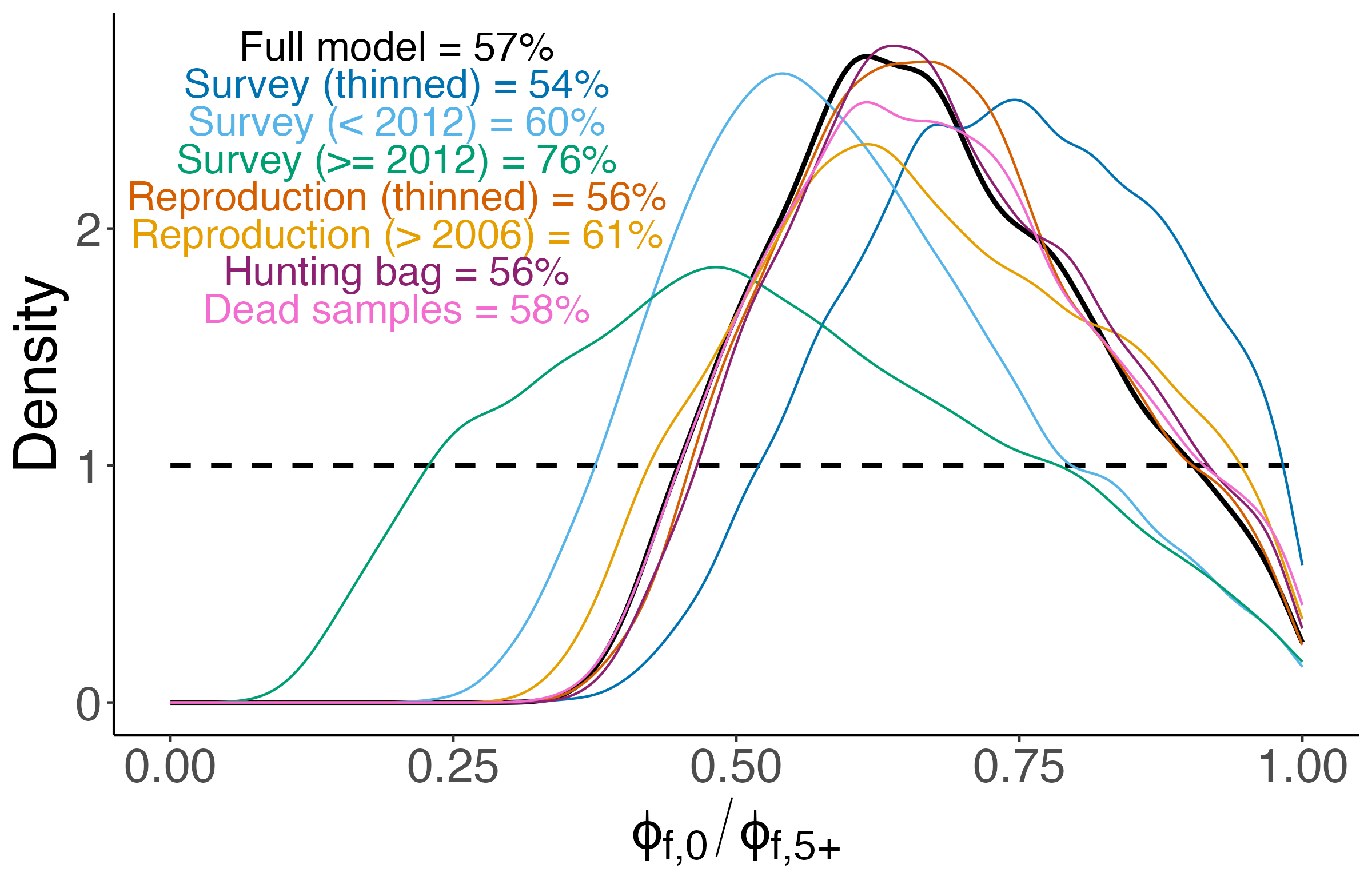}}
\subfloat[][]{
    \includegraphics[width=0.49\linewidth]{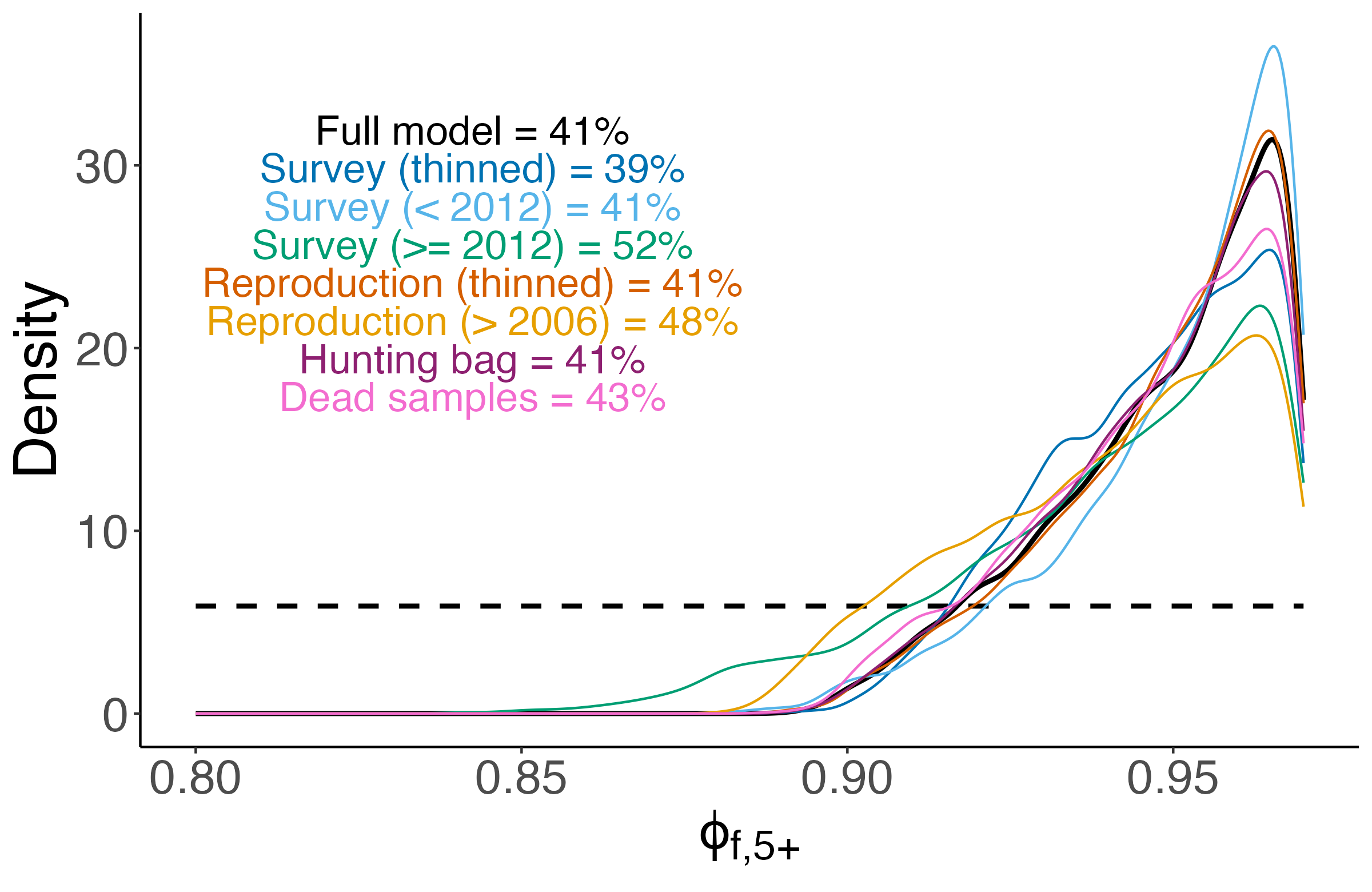}} \\
\subfloat[][]{
    \includegraphics[width=0.49\linewidth]{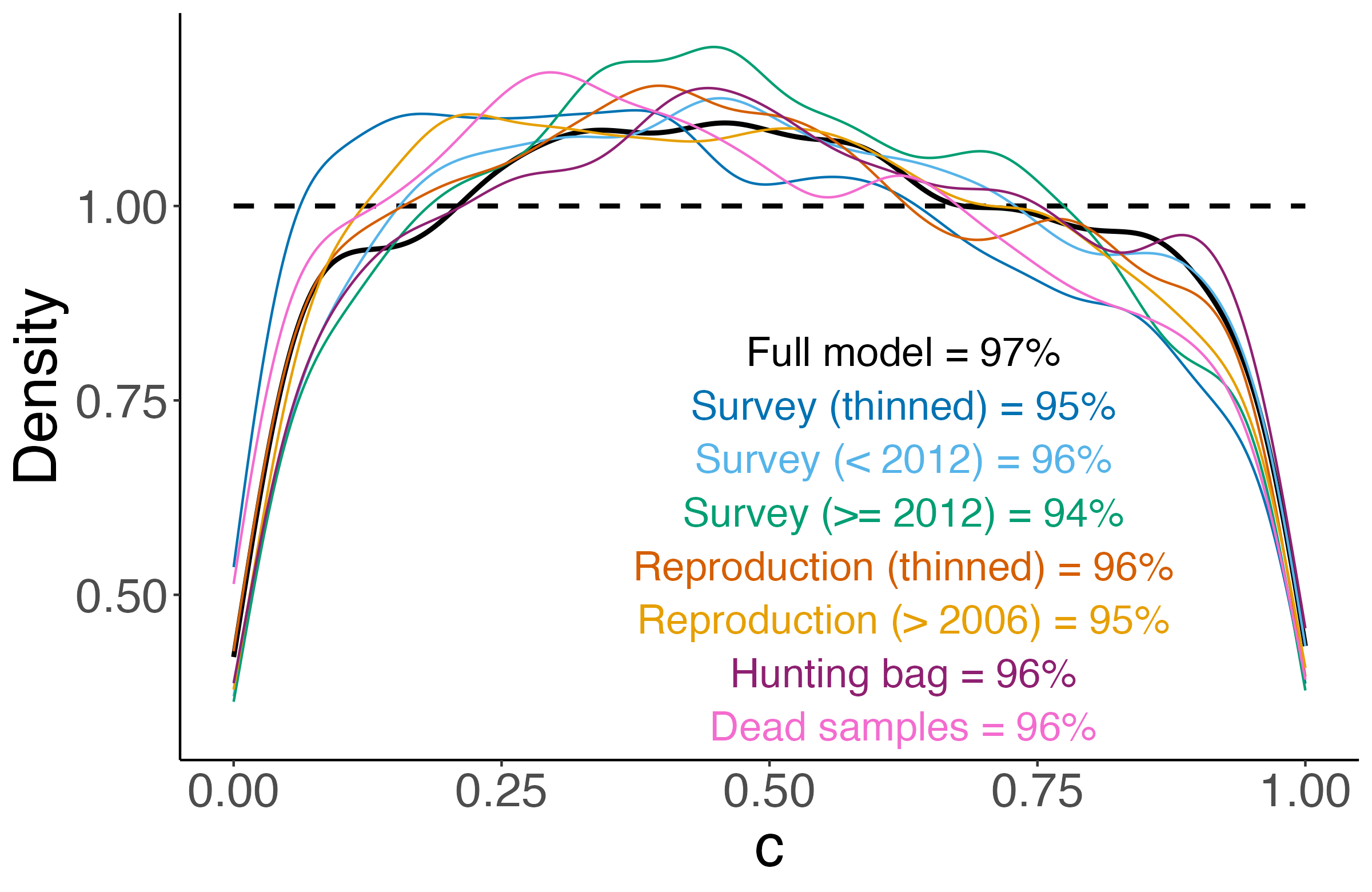}} 
\subfloat[][]{
    \includegraphics[width=0.49\linewidth]{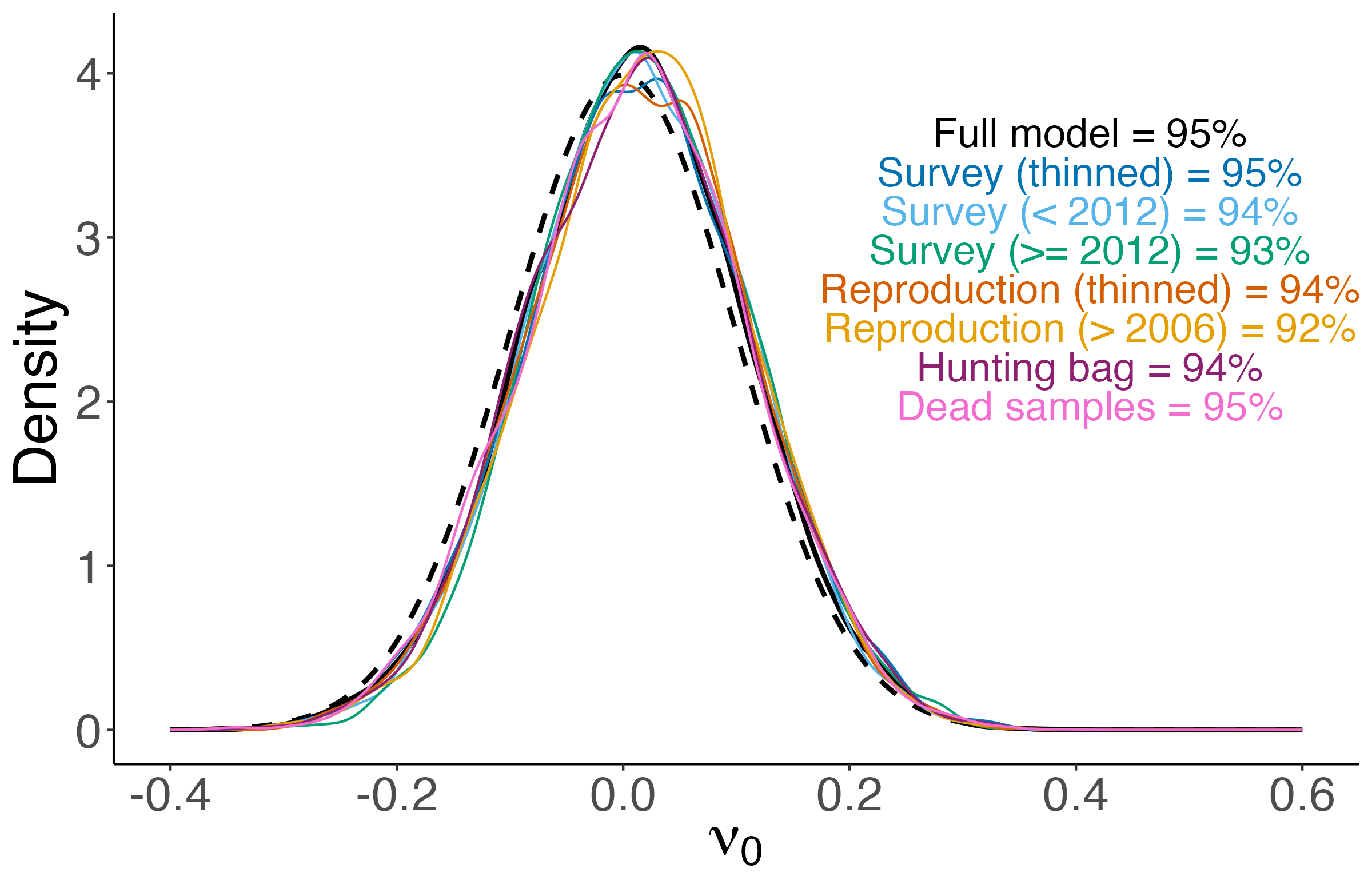}} 
\caption{\setstretch{1.0} Prior and posterior distributions and overlaps for survival parameters. Dashed lines represent prior distributions.}
\label{fig:survival_comp}
\end{figure}

\begin{figure}[H]
\centering
\captionsetup[subfigure]{position=top,justification=raggedright,singlelinecheck=false}
\subfloat[][]{
    \includegraphics[width=0.49\linewidth]{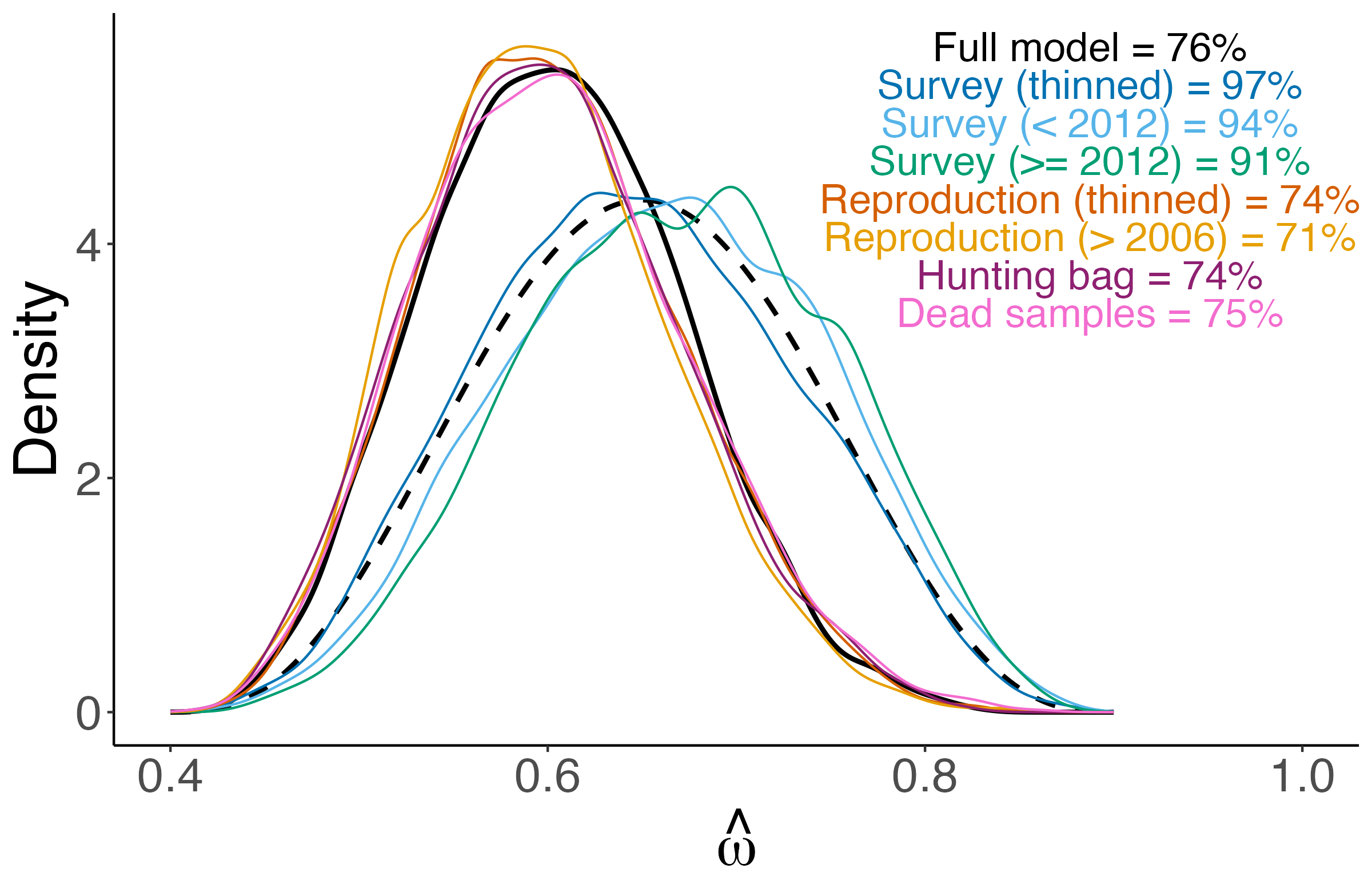}}
\subfloat[][]{
    \includegraphics[width=0.49\linewidth]{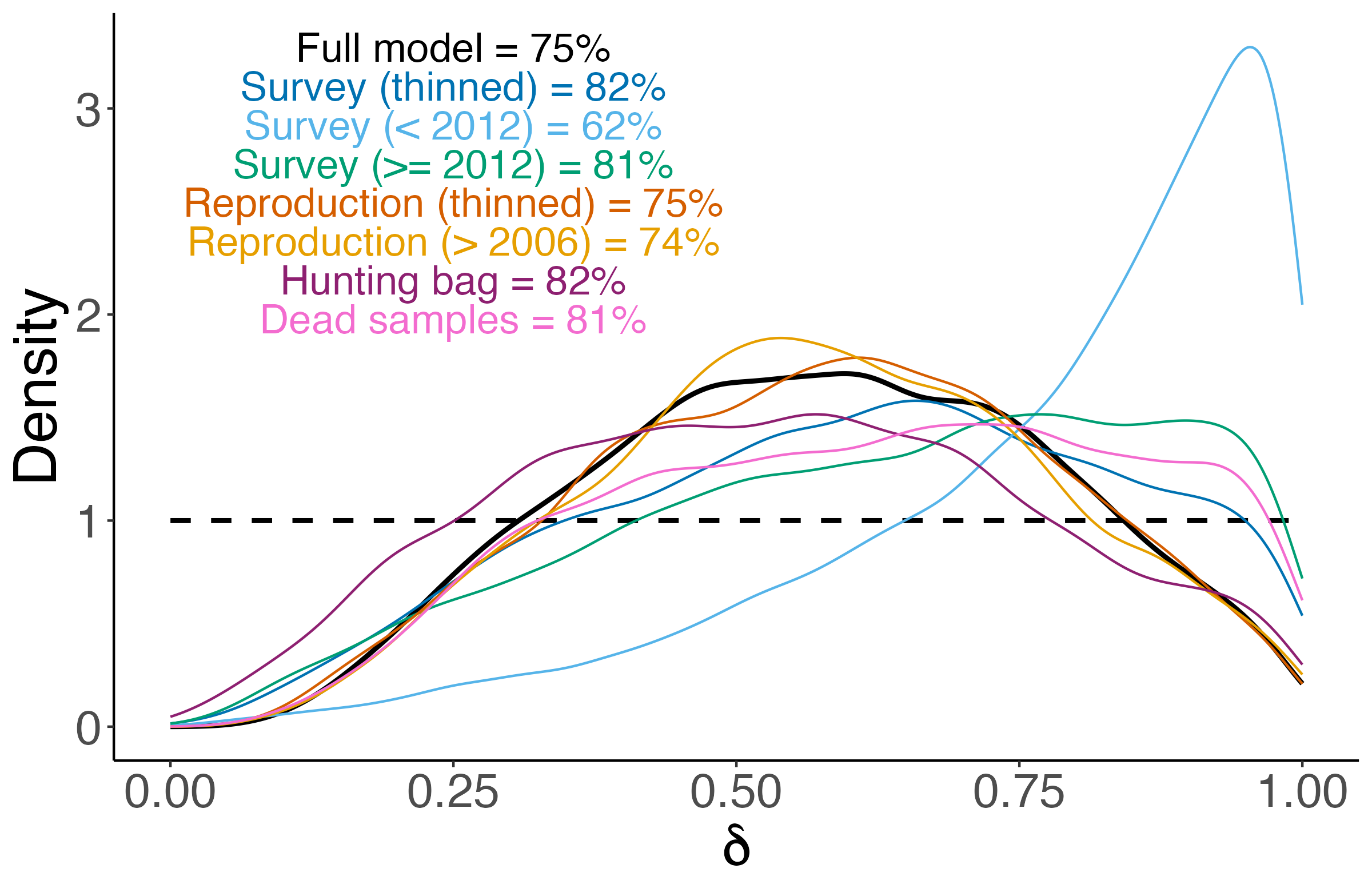}} \\
\subfloat[][]{
    \includegraphics[width=0.49\linewidth]{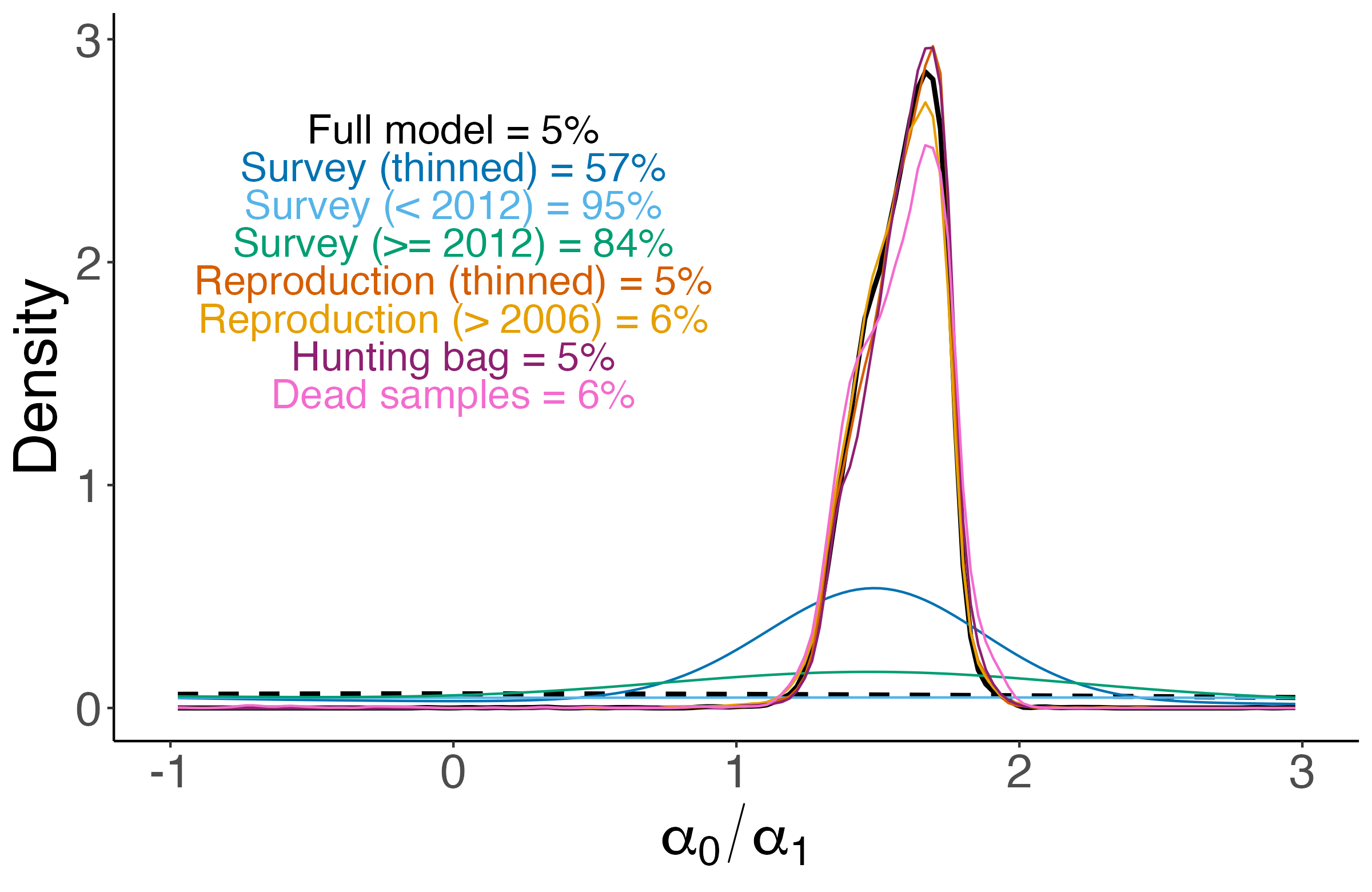}} 
\subfloat[][]{
    \includegraphics[width=0.49\linewidth]{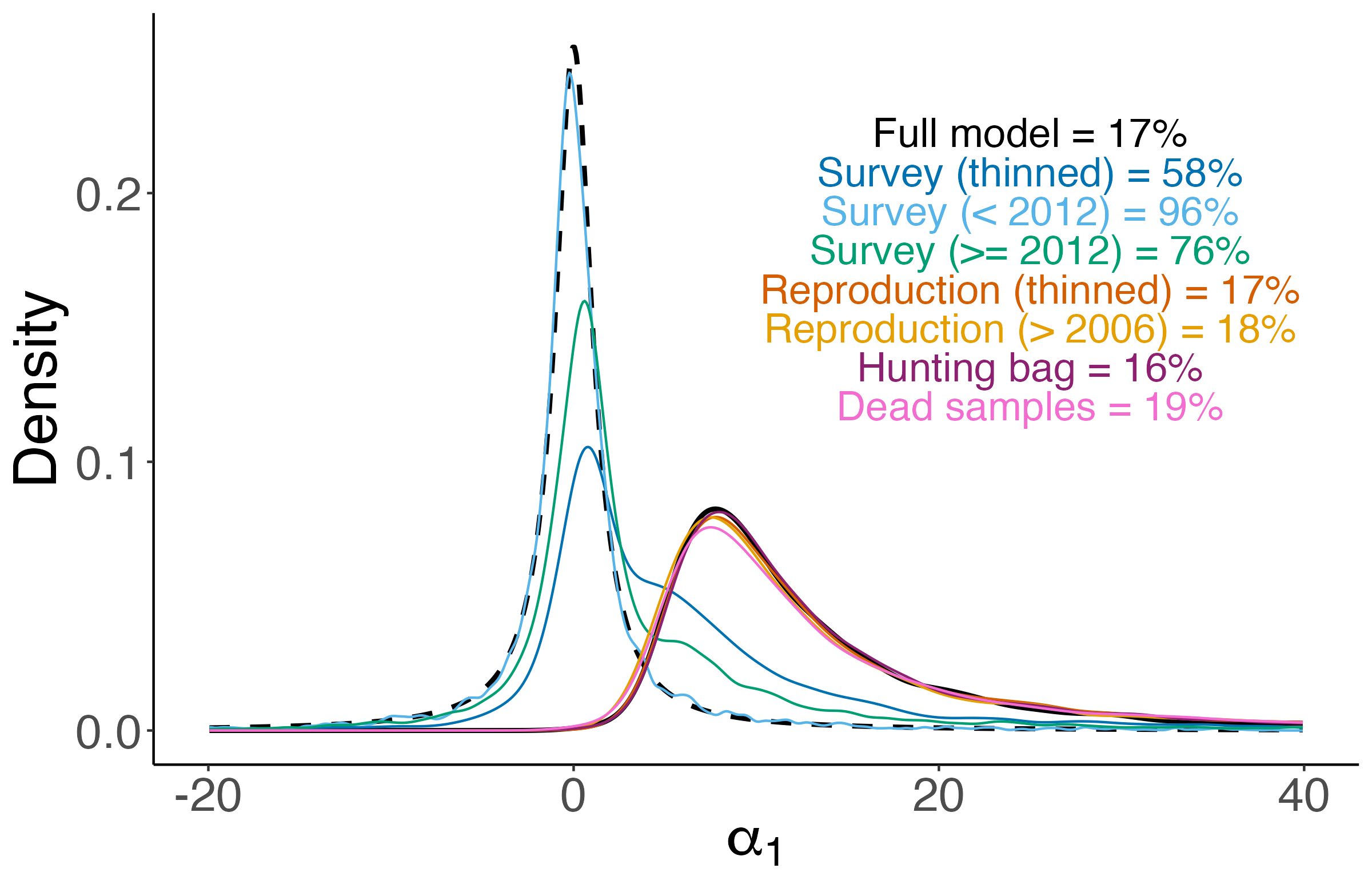}} \\
\subfloat[][]{
    \includegraphics[width=0.49\linewidth]{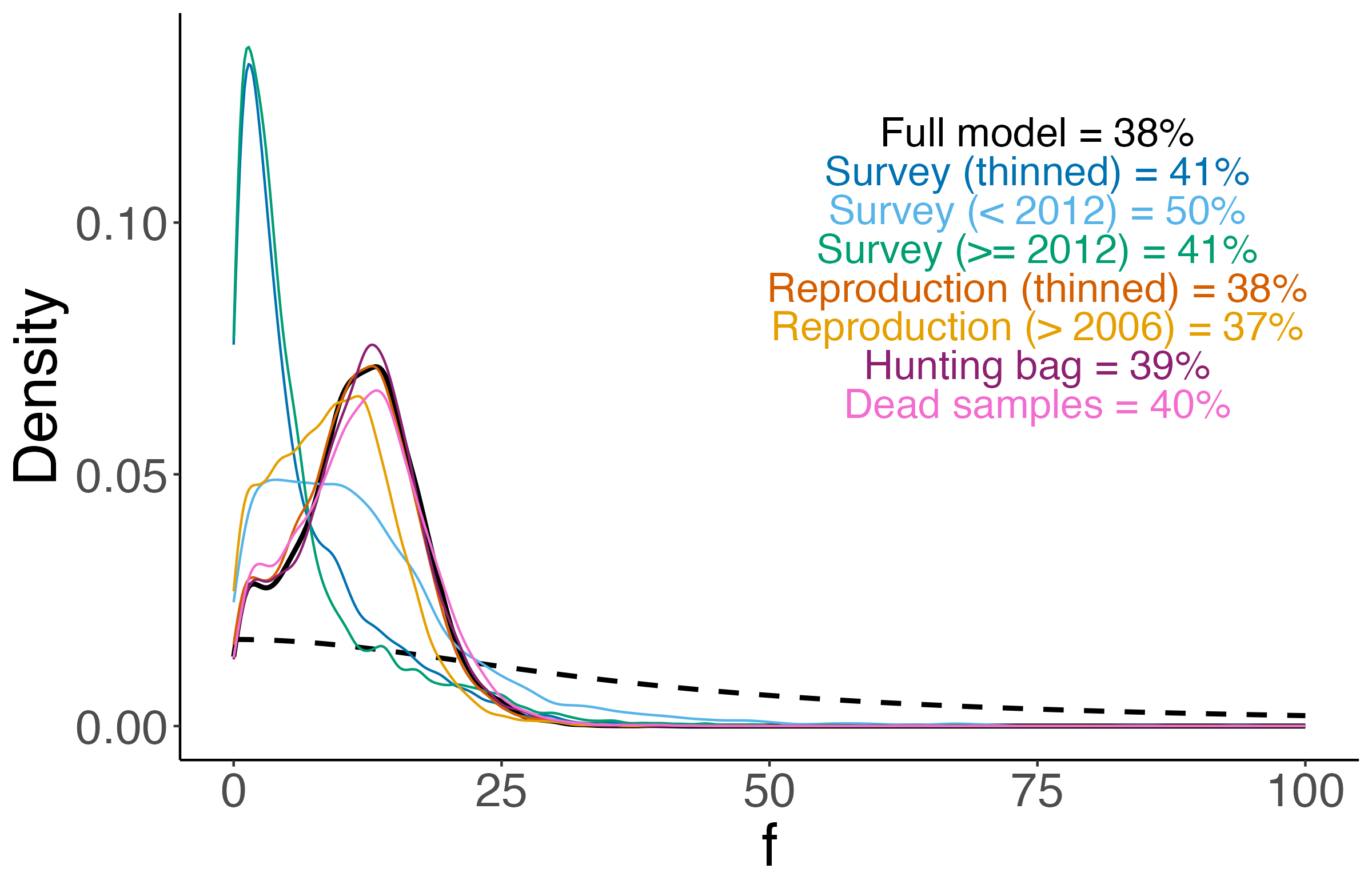}} 
\subfloat[][]{
    \includegraphics[width=0.49\linewidth]{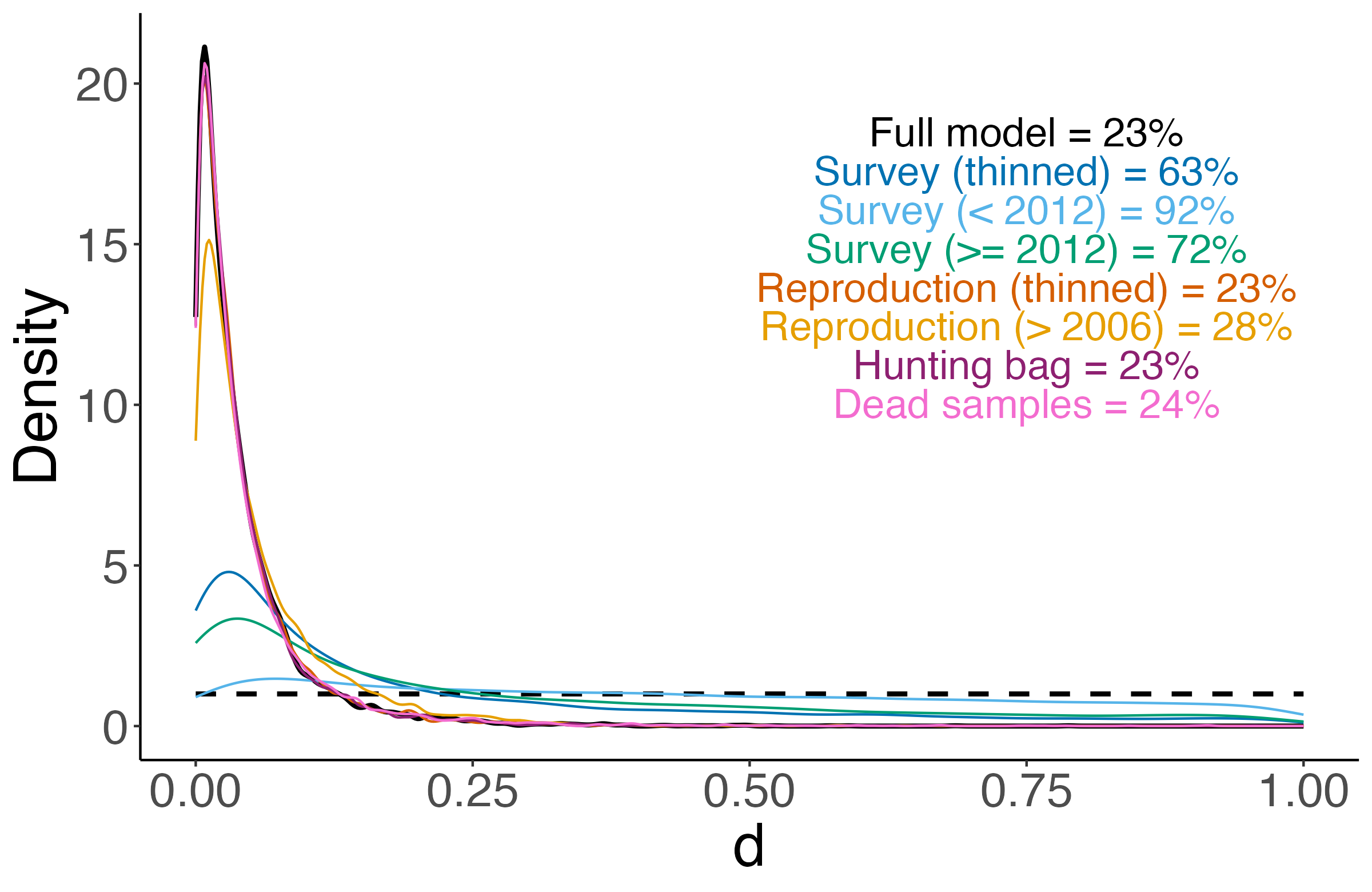}} 
\caption{\setstretch{1.0} Prior and posterior distributions and overlaps for haul-out parameters. Dashed lines represent prior distributions.}
\label{fig:haulout_comp}
\end{figure}

\begin{figure}[H]
\centering
\captionsetup[subfigure]{position=top,justification=raggedright,singlelinecheck=false}
\subfloat[][]{
    \includegraphics[width=0.49\linewidth]{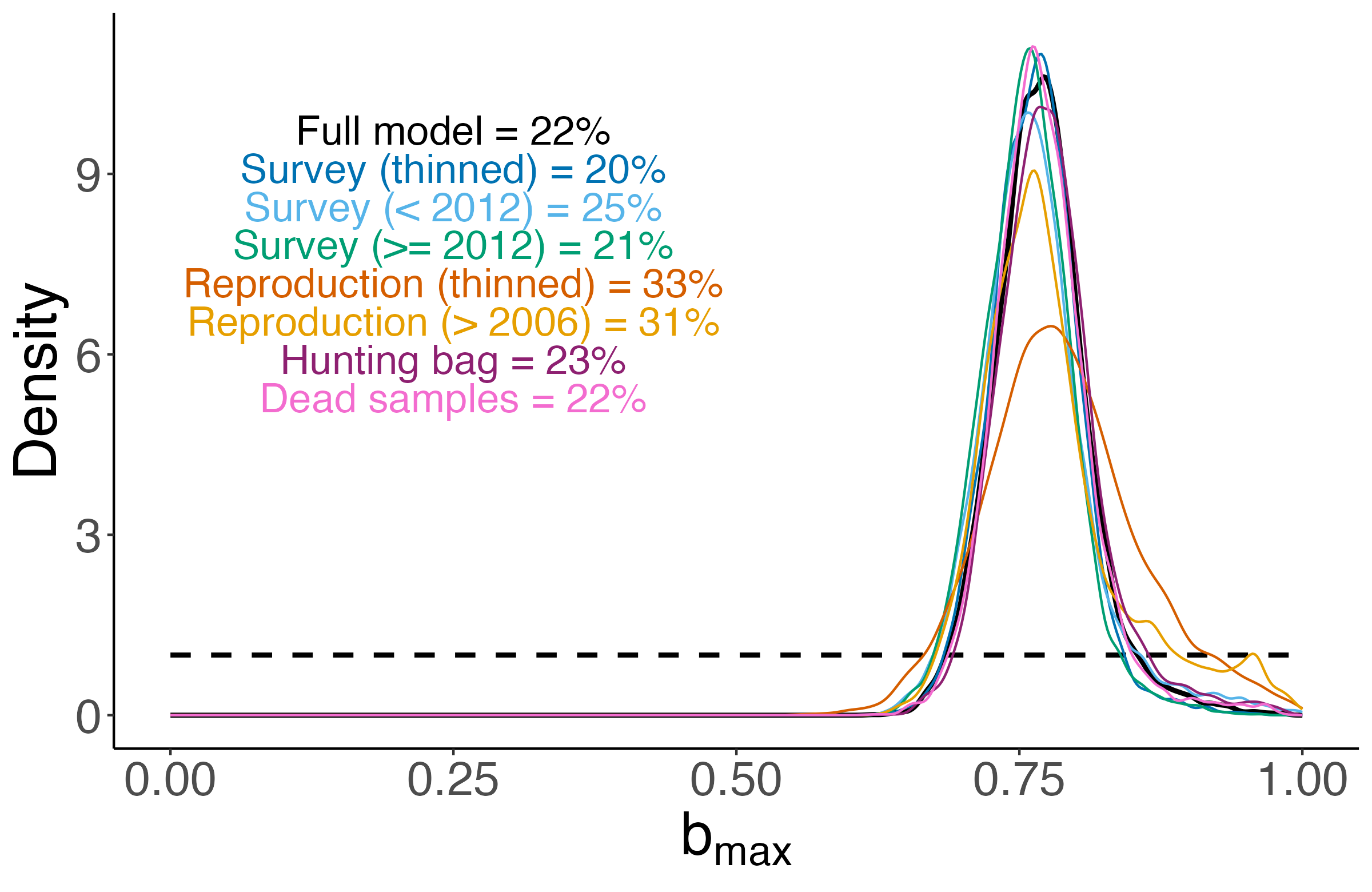}}
\subfloat[][]{
    \includegraphics[width=0.49\linewidth]{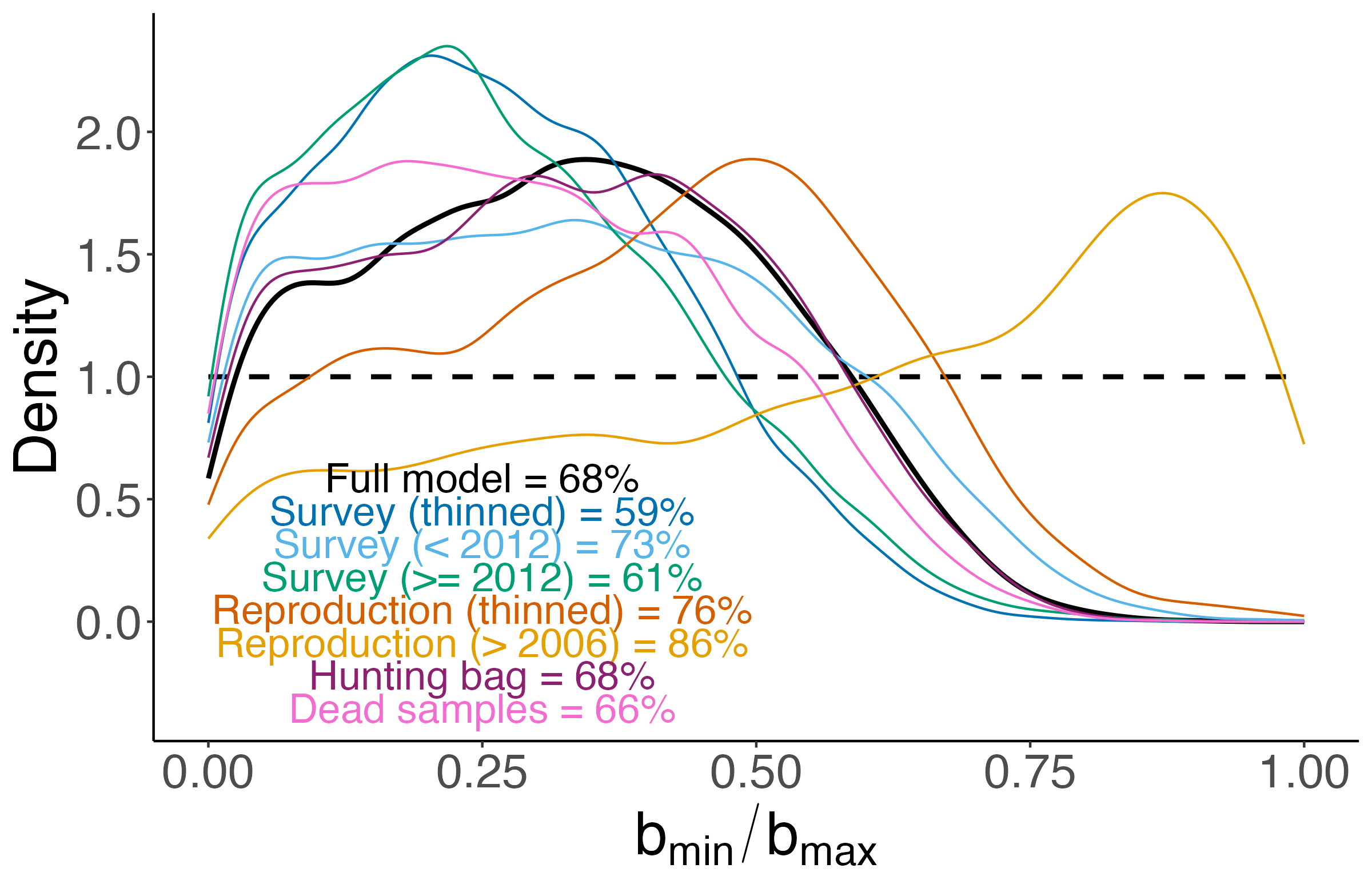}} \\
\subfloat[][]{
    \includegraphics[width=0.49\linewidth]{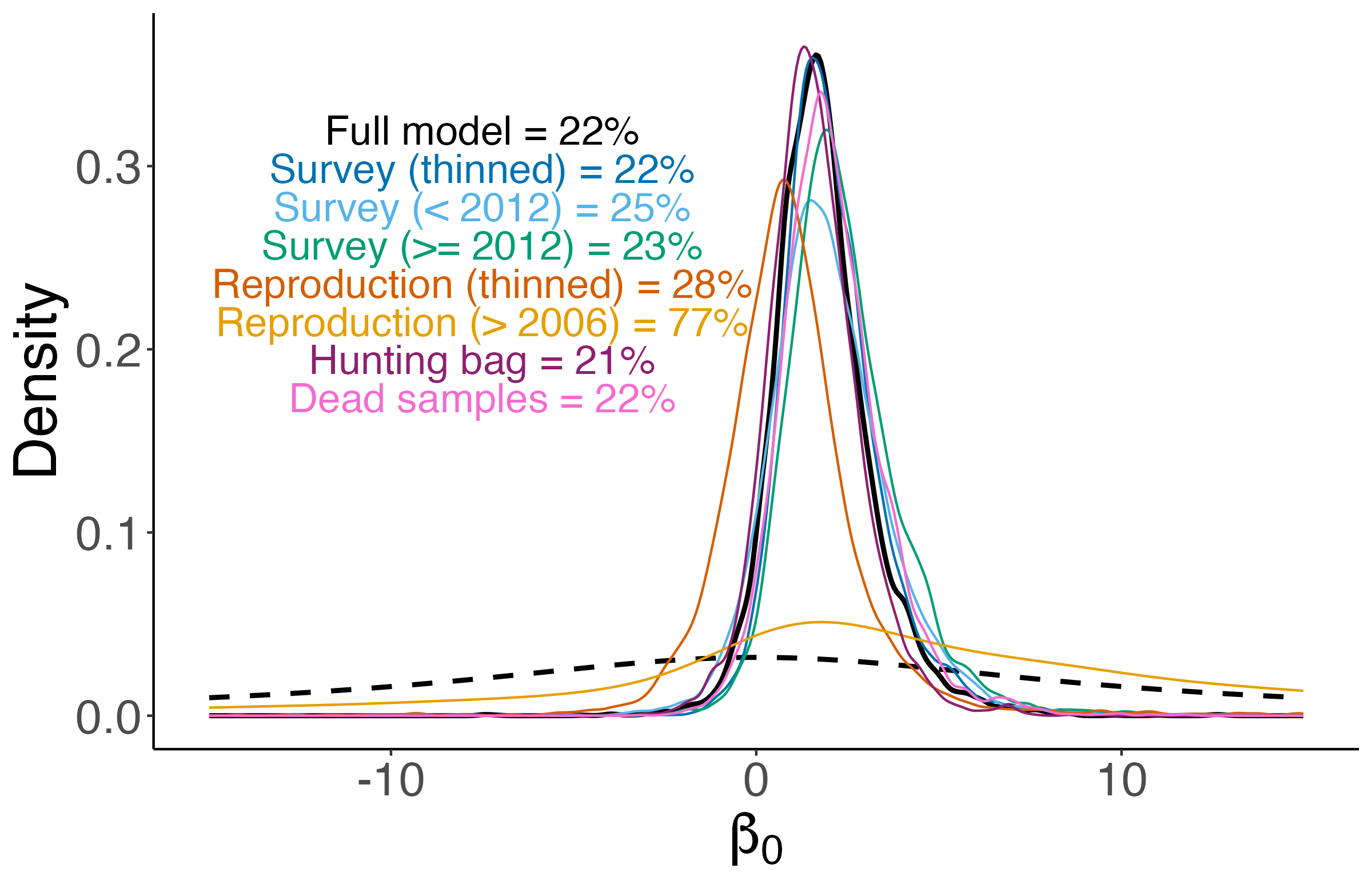}} 
\subfloat[][]{
    \includegraphics[width=0.49\linewidth]{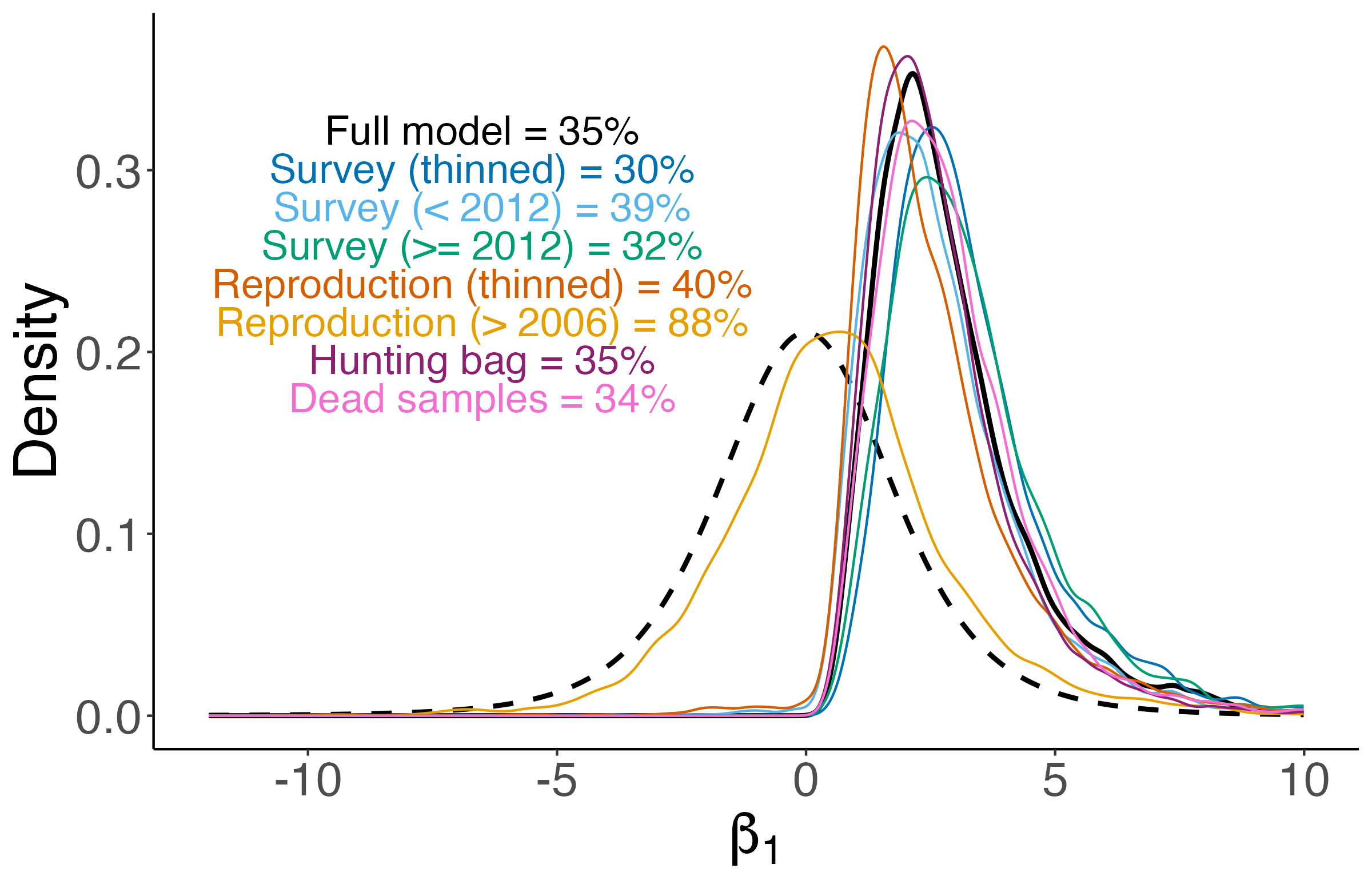}} \\
\subfloat[][]{
    \includegraphics[width=0.33\linewidth]{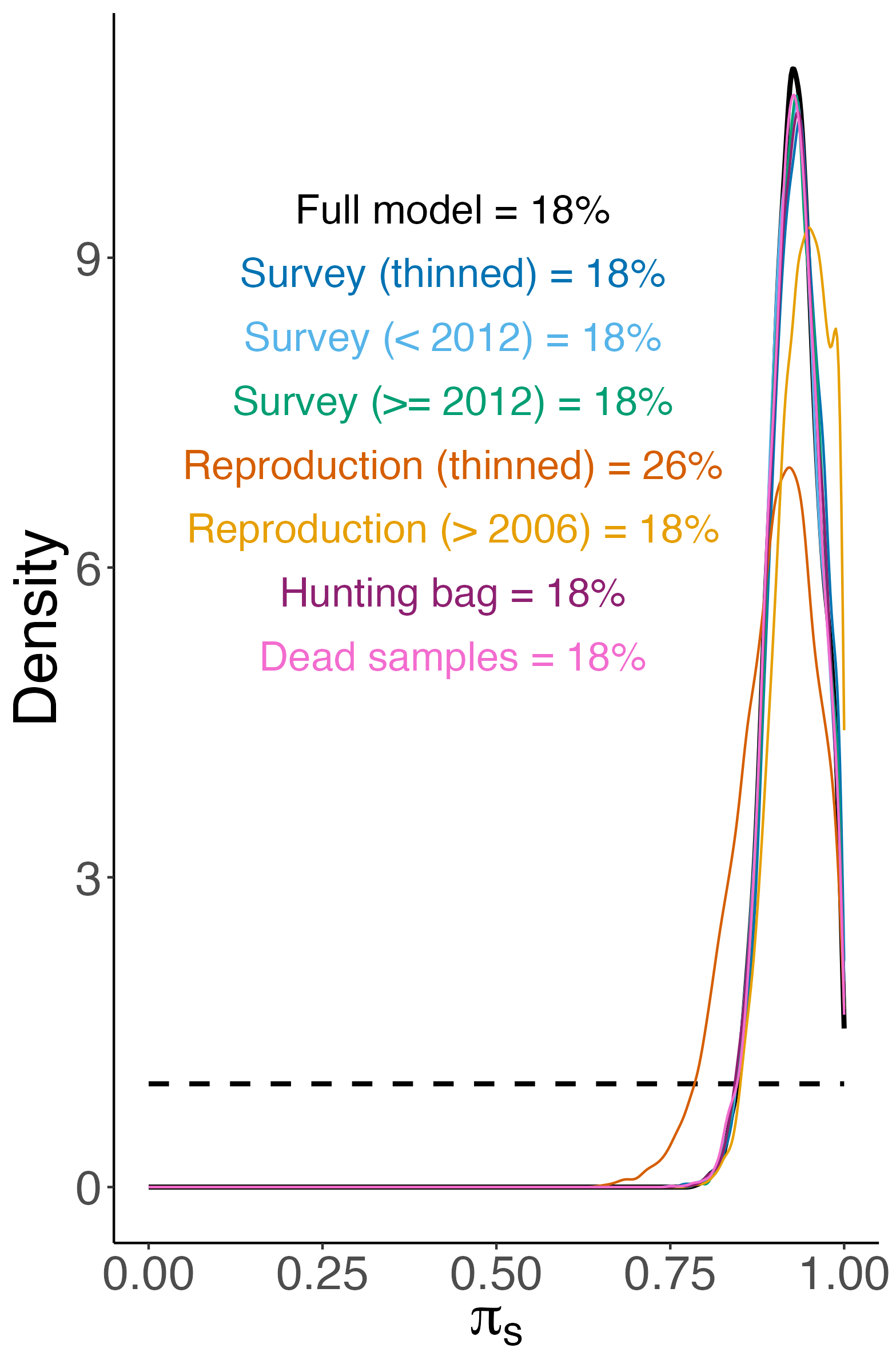}} 
\subfloat[][]{
    \includegraphics[width=0.33\linewidth]{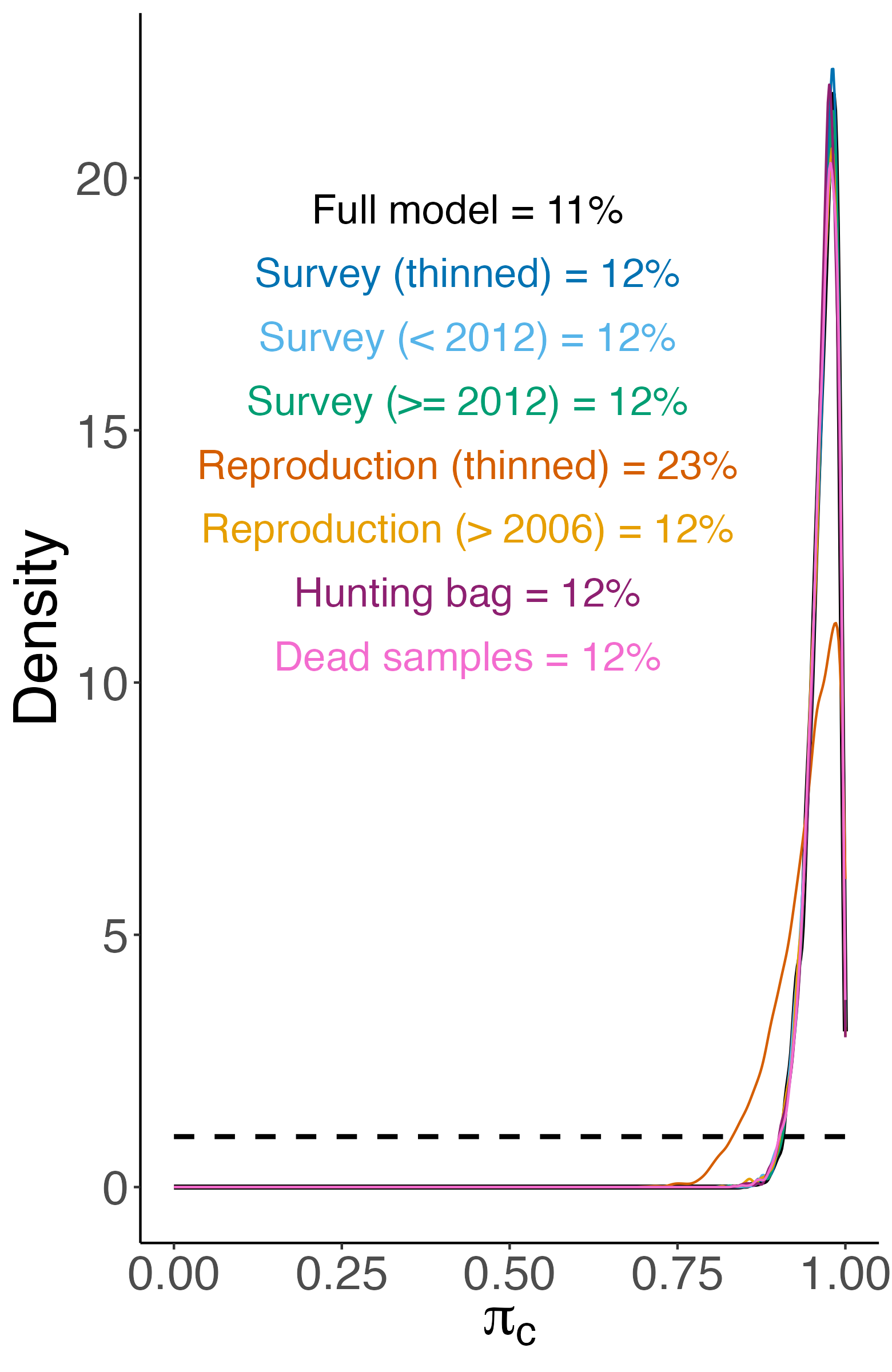}}
\subfloat[][]{
    \includegraphics[width=0.33\linewidth]{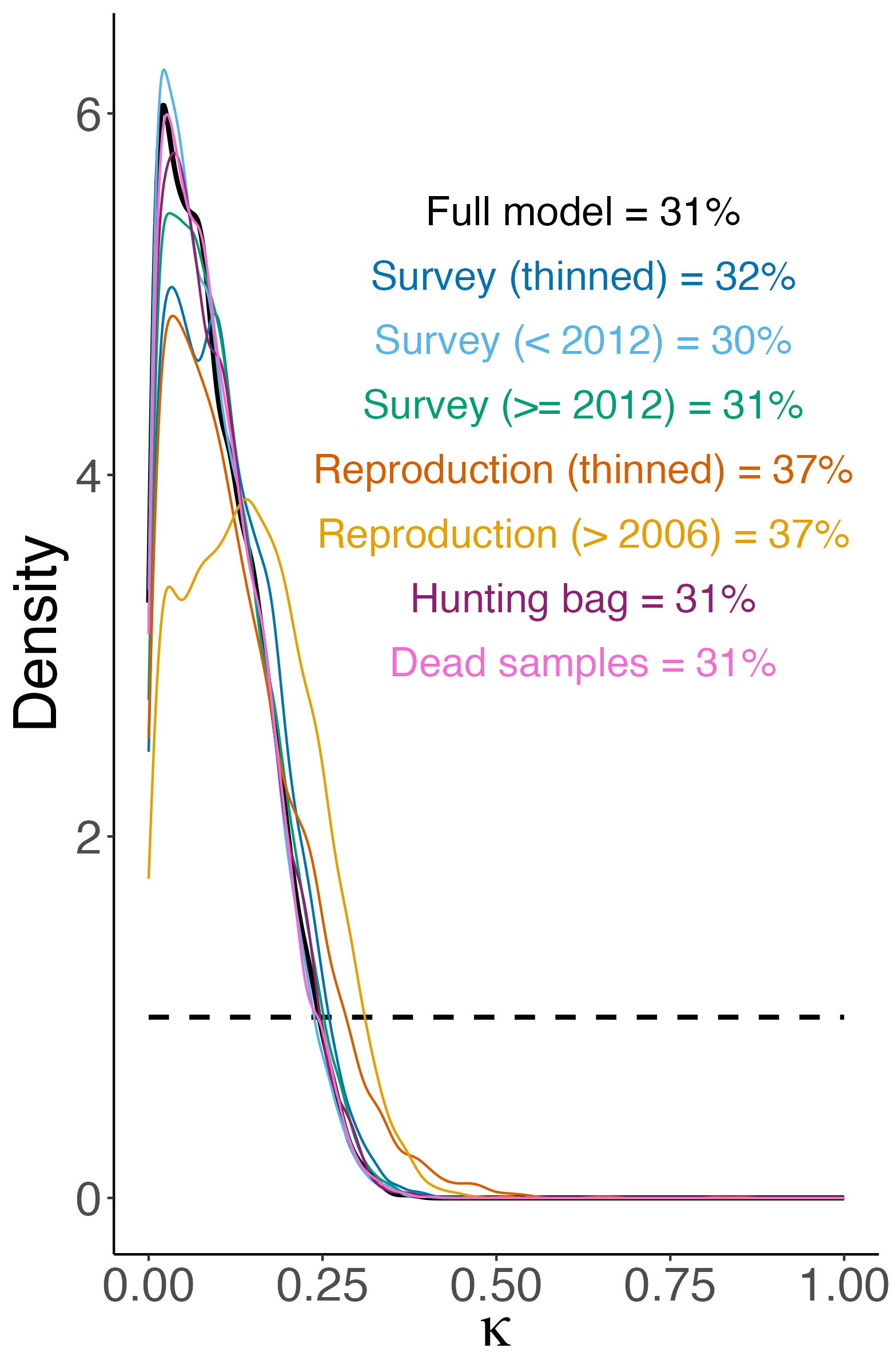}} 
\caption{\setstretch{1.0} Prior and posterior distributions and overlaps for reproduction parameters. Dashed lines represent prior distributions.}
\label{fig:reproduction_comp}
\end{figure}

\begin{figure}[H]
\centering
\captionsetup[subfigure]{position=top,justification=raggedright,singlelinecheck=false}
\subfloat[][]{
    \includegraphics[width=0.49\linewidth]{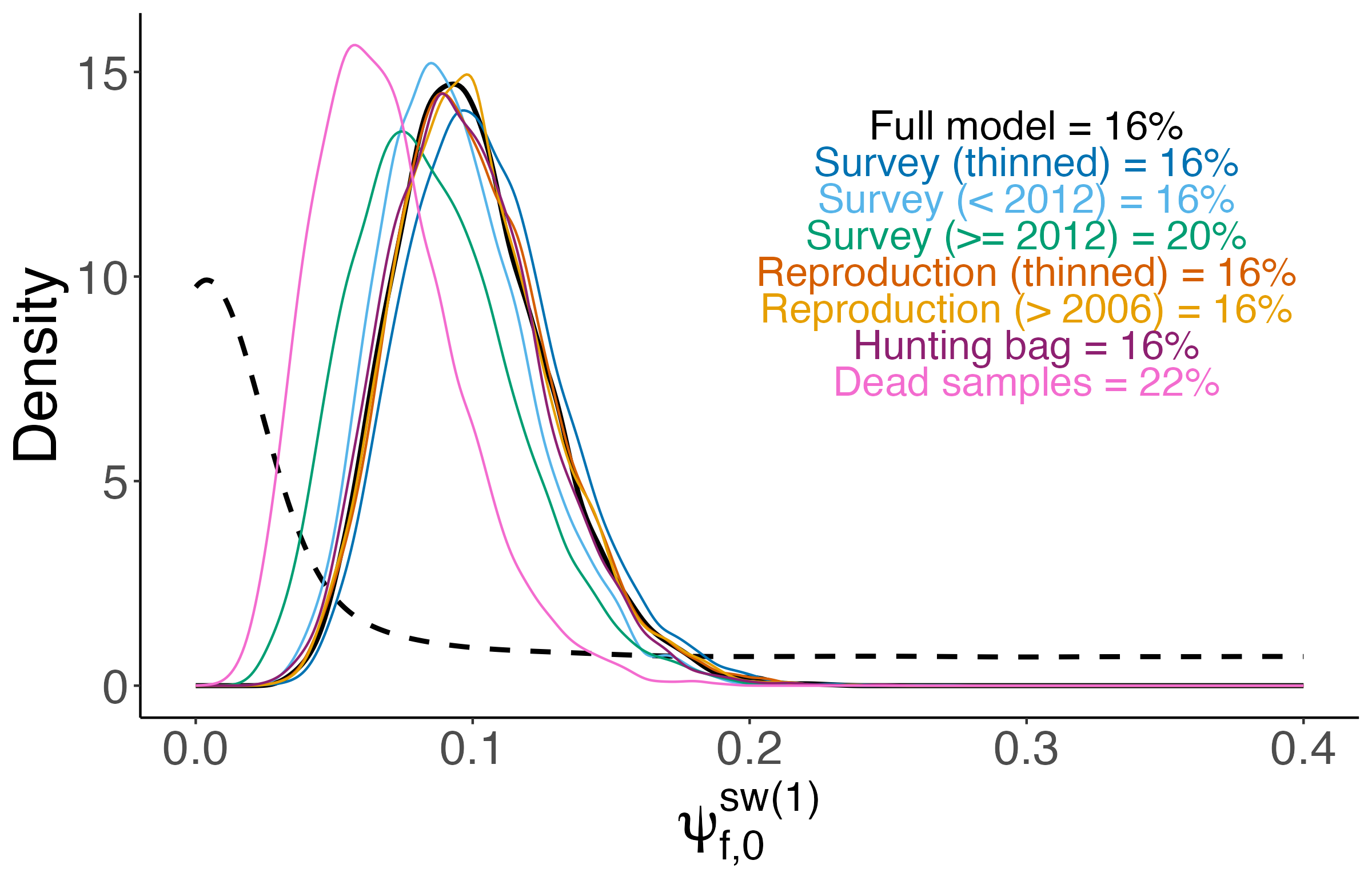}}
\subfloat[][]{
    \includegraphics[width=0.49\linewidth]{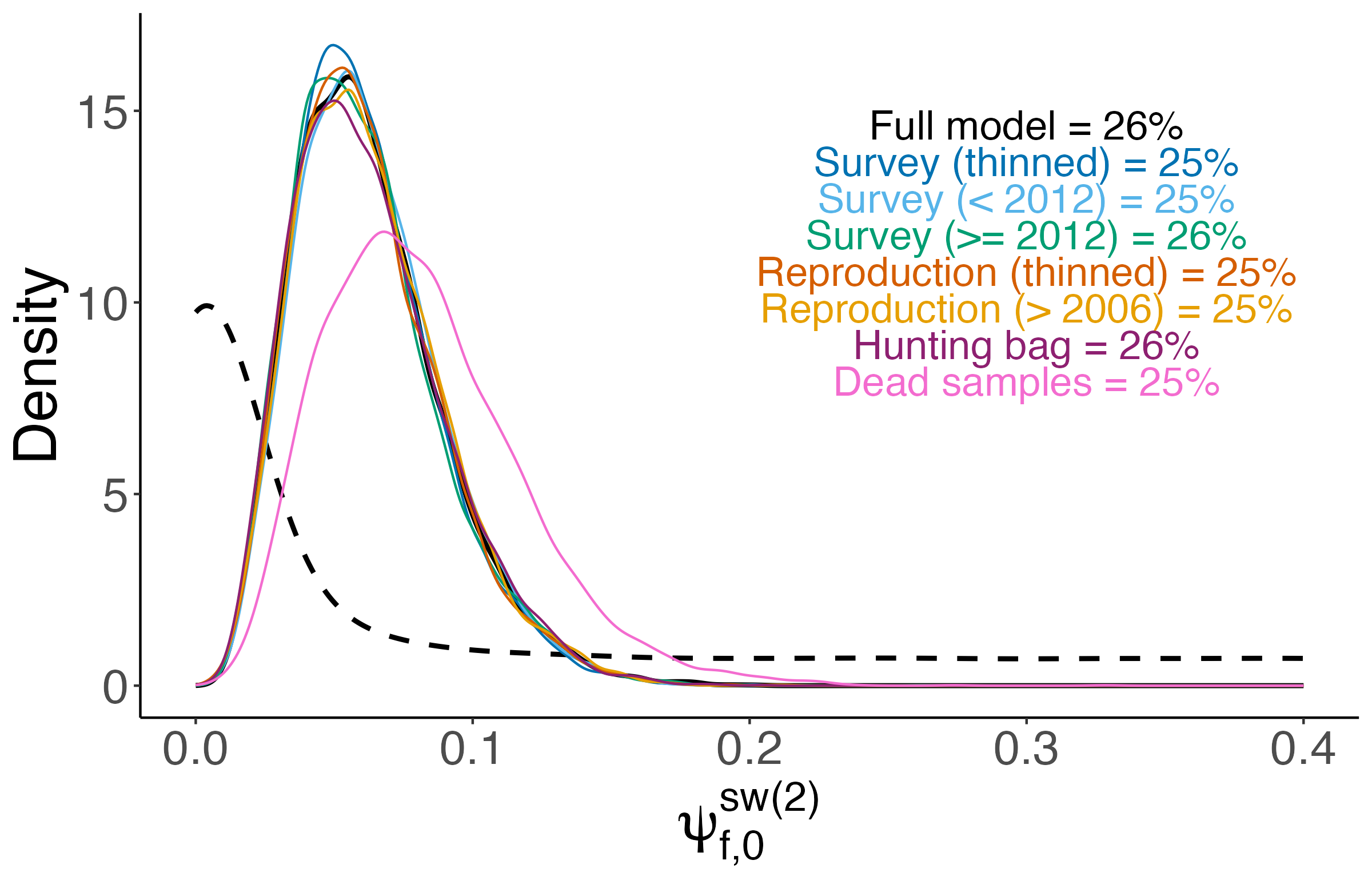}} \\
\subfloat[][]{
    \includegraphics[width=0.49\linewidth]{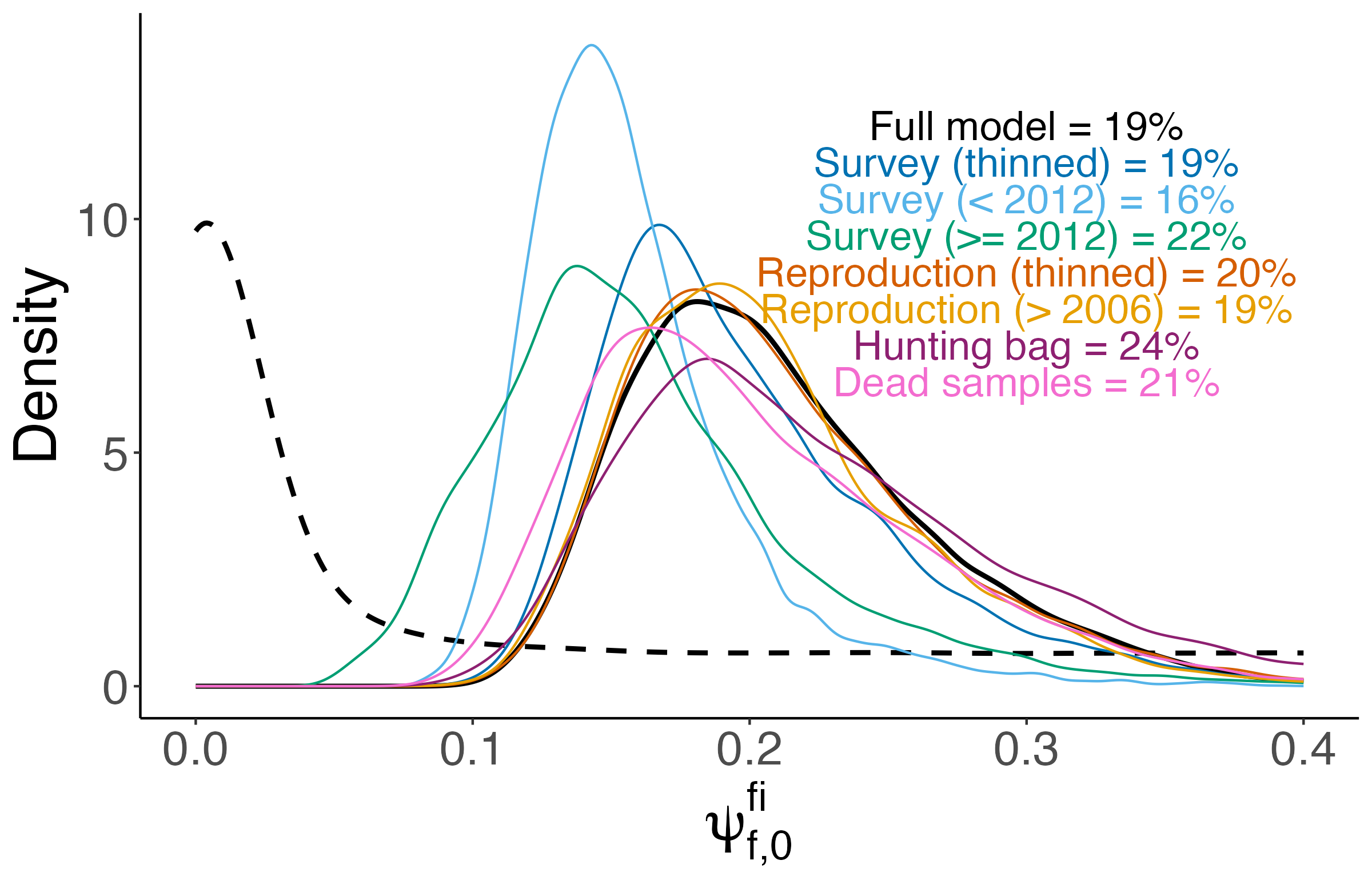}} 
\subfloat[][]{
    \includegraphics[width=0.49\linewidth]{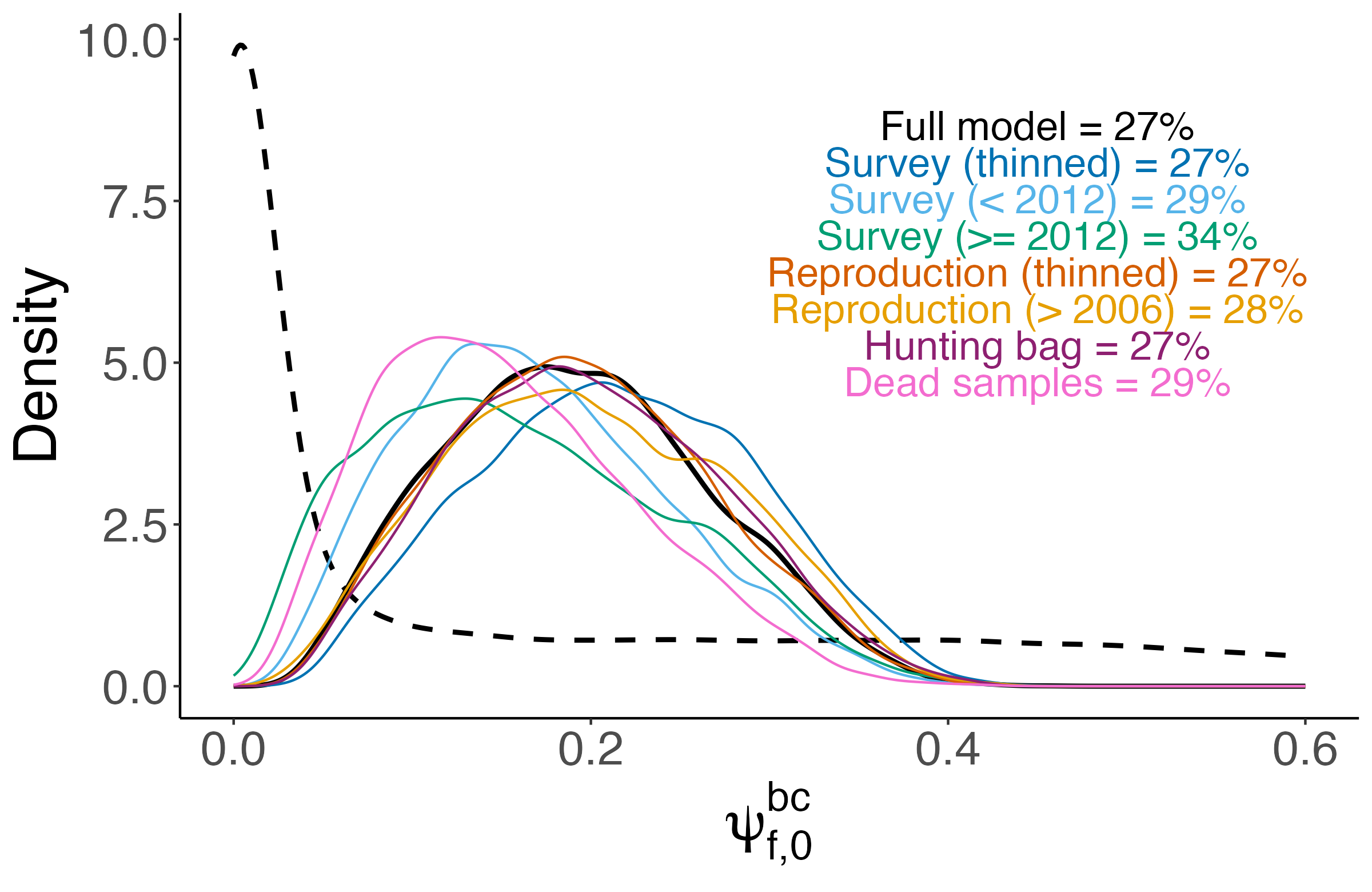}} 
\caption{\setstretch{1.0} Prior and posterior distributions and overlaps for hunting and bycatch bias parameters. Dashed lines represent prior distributions. Only parameters related to pups are shown due to space constraints.}
\label{fig:bias_comp}
\end{figure}

\begin{figure}[H]
\centering
\captionsetup[subfigure]{position=top,justification=raggedright,singlelinecheck=false}
\subfloat[][]{
    \includegraphics[width=0.49\linewidth]{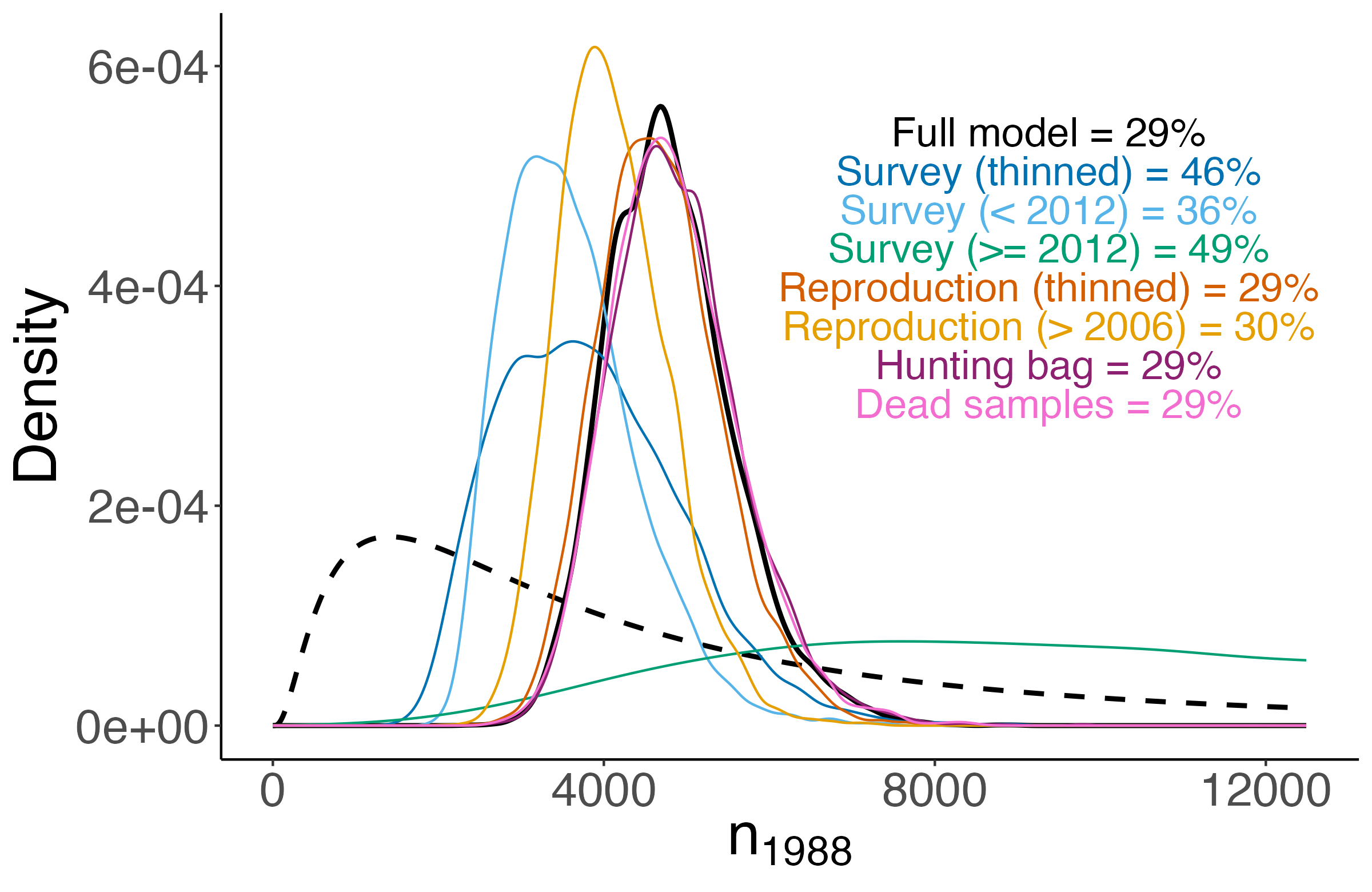}}
\subfloat[][]{
    \includegraphics[width=0.49\linewidth]{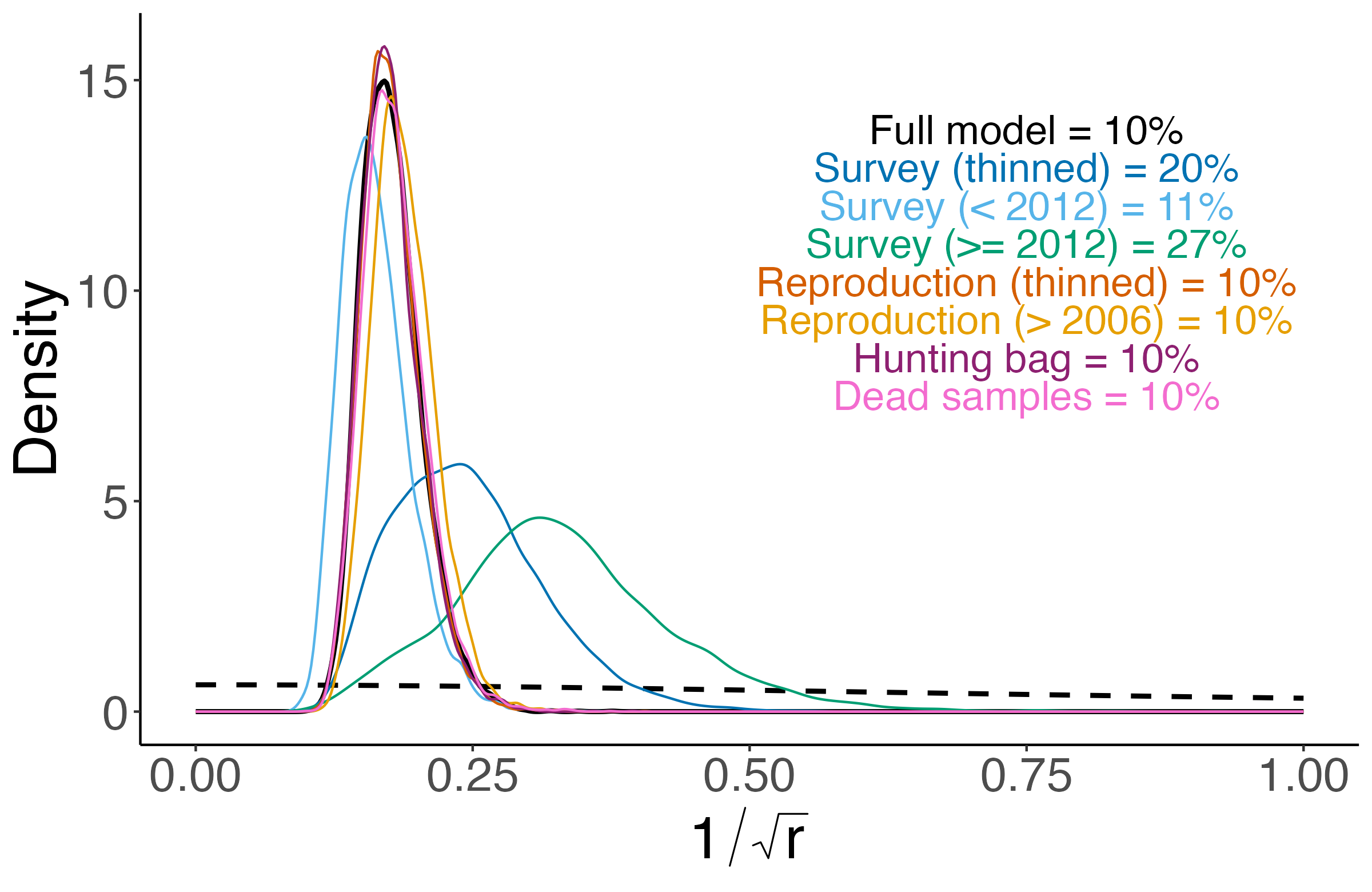}} 
\caption{\setstretch{1.0} Prior and posterior distributions and overlaps for (a) the initial population size and (b) the variance parameter for aerial surveys. Dashed lines represent prior distributions.}
\label{fig:popsize_comp}
\end{figure}

\begin{figure}[H]
\centering
\captionsetup[subfigure]{position=top,justification=raggedright,singlelinecheck=false}
\subfloat[][]{
    \includegraphics[width=0.4\linewidth]{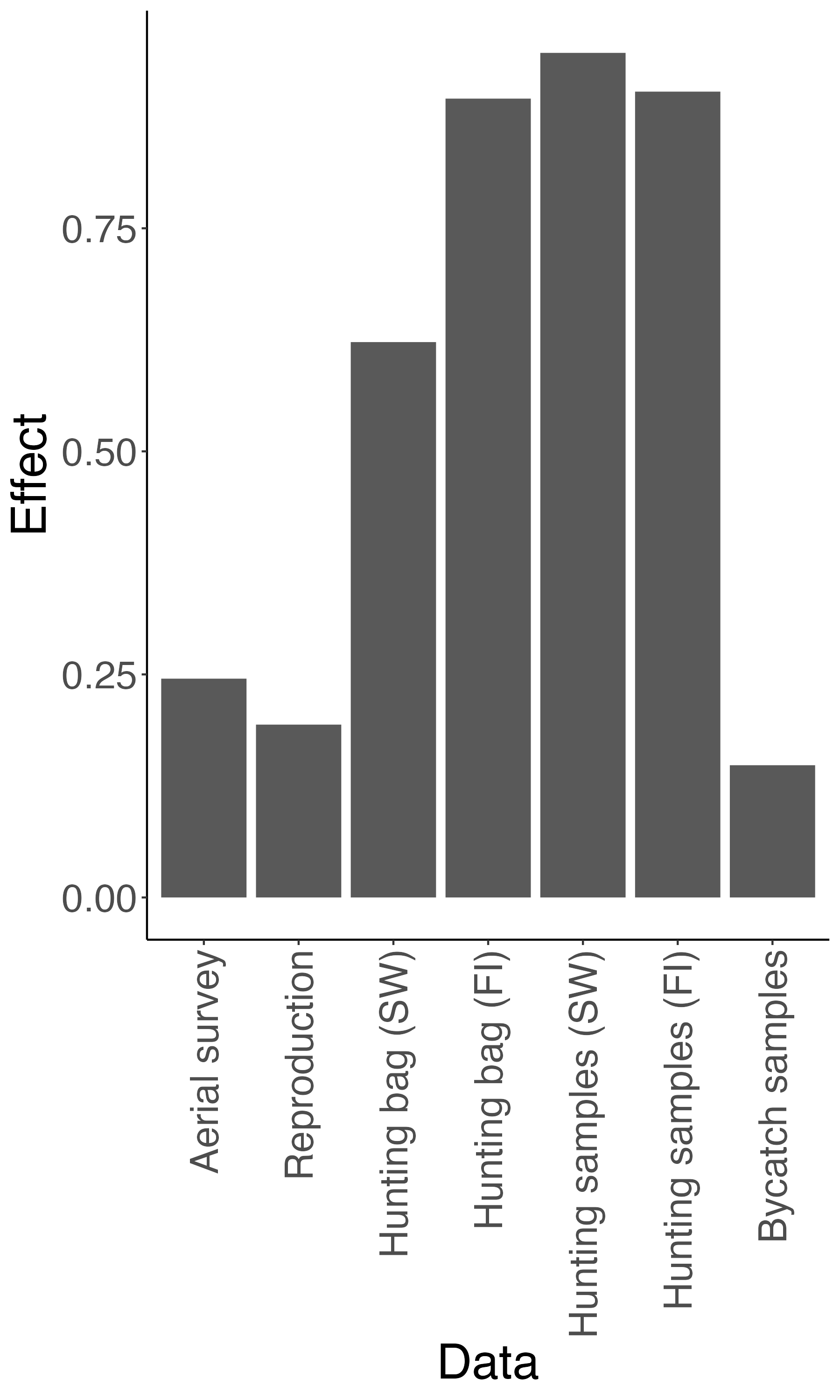}}
\subfloat[][]{
    \includegraphics[width=0.4\linewidth]{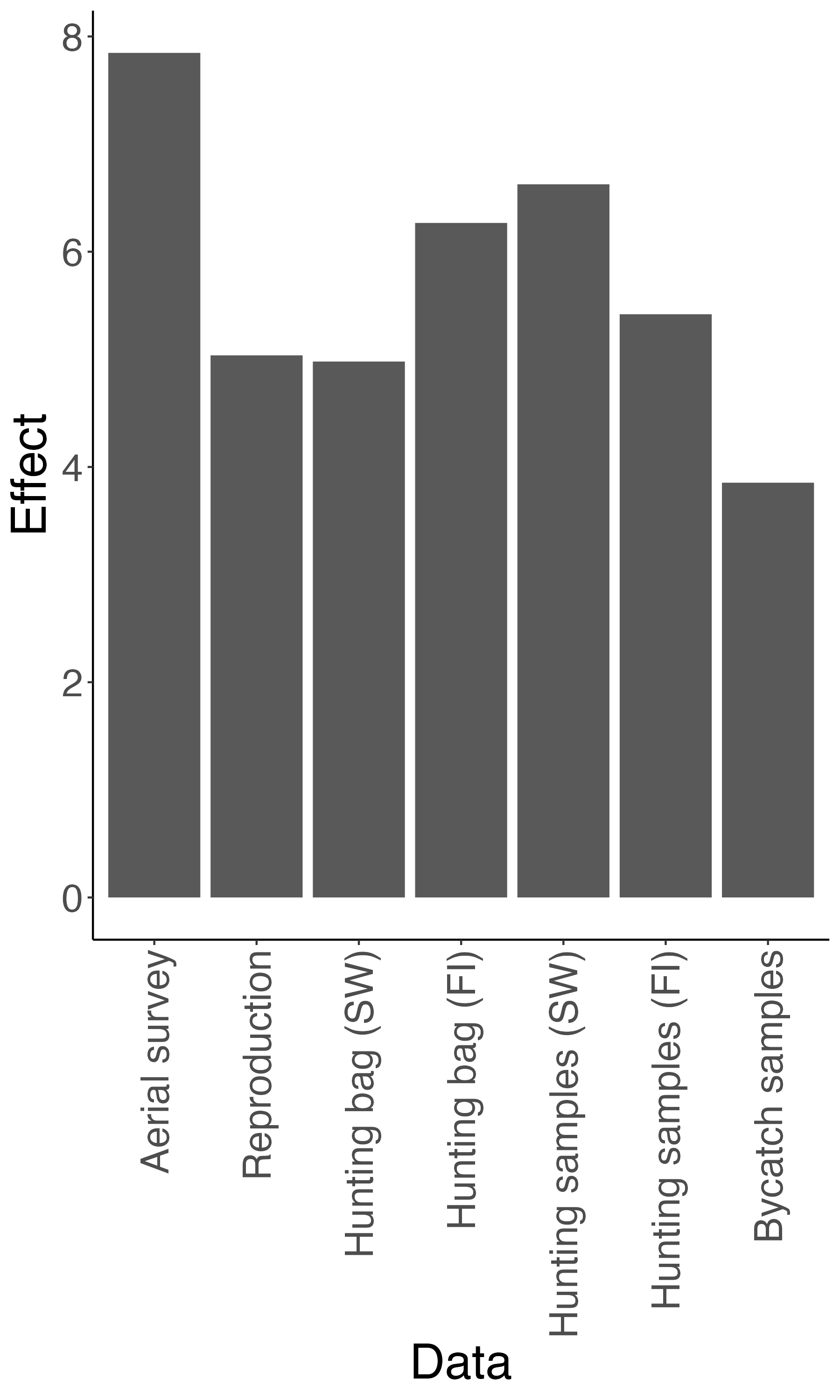}} 
\caption{\setstretch{1.0} Pareto-k statistics for different data types, measuring the relative influence of each data point on the posterior density. (a) The average k-statistic per data point and (b) the sum of k-statistics for each data type, following \citet{tinker2024}.}
\label{fig:pareto-k}
\end{figure}

\begin{figure}[H]
\begin{center}
 \includegraphics[width=0.9\linewidth]{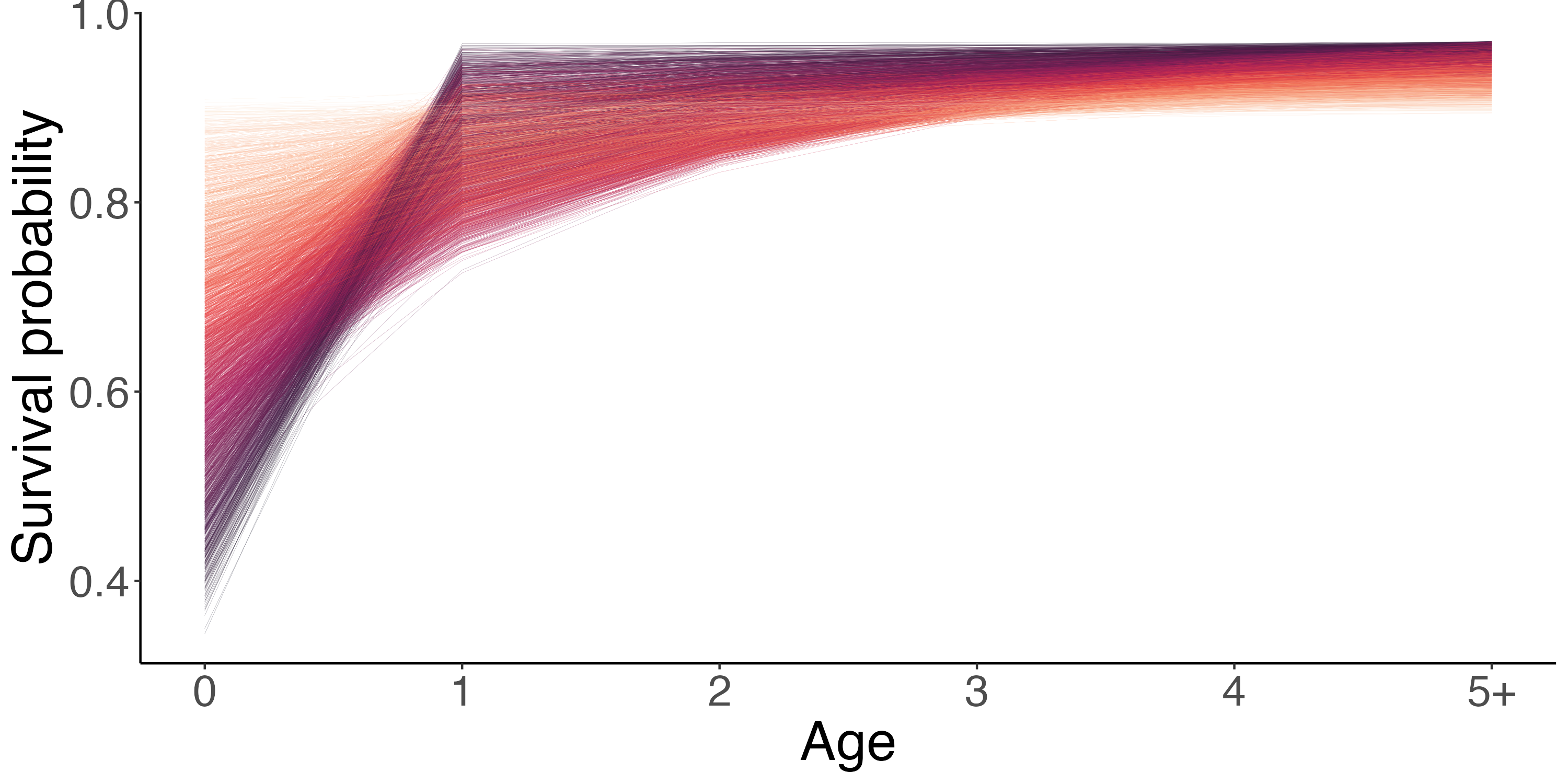}
 \caption{\setstretch{1.0} Posterior samples of age-specific survival probabilities. Samples are colored to highlight posterior correlations.}
 \label{fig:survival_corr}
\end{center}
\end{figure}

\begin{landscape}
\begin{longtable}{ |p{1.9cm} | p{1.66cm} p{1.66cm} p{1.66cm} p{1.66cm} p{1.66cm} p{1.66cm} p{1.66cm} | p{1.66cm} | p{1.66cm} p{1.66cm} | }
\caption{\setstretch{1.0} Posterior overlaps of each model fit with the posterior distribution of the full model. The models are (1) thinned aerial survey data, (2) pre-2012 survey data, (3) post-2012 survey data, (4) thinned reproduction data, (5) post-2007 reproduction data, (6) thinned harvest total data, (7) thinned dead sample data, (8) linearly decreasing reproductive failure rate, (9) left-shifted prior for $\hat{\omega}$, and (10) right-shifted prior for $\hat{\omega}$. Parameter descriptions are given in \ref{app:priors}: Table \ref{table:priors}. }\label{table:post_post_overlap} \\
    \hline
    Parameter & 1 & 2  & 3 & 4 & 5 & 6 & 7 & 8 & 9 & 10 \\
    \hline
    $n_{1988}$ & 0.54 & 0.41 & 0.22 & 0.93 & 0.62 & 0.96 & 0.97 & 0.98 & 0.75 & 0.74 \\
    $\phi_{f,5+}$ & 0.92 & 0.94 & 0.87 & 0.98 & 0.85 & 0.98 & 0.95 & 0.99 & 0.99 & 0.98 \\
    $\phi_{f,0}/\phi_{f,5+}$ & 0.81 & 0.79 & 0.64 & 0.97 & 0.92 & 0.97 & 0.97 & 0.97 & 0.98 & 0.98 \\
    $\nu_{0}$ & 0.98 & 0.98 & 0.97 & 0.98 & 0.96 & 0.98 & 0.98 & 0.98 & 0.99 & 0.98  \\
    $\nu_{1+}$ & 0.98 & 0.98 & 0.96 & 0.98 & 0.98 & 0.98 & 0.98 & 0.98 & 0.99 & 0.98 \\
    $c$ & 0.96 & 0.99 & 0.97 & 0.99 & 0.98 & 0.99 & 0.97 & 0.98 & 0.98 & 0.99 \\
    $p_Q^\text{fi}$ & 0.75 & 0.80 & 0.92 & 0.98 & 0.95 & 0.55 & 0.96 & 0.98 & 0.98 & 0.96 \\
    $p_Q^\text{sw(1)}$ & 0.90 & 0.51 & 0.67 & 0.98 & 0.92 & 0.90 & 0.95 & 0.98 & 0.92 & 0.89 \\
    $p_Q^\text{sw(2)}$ & 0.91 & 0.60 & 0.72 & 0.98 & 0.94 & 0.50 & 0.94 & 0.97 & 0.91 & 0.91 \\
    $\sigma_{\text{E}_\text{fi}}$ & 0.93 & 0.84 & 0.82 & 0.97 & 0.95 & 0.27 & 0.95 & 0.98 & 0.98 & 0.99 \\
    $\sigma_{\text{E}_\text{sw(1)}}$ & 0.98 & 0.98 & 0.96 & 0.98 & 0.97 & 0.88 & 0.98 & 0.98 & 0.98 & 0.99 \\
    $\sigma_{\text{E}_\text{sw(2)}}$ & 0.97 & 0.96 & 0.96 & 0.98 & 0.97 & 0.50 & 0.95 & 0.98 & 0.98 & 0.98 \\
    $\boldsymbol{\psi}^\text{bc}$ & 0.88-0.96 & 0.87-0.95 & 0.82-0.94 & 0.98-0.99 & 0.94-0.98 & 0.97-0.98 & 0.78-0.95 & 0.97-0.99 & 0.97-0.98 & 0.97-0.99 \\
    $\boldsymbol{\psi}^\text{sw(1)}$ & 0.88-0.97 & 0.83-0.97 & 0.69-0.92 & 0.97-0.99 & 0.97-0.99 & 0.95-0.98 & 0.56-0.93 & 0.98-0.99 & 0.97-0.99 & 0.98-0.99 \\
    $\boldsymbol{\psi}^\text{sw(2)}$ & 0.88-0.98 & 0.86-0.98 & 0.70-0.98 & 0.97-0.99 & 0.97-0.98 & 0.83-0.98 & 0.66-0.94 & 0.98-0.99 & 0.97-0.99 & 0.97-0.99 \\
    $\boldsymbol{\psi}^\text{fi}$ & 0.77-0.98 & 0.52-0.87 & 0.60-0.84 & 0.97-0.99 & 0.96-0.98 & 0.81-0.93 & 0.71-0.93 & 0.96-0.98 & 0.97-0.98 & 0.95-0.97 \\
    $\hat{\omega}$ & 0.78 & 0.70 & 0.67 & 0.97 & 0.94 & 0.96 & 0.97 & 0.98 & 0.68 & 0.62 \\
    $\delta$ & 0.91 & 0.61 & 0.84 & 0.98 & 0.97 & 0.90 & 0.90 & 0.98 & 0.97 & 0.95 \\
    $\alpha_0 / \alpha_1$ & 0.38 & 0.22 & 0.26 & 0.97 & 0.97 & 0.95 & 0.92 & 0.97 & 0.92 & 0.88 \\
    $\alpha_1$ & 0.50 & 0.14 & 0.33 & 0.98 & 0.96 & 0.97 & 0.96 & 0.98 & 0.95 & 0.94 \\
    $f$ & 0.57 & 0.79 & 0.51 & 0.97 & 0.83 & 0.98 & 0.96 & 0.98 & 0.88 & 0.87 \\
    $d$ & 0.58 & 0.28 & 0.49 & 0.98 & 0.89 & 0.98 & 0.98 & 0.98 & 0.90 & 0.93  \\
    $b_\text{max}$ & 0.95 & 0.90 & 0.88 & 0.76 & 0.85 & 0.95 & 0.97 & 0.80 & 0.96 & 0.97 \\
    $b_\text{min}/b_\text{max}$ & 0.84 & 0.94 & 0.83 & 0.83 & 0.54 & 0.98 & 0.91 & 0.99 & 0.98 & 0.99 \\
    $\beta_0$ & 0.93 & 0.90 & 0.84 & 0.69 & 0.40 & 0.92 & 0.91 & 0.96 & 0.97 & 0.97 \\
    $\beta_1$ & 0.87 & 0.92 & 0.87 & 0.86 & 0.44 & 0.95 & 0.95 & 0.96 & 0.97 & 0.97 \\
    $K$ & 0.98 & 0.98 & 0.98 & 0.98 & 0.98 & 0.98 & 0.98 & 0.98 & 0.98 & 0.98 \\
    $\theta_0$ & 0.99 & 0.96 & 0.98 & 0.75 & 0.97 & 0.98 & 0.98 & 0.77 & 0.98 & 0.98  \\
    $1 / \sqrt{r}$ & 0.45 & 0.77 & 0.19 & 0.98 & 0.85 & 0.97 & 0.96 & 0.98 & 0.95 & 0.90 \\
    $\pi_\text{s}$ & 0.95 & 0.98 & 0.98 & 0.77 & 0.86 & 0.98 & 0.99 & 0.93 & 0.98 & 0.98 \\
    $\pi_\text{c}$ & 0.98 & 0.98 & 0.99 & 0.72 & 0.98 & 0.97 & 0.98 & 0.95 & 0.98 & 0.98 \\
    $\kappa$ & 0.92 & 0.98 & 0.97 & 0.90 & 0.77 & 0.98 & 0.99 & 0.91 & 0.98 & 0.98 \\
    \hline
\end{longtable}
\end{landscape}

\bibliographystyle{inter_research}
\bibliography{ringed_seal_citations}
\setcounter{page}{1}

\section{Prior distributions}\label{app:priors}

\subsubsection*{Initial population}

Estimates of the basking population size of ringed seals in the Bothnian Bay during 1975-1987 range form 2,000-3,000 \citep{Harkonen1998} and literature suggests that the basking population may represents about 60\%-65\% of the total population size \citep{Harkonen1992, born2002, Kelly2010, Sundqvist2012}. We thus gave a wide $\text{Log-Gaussian}(\log(\frac{2,500}{0.65}), 1)$ prior for the population size in 1988, which corresponds to a prior median of 3,846 (and 95\% CI: 542-27,304). The initial demographic structure of the population was assumed to be the stable stage distribution implied by the vital rates in 1988, which is given by the leading eigenvector of the corresponding Leslie matrix \citep{caswell2001}. The density-dependent birth rate in the beginning of 1988 was estimated using the population size in the same year, and the leading eigenvector was computed using power iteration.

\subsubsection*{Mortality}

Following \citet{thomas2019}, we restricted the survival probability of female adults, $\phi_{f,5+}$, to the interval 0.80-0.97, and gave a uniform prior over this interval. Although the bounds of this interval were determined based on studies of grey seals, estimates of adult survival for ringed seals consistently fall within this range \citep{mclaren1958, durant1986, Kelly1988, harding2007, Reimer2019, koivuniemi2019}.

Because previous estimates for pup survival are highly variable \citep{smith1970, Kelly1988, kokko1998}, but are unequivocally lower than that of adults, we gave a $\text{Uniform}(0,1)$ prior to the ratio $\phi_{f,0} / \phi_{f,5+}$.
In our age-specific natural mortality model given in Equation \eqref{eq:natural_mortality}, a value of $c < 1$ implies that the natural mortality rate declines rapidly during the initial years of a seals life, and more slowly as seals approach maturity. In contrast, $c > 1$ implies that the mortality rate declines slowly when seals are young, and accelerates as seals grow older. Since the survival probability of ringed seals is known to stabilize once pups complete their first year \citep{kokko1998}, we gave $c$ a Uniform(0,1) prior.

We parameterized male mortality rates in terms of deviations from female rates. Specifically,
\begin{equation}
    \log(\mu_{m,0}) = \log(\mu_{f,0}) + \nu_0
\end{equation}
\begin{equation}
    \log(\mu_{m,1+}) = \log(\mu_{f,1+}) + \nu_{1+}
\end{equation}
where $\nu_0$ is the difference between the mortality rates of male and female pups on a log scale, and $\nu_{1+}$ is the same quantity for sub-adult and adult age groups. As the sex ratio of ringed seals is known to be close to parity \citep{mclaren1958, Helle1979, Lydersen1987}, $\nu$ must be relatively small. We therefore gave a N(0, $0.1^2$) prior which implies that, \textit{a priori}, the mortality rates of similarly aged males and females differ by less than a factor of 1.2 with 95\% probability. For a female survival probability of 0.5, this corresponds to a central 95\% interval of about 0.43-0.57 for a male of the same age. For a female survival probability of 0.95, the same interval for males is about 0.94-0.96.

\subsubsection*{Hunting and bycatch}

We parameterized the age and sex specific median harvest rates, $\hat{E}_{s,a}$, as the product of a global scaling parameter, $\hat{E}$, and a vector of age-specific weights, $\boldsymbol{\psi}$, defined over the unit simplex, so that
\begin{equation}
    \hat{E}_{s,a} = \hat{E}\psi_{s,a}.
\end{equation}
Decoupling the magnitude of the harvest rate from its allocation to different demographic groups significantly improves posterior sampling performance.
Using \ref{app:hunting_probability}: Equation~\eqref{H_tot}, the scaling parameter $\hat{E}$ can be reparametrized as
\begin{equation}
    \hat{E} = -k^{-1} \log(1-p_Q)
\end{equation}
where $p_Q=H/Q$ and $k$ is a constant. We gave Uniform(0,1) priors to $p_Q^\text{fi}$, $p_Q^\text{sw(1)}$ and $p_Q^\text{sw(2)}$, corresponding respectively to Finnish hunting, Swedish hunting in the spring, and Swedish hunting in the fall. Considering population size estimates prior to 1988 and historical average ice extent, and further assuming approximately unbiased hunting, we set $k=5$ for Finnish hunting and $k=100$ for Swedish hunting. 

Annual variation in harvest rate was modeled as
\begin{equation}
    E_{s,a,t}=\hat{E}_{s,a} e^{\epsilon_t}
\end{equation} 
where $\epsilon_t \sim N(0, \sigma_E)$ (see Equation~\eqref{eq:hunting_effort}). For each of $\sigma_{E_\text{sw(1)}}$, $\sigma_{E_\text{sw(2)}}$ and $\sigma_{E_\text{fi}}$, we gave weakly informative Cauchy$_+(0,0.1)$ priors.

Our model includes four "bias vectors" for Finnish hunting, Swedish hunting in the spring and fall,  and bycatch (Section~\ref{sec:By-catch_samples} and 
\ref{sec:Hunting_bag_and_hunting_samples}), which we denote by $\boldsymbol{\psi}_\text{fi}$, $\boldsymbol{\psi}_\text{sw(1)}$, $\boldsymbol{\psi}_\text{sw(2)}$ and $\boldsymbol{\psi}_\text{bc}$, respectively. We gave logistic-Gaussian priors to these bias vectors as this choice offers flexibility in incorporating prior information in the form of correlations between the parameter values for different age and sex classes. The logistic-Gaussian distribution for the 12-dimensional vector $\boldsymbol{\psi}$ involves applying a logistic transformation to an 11-dimensional multivariate Gaussian distribution, so that
\begin{equation}
    \psi_i = \frac{e^{g_i}}{1+\sum_{j=1}^{11} e^{g_j}}
    \label{eq:multivariate_logit}
\end{equation}
where $\mathbf{g} =[g_1,\dots,g_{11}]^T \sim \mathcal{N}\big(\boldsymbol{\mu}, \ \mathbf{\Sigma} \big)$.

For all three bias vectors in our model, we assumed the $\mathbf{0}$ vector as the expectation for $\mathbf{g}$. We defined the covariance matrix as $\boldsymbol{\Sigma} = \sigma^2 \boldsymbol{\mathcal{P}}$, where $\sigma$ is the standard deviation along each component and $\boldsymbol{\mathcal{P}}$ is a matrix of correlations between bias parameters in logit space.

Due to their ecological differences, we expected differences to be greatest between pups, non-breeding sub-adults and breeding adults. In contrast, we expected that the parameters of male and female seals of the same age, as well as among the sub-adult age groups, are relatively more similar. We therefore assumed zero correlation between the parameters of pups, the sub-adult age groups, and adults, and defined the correlation between each demographic class through the matrix
\begin{equation}
\boldsymbol{\tilde{\mathcal{P}}} = 
    \begin{bmatrix}
        \boldsymbol{\mathcal{P'}} & \varrho \boldsymbol{\mathcal{P'}} \\
        \varrho \boldsymbol{\mathcal{P'}} & \boldsymbol{\mathcal{P'}} \\
    \end{bmatrix}
\end{equation}
where,
\begin{equation}
\boldsymbol{\mathcal{P'}} = 
    \begin{bmatrix}
        1 & 0 & 0 & 0 & 0 & 0 \\
        0 & 1 & \varrho & \varrho^2 & \varrho^3 & 0 \\
        0 & \varrho & 1 & \varrho & \varrho^2 & 0 \\
        0 & \varrho^2 & \varrho & 1 & \varrho & 0 \\
        0 & \varrho^3 & \varrho^2 & \varrho & 1 & 0 \\
        0 & 0 & 0 & 0 & 0 & 1
    \end{bmatrix}
\end{equation}
is the correlation matrix for seals of the same sex, and $\varrho$ is a correlation coefficient.

To determine the value of $\varrho$, we considered the relative bias between two demographic classes $i$ and $j$, which is given by
\begin{equation}
    \frac{\psi_i}{\psi_j} = e^{g_i-g_j}
\end{equation}
Since the joint marginal distribution of any $g_i$ and $g_j$ is also Gaussian distributed, the conditional marginal distribution for the log relative bias is
\begin{equation}
    \log\left(\frac{\psi_i}{\psi_j}\right) \bigg| (g_i = x) \sim \mathcal{N}\bigg(x(1-\varrho), \ \sigma \sqrt{1-\varrho^2}\bigg)
\end{equation}
Conditional on $g_i$, we assumed that the chances of becoming hunted or bycaught differ by less than a factor of four between opposite sexes of the same age and between sub-adults of the same sex with a one year age difference with 95\% prior probability. Assuming $g_i > g_j$ without loss of generality, smaller values of $g_i$ imply a smaller correlation coefficient $\varrho$, and hence a less informative prior. Therefore, we chose a value of $g_i$ that was two standard deviations below its expected value (i.e. $x = -2\sigma$):
\begin{equation}
    \log(4) = -2\sigma(1 - \varrho) + 2\sigma\sqrt{1-\varrho^2}
\end{equation}

We chose a large standard deviation of $\sigma=5$ for all components of $\mathbf{g}$ (note that $\sigma$ is on a logarithmic scale in Equation \eqref{eq:multivariate_logit}), which yields a correlation coefficient of $\varrho = 0.9887$.

Since the logistic transformation in Equation \eqref{eq:multivariate_logit} requires only 11 degrees of freedom for the 12 demographic groups in our model, we arbitrarily set $g_{12}=0$, and computed the conditional covariance matrix:
\begin{equation}
    \boldsymbol{\mathcal{P}} = 
    \boldsymbol{\tilde{\mathcal{P}}}_{\mathbf{k},\mathbf{k}} - 
    \sigma^{-2}
    \boldsymbol{\tilde{\mathcal{P}}}_{\mathbf{k},12}
    \boldsymbol{\tilde{\mathcal{P}}}_{\mathbf{k},12}^\mathbf{T}
\end{equation}
Here, $\mathbf{k}=\{1,\dots,11\}$, $\boldsymbol{\tilde{\mathcal{P}}}_{\mathbf{k},\mathbf{k}}$ is a matrix composed of the first 11 rows and columns of $\boldsymbol{\tilde{\mathcal{P}}}$ and $\boldsymbol{\tilde{\mathcal{P}}}_{\mathbf{k},12}$ is a vector composed of the first 11 elements of the 12th column of $\boldsymbol{\tilde{\mathcal{P}}}$.

\subsubsection*{Haul-out}

Estimates for the hauled-out proportion of sub-adult and adult seals during the moulting period range from 50\% to 84\% \citep{Harkonen1992}, with estimates generally falling closer to 60\%-65\% \citep{born2002, Kelly2010}. We therefore gave an informative, scaled $\text{Beta}(4, 4)$ prior within the interval 0.4-0.9 to $\hat{\omega}$, which has an expectation of 0.65 (and 95\% CI: 0.49-0.81).

In Equation \eqref{eq:haul_out_eq}, we transformed the ice cover covariate such that it had zero mean and unit standard deviation. Following \citep{gelman2008}, we then gave a weakly informative Cauchy(0, 1.25) prior for the slope parameter $\alpha_1$ of the logistic equation. Due to the high correlation between $\alpha_0$ and $\alpha_1$, we parametrized $\alpha_0$ in terms of the ratio $\alpha_0 / \alpha_1$, and gave this ratio a weakly informative Cauchy(0, 5) prior.
For $\delta$ and $d$ we gave Uniform(0, 1) priors. For the half-saturation constant $f$, we gave a weakly informative Cauchy$_+(0, 37)$ prior, where the scale parameter was set to be equal to the total area of the Bothnian Bay (in 000's).

\subsubsection*{Reproduction}

We gave uniform priors to the maximum birth rate $b_\text{max}$ and the ratio $b_\text{min}/b_\text{max}$. In Equation \eqref{eq:logistic_preg_rate}, the time covariate was transformed to have zero mean and unit standard deviation. Following \citet{gelman2008}, we gave a weakly informative Cauchy(0,10) prior to the intercept $\beta_0$ of the logistic equation.
Due to small sample sizes in years prior to 2018, the rate parameter, $\beta_1$, of the logistic pregnancy rate function required a more informative prior that restricted its maximum value. 
A maximum value for $\beta_1$ can be determined on theoretical grounds by considering the greatest possible increase in the birth rate in a single year.

Consider a population growing at a rate $\lambda$ per year, where a fraction $u$ of adult females suffer from uterine occlusions caused by organochlorine contamination, and are thus infertile \citep{helle1980, Kauhala2019}. The remaining fraction $1-u$ of healthy individuals give birth at the maximum birth rate $b_\text{max}$. To estimate an upper bound on the annual increase in birth rates, we consider a scenario in which the Baltic Sea instantaneously becomes free of all contaminants such that all subsequent cohorts born into the population consist entirely of healthy individuals. Since uterine pathologies tend to be permanent \citep{Kauhala2019}, the birth rate will increase as infertile adults die out and become replaced by young, healthy seals. Denoting by $\phi$ the probability that an adult female survives the year, the discrete-time dynamics of the birth rate can be modelled as
\begin{equation}
    b_{t+1} = b_\text{max}-\frac{\phi}{\lambda}(b_\text{max} - b_t)
\end{equation}
with the annual change in birth rate given by
\begin{equation}
    b_{t+1}-b_{t} = (1-\frac{\phi}{\lambda})(b_\text{max}-b_t).
    \label{eq:birth_rate_dynamics}
\end{equation}
We can ensure that the maximum rate of change along the logistic curve in Equation \eqref{eq:logistic_preg_rate} does not exceed the change in pregnancy rate implied by Equation \eqref{eq:birth_rate_dynamics} by satisfying the inequality
\begin{equation}
    \beta_1 < 4(1-\frac{\phi}{\lambda})\frac{(b_\text{max}-b_t)}{(b_\text{max}-b_\text{min})}
    \label{eq:beta_1_inequality}
\end{equation}
An appropriate upper bound on $\beta_1$ can therefore be determined by finding the maximum plausible value of the right hand side of this inequality. The ratio $\frac{\phi}{\lambda}$ is minimized when the maximum possible growth rate is achieved for a given adult survival probability $\phi$. We therefore consider a scenario in which pup and sub-adult survival probabilities are as high as that of adults. With these assumptions, the well known Euler-Lotka Equation
\begin{equation}
    \frac{b}{2} \sum_{i=5}^\infty \bigg(\frac{\phi}{\lambda}\bigg)^i = 1
\end{equation}
implies that, when constant, $b$ is uniquely determined by the ratio $\frac{\phi}{\lambda}$. Using this result as an approximation, Equation \eqref{eq:beta_1_inequality} can be written in terms of the ratio $\frac{\phi}{\lambda}$ as
\begin{equation}
    \beta_1 < \frac{4}{(b_\text{max}-b_\text{min})}
    (1-\frac{\phi}{\lambda})
    \bigg[b_\text{max}-\frac{2(1-\phi/\lambda)}{(\phi/\lambda)^5}\bigg]
\end{equation}
Assuming minimum and maximum pregnancy rates of $0.25$ \citep{helle1980, Kauhala2019} and $0.91$ \citep{Lydersen1987, harding2007} respectively, the right hand side of this inequality obtains a global maximum of about $0.35$, implying a maximum increase in the birth rate of about 0.06 per year. We therefore gave a Student-t$_{\nu = 5}$(0, 1.8) prior to $\beta_1$ on a scale in which the year covariate was transformed to have unit standard deviation. The scale parameter was chosen to equal half of the theoretical maximum of $0.35$.

\subsubsection*{Density dependence} 

\ref{app:density_dependence}: Equation \eqref{eq:p_0} implies that $e^{-\theta_0}$ cannot be less than $b_\text{max}$. We therefore gave a Uniform($b_\text{max}$, 1) prior to $e^{-\theta_0}$. Instead of assigning a prior distribution to $\theta_1$, we parametrized our model in terms of the carrying capacity, $K$ (see \ref{app:density_dependence}: Equation \eqref{eq:carrying_capacity}). Estimates of the total number of Baltic ringed seals at the turn of the 20$^\text{th}$ century range from 50,000 to 450,000 \citep{durant1986, Harding1999, Kokko1999}. However, estimates on the lower end of this range are based on very conservative assumptions, including the absence of density dependent regulation. The carrying capacity of the Baltic Sea is therefore highly likely to exceed 200,000 ringed seals \citep{Harding1999}. We assumed that, in the absence of hunting pressure, the carrying capacity of the Baltic Sea as a whole is likely to be between 200,000 and 450,000 ringed seals. Based on the distribution of sea ice across the Baltic Sea in late April, about 75\% of the Baltic ringed seal population may be present in the Bothnian Bay during the aerial surveys. We therefore gave a Gamma(18, 8e-5) prior to the carrying capacity, $K$, of the Bothnian Bay, which has an expectation of 225,000 (and 95\% CI: 133,000 - 340,000).

\subsubsection*{Observation model parameters}

For all parameters relating to observations of reproductive success, namely $\pi_s$, $\pi_c$ and $\kappa$, we give uniform priors. For the overdispersion parameter of the negative binomial model, we penalized small values that lead to large overdispersion, favoring the simpler Poisson observation model. Therefore, we gave a weakly informative Cauchy$_+(0,1)$ prior to $1/\sqrt{r}$. When the expectation of a negative binomial distribution is large, $1/\sqrt{r}$ is approximately equal to the coefficient of variation. For all parameters related to sampling bias from the Swedish harvests during the spring, namely $x_0$, $x_1$ and $q_{s,a}$ (\ref{app:sampling_bias_sw}), we gave Uniform(0, 1) priors. 

\newpage

\begin{landscape}
\begin{singlespace}
\begin{longtable}{ |p{12cm} p{1.5cm} p{4.5cm} p{3.5cm}|  }
\caption{\setstretch{1.0} Prior distributions and posterior estimates for model parameters. Posterior estimates are provided for the untransformed parameters.}\label{table:priors} \\
    \hline
    Parameter name & Symbol & Prior & Posterior median (95\% CI) \\
    \hline
    Initial population size & $n_{1988}$ & $n_{1988} \sim \text{log-N}(8.25,1)$ & 4,747 (3,516-6,559) \\
    Survival prob. of female adults & $\phi_{f,5+}$ & $\phi_{f,5+} \sim U(0.80,0.97)$ & 0.95 (0.91-0.97) \\
    Survival prob. of female pups & $\phi_{f,0}$ & $\frac{\phi_{f,0}}{\phi_{f,5+}} \sim U(0,1)$ & 0.63 (0.42-0.87) \\
    Deviation between mortality rates of male and female pups & $\nu_{0}$ & $\nu_{0} \sim N(0,0.1^2)$ & 0.01 (-0.18-0.20) \\
    Deviation between mortality rates of 1+ yo male and females & $\nu_{1+}$ & $\nu_{1+} \sim N(0,0.1^2)$ & 0.00 (-0.19-0.18) \\
    Shape param. for survival curve & $c$ & $c \sim U(0,1)$ & 0.49 (0.03-0.97) \\
    Scale of FI harvest rate & $\hat{E}_\text{fi}$ & $p_Q^\text{fi} \sim U(0,1)$ & 0.45 (0.26-0.82) \\
    Scale of spring SW harvest rate & $\hat{E}_\text{sw(1)}$ & $p_Q^\text{sw(1)} \sim U(0,1)$ & 2e-3 (6e-4-4e-3) \\
    Scale of fall SW harvest rate & $\hat{E}_\text{sw(2)}$ & $p_Q^\text{sw(2)} \sim U(0,1)$ & 1e-3 (4e-4-4e-3) \\
    SD for FI harvest rate & $\sigma_{\text{E}_\text{fi}}$ & $\sigma_{\text{E}_\text{fi}} \sim \text{Cauchy}_+(0,0.1)$ & 0.48 (0.19-1.29) \\
    SD for spring SW harvest rate & $\sigma_{\text{E}_\text{sw(1)}}$ & $\sigma_{\text{E}_\text{sw(1)}} \sim \text{Cauchy}_+(0,0.1)$ & 1.0 (0.6-2.0) \\
    SD for fall SW harvest rate & $\sigma_{\text{E}_\text{sw(2)}}$ & $\sigma_{\text{E}_\text{sw(2)}} \sim \text{Cauchy}_+(0,0.1)$ & 1.2 (0.6-3.3) \\
    Bias vector for bycatch & $\boldsymbol{\psi}^\text{bc}$ & $\boldsymbol{\psi}^\text{bc} \sim \text{logit-N}(\mathbf{0},\mathbf{\Sigma})$ & See Figure~\ref{fig:bycatch_bias}. \\
    Bias vector for spring SW hunting & $\boldsymbol{\psi}^\text{sw(1)}$ & $\boldsymbol{\psi}^\text{sw(1)} \sim \text{logit-N}(\mathbf{0},\mathbf{\Sigma})$ & See Figure~\ref{fig:sw_bias_spring}. \\
    Bias vector for fall SW hunting & $\boldsymbol{\psi}^\text{sw(2)}$ & $\boldsymbol{\psi}^\text{sw(2)} \sim \text{logit-N}(\mathbf{0},\mathbf{\Sigma})$ & See Figure~\ref{fig:sw_bias_fall}. \\
    Bias vector for FI hunting & $\boldsymbol{\psi}^\text{fi}$ & $\boldsymbol{\psi}^\text{fi} \sim \text{logit-N}(\mathbf{0},\mathbf{\Sigma})$ & See Figure~\ref{fig:fi_bias}. \\
    Baseline haul-out prob. for 1+ yo & $\hat{\omega}$ & $\frac{\hat{\omega}-0.4}{0.5} \sim \text{Beta}(4,4)$ & 0.60 (0.48-0.74) \\
    Ratio of pup to adult haul-out & $\delta$ & $\delta \sim U(0,1)$ & 0.58 (0.20-0.94) \\
    Intercept for haul-out model & $\alpha_0$ & $\frac{\alpha_0}{\alpha_1} \sim \text{Cauchy}(0,5)$ & 17.6 (6.2-209.7) \\
    Slope for haul-out model & $\alpha_1$ & $\alpha_1 \sim \text{Cauchy}(0, 1.25)$ & 11.0 (4.5-130.7) \\
    Half-saturation const. & $f$ & $f \sim \text{Cauchy}_+(0, 37)$ & 11.9 (0.9-22.9)  \\
    Ratio btw. asymptotes of movement rate into the water & $d$ & $d \sim U(0,1)$ & 0.03 (0.00-0.19)  \\
    Max. birth rate & $b_\text{max}$ & $b_\text{max} \sim U(0,1)$ & 0.77 (0.69-0.87) \\
    Min. birth rate & $b_\text{min}$ & $\frac{b_\text{min}}{b_\text{max}} \sim U(0,1)$ & 0.25 (0.02-0.51) \\
    Intercept for logistic birth rate & $\beta_0$ & $\beta_0 \sim  \text{Cauchy}(0, 10)$ & 1.7 (-0.5-5.0) \\
    Slope for logistic birth rate & $\beta_1$ & $\beta_1 \sim \text{t}_{\nu=5}(0,1.8)$ & 2.6 (0.9-7.5) \\
    Carrying capacity & $K$ & $K \sim \text{Gamma}(18,$8e-5$)$ & 2.2e5 (1.3e5-3.4e5) \\
    Base pregnancy failure rate & $\theta_0$ & $e^{-\theta_0} \sim U(b_\text{max},1)$ & 0.10 (0.01-0.25) \\
    Dispersion parameter & $r$ & $\frac{1}{\sqrt{r}} \sim \text{Cauchy}_+(0,1)$ & 33.0 (17.1-56.6) \\
    Placental scar detection prob. & $\pi_\text{s}$ & $\pi_\text{s} \sim U(0, 1)$ & 0.93 (0.85-0.99) \\
    CA detection prob. & $\pi_\text{c}$ & $\pi_\text{c} \sim U(0,1)$ & 0.97 (0.91-1.00) \\
    CA prob. for non-pregnant seals & $\kappa$ & $\kappa \sim U(0,1)$ & 0.09 (0.00-0.26) \\
    Sampling prob. for seals $<$100cm & $x_0$ & $x_0 \sim U(0,1)$ & See \ref{app:sampling_bias_sw}. \\
    Sampling prob. for seals $>$100cm & $x_1$ & $x_1 \sim U(0,1)$ & See \ref{app:sampling_bias_sw}. \\
    Probability that seal is $>$100cm & $q_{s,a}$ & $q_{s,a} \sim U(0,1) \ \ \forall s,a$ & See \ref{app:sampling_bias_sw}. \\
    \hline
\end{longtable}
\end{singlespace}
\end{landscape}

\bibliographystyle{inter_research}
\bibliography{ringed_seal_citations}
\setcounter{page}{1}

\section{Hunting mortality}\label{app:hunting_probability}

For both Finnish hunting and Swedish hunting during the spring and fall, the basic form of the system of ordinary differential equations (ODE) \eqref{eq:dHi} and \eqref{eq:dni} can be expressed as
\begin{equation}
    \varepsilon \frac{dH_i}{d\tau} = E_i \varepsilon(Q - H(\tau)) \tilde{n}_i(\tau)
\end{equation}
\begin{equation}
    \frac{d\tilde{n}_i}{d\tau} = - \big[ E_i \varepsilon(Q - H(\tau)) + \mu_i \big] \tilde{n}_i(\tau)
\end{equation}
where $\tau$ is the time since the onset of the hunting season, $H(\tau)$ is the total harvest at time $\tau$, $\tilde{n}_i(\tau)$ is the number (or density on ice) of seals of class $i$ that are still alive at time $\tau$, and $\varepsilon$ is a constant governing the magnitude of hunting relative to the population size, often called a \textit{timescale parameter} \citep{kuehn2015}. If we assume that the number of hunting licenses is small, we can take the limit $\varepsilon \to 0$ to decouple the second ODE from the first. In other words, we assume that the dynamics of hunting and natural mortality unfold over different timescales, and approximate the joint dynamics of the population and the harvests as though the population was unaffected by hunting.
\begin{equation}
    \frac{dH_i}{d\tau} = E_i (Q - H(\tau)) \tilde{n}_i(\tau)
\end{equation}
\begin{equation}
    \frac{d\tilde{n}_i}{d\tau} = - \mu_i \tilde{n}_i(\tau)
\end{equation}
The solution of the second ODE is simply an exponential decay with the initial condition $\tilde{n}_i(0)$, corresponding to the number (or density on ice) of seals of class $i$ at the beginning of the hunting season. Substituting this solution into the first ODE yields the non-autonomous ODE,
\begin{equation}
    \frac{dH_i}{d\tau} = E_i (Q - H(\tau)) \tilde{n}_i(0)e^{-\mu_i \tau}
    \label{fi_diff_class}
\end{equation}
The total rate of hunting is given by a sum of Equation \eqref{fi_diff_class} over all demographic classes,
\begin{equation}
    H'(\tau) = \sum_i E_i (Q - H(\tau)) \tilde{n}_i(0)e^{-\mu_i \tau}
    \label{fi_diff_tot}
\end{equation}
which is separable, and has the explicit time-dependent solution for the harvest totals,
\begin{equation}
    H(\tau) = Q \left(1 - \exp\left(- \sum_{i} E_i \tilde{n}_i(0) \frac{1-e^{-\mu_i \tau}}{\mu_i}\right)\right)
    \label{H_tot}
\end{equation}

To estimate the demographic composition of the harvests, we first divide Equation \eqref{fi_diff_class} by Equation \eqref{fi_diff_tot},
\begin{equation}
   \frac{1}{{H'}(\tau)}\frac{dH_i}{d\tau} = 
   \frac{E_i \tilde{n}_i(0) e^{-\mu_i \tau}}{\sum_{i,j} E_i \tilde{n}_i(0) e^{-\mu_i \tau}} 
   \label{fi_div}
\end{equation}
After separating the variables and integrating both sides,
\begin{equation}
   H_i(\tau) = \int_0^{\tau} \frac{E_i \tilde{n}_i(0) e^{-\mu_i \tau'}}{\sum_{j} E_j \tilde{n}_j(0) e^{-\mu_j \tau'}} H'(\tau')d\tau' 
\end{equation}
Dividing both sides by the total harvests and evaluating at a time corresponding to the length of the hunting season, $\tau^*$, yields an expression for the expected demographic composition of the harvests
\begin{equation}
   \frac{H_i(\tau^*)}{H(\tau^*)} = \int_0^{\tau^*} g_i(\tau) h(\tau)d\tau 
\end{equation}
where
\begin{equation}
   g_i(\tau) = \frac{E_i \tilde{n}_i(0) e^{-\mu_i \tau}}
   {\sum_{j} E_j \tilde{n}_j(0) e^{-\mu_j \tau}}
\end{equation}
\begin{equation}
   h(\tau) = \frac{H'(\tau)}{H(\tau^*)}
\end{equation}
Since $H'(\tau)$ integrates to $H(\tau^*)$ over the interval $[0,\tau^*]$, $h(\tau)$ is a valid probability density function defined in this interval, and represents the conditional probability that a randomly sampled seal from the harvests was hunted within the time interval $[\tau, \tau+d\tau]$. The composition of the harvest is therefore an expectation of $g_i(\tau)$ over the distribution $h(\tau)$:
\begin{equation}
   \frac{H_i(\tau^*)}{H(\tau^*)} = E_{h(\tau)}[g_i(\tau)]
\end{equation}
Since all three hunting seasons in our model are shorter than two months, and the mortality rates of all age and sex groups are expected to be relatively low, the Taylor expansions of $g_i(\tau)$ will be dominated by the first order terms throughout the hunting season. Thus,
\begin{equation}
   \frac{H_i(\tau^*)}{H(\tau^*)} \approx g_i(E_{h(\tau)}[\tau]) = 
   \frac{E_i \tilde{n}_i(0) e^{-\mu_i \hat{\tau}}}
   {\sum_{j} E_j \tilde{n}_j(0) e^{-\mu_j \hat{\tau}}}
\end{equation}
where $\hat{\tau} = E_{h(\tau)}[\tau]$ indicates the average time of hunting.

While the evaluation of $\hat{\tau}$ requires numerical integration which can be computationally intensive, empirical records of hunting times can be regarded as samples from the distribution $h(\tau)$. Therefore, a reasonable approximation of $\hat{\tau}$ can be obtained by calculating the empirical average time of hunting in each year. 

Using the derivations outlined above, and the notation and indexing used in Section~\ref{sec:survival_and_mortality}, we can express the probabilities that seals become hunted in Finland, Sweden during the spring or Sweden during the fall as
\begin{equation}
    \rho^{H^\text{fi}}_\text{s,a,t} = 
    \frac{E^\text{fi}_{s,a,t} \omega_{s,a,t} \phi_{s,a}^{\hat{\tau}_\text{fi}}}
    {\sum_{i,j} E^\text{fi}_{i,j,t} \omega_{s,a,t} \phi_{i,j}^{\hat{\tau}_\text{fi}} n_{i,j,t}} 
    \bigg[ 
    1 - \exp\left(\sum_{i,j} E^\text{fi}_{i,j,t} 
    \frac{\omega_{s,a,t} \phi_{i,j}^{\tau^0_\text{fi}} n_{i,j,t}}{C_t} 
    \frac{1-\phi_{i,j}^{\tau_\text{fi}}}{\log(\phi_{i,j})}\right) 
    \bigg] Q^\text{fi}_t
    \label{rho_fi_full}
\end{equation}
\begin{equation}
    \rho^{H^\text{sw(1)}}_{s,a,t} = 
    \frac{E^\text{sw(1)}_{s,a,t} \phi_{s,a}^{\hat{\tau}_\text{sw(1)}}}{\sum_{i,j} E^\text{sw}_{i,j,t} \phi_{i,j}^{\hat{\tau}_\text{sw(1)}} 
    n_{i,j,t}} 
    \bigg[ 
    1 - \exp\left(\sum_{i,j} E^\text{sw(1)}_{i,j,t} \phi_{i,j}^{\tau^0_\text{sw(1)}} n_{i,j,t} 
    \frac{1-\phi_{i,j}^{\tau_\text{sw(1)}}}{\log(\phi_{i,j})}\right) 
    \bigg] Q^\text{sw}_t
    \label{rho_sw_spring_full}
\end{equation}
\begin{equation}
    \rho^{H^\text{sw(2)}}_{s,a,t} = 
    \frac{E^\text{sw(2)}_{s,a,t} \phi_{s,a}^{\hat{\tau}_\text{sw(2)}} 
    (1-\rho^{H^\text{sw(1)}}_{s,a,t}-\rho^{H^\text{fi}}_{s,a,t})}
    {\sum_{i,j} E^\text{sw}_{i,j,t} \phi_{i,j}^{\hat{\tau}_\text{sw(2)}} n^*_{i,j,t}} 
    \bigg[ 
    1 - \exp\left(\sum_{i,j} E^\text{sw(2)}_{i,j,t} \phi_{i,j}^{\tau^0_\text{sw(2)}} n^*_{i,j,t} 
    \frac{1-\phi_{i,j}^{\tau_\text{sw(2)}}}{\log(\phi_{i,j})}\right) 
    \bigg] \tilde{Q}^\text{sw}_t
    \label{rho_sw_fall_full}
\end{equation}
Here, $\tau^0$ and $\hat{\tau}$ respectively denote the time until the beginning, and the middle of the hunting season as measured from the census point, and $\tau$ denotes the length of the hunting season. With a post-breeding census corresponding approximately to March 1, we assume $\tau^0_\text{fi} \approx 1.5/12$, $\tau^0_\text{sw(1)} \approx 2/12$ and $\tau^0_\text{sw(2)} \approx 6/12$. Finnish hunting is assumed to take place before the end of May when all sea ice has typically melted in the Bothnian Bay (i.e. $\tau_\text{fi} \approx 1.5/12$), and the spring and fall hunting seasons in Sweden are each assumed to take place over a two month period (i.e. $\tau_\text{sw(1)} = \tau_\text{sw(2)} \approx 2/12$) (Figure~\ref{fig:lifecycle}). For fall hunting in Sweden, $n^*_{i,j,t} = n_{i,j,t} - H^\text{fi}_{i,j,t} - H^\text{sw(1)}_{i,j,t}$ denotes the number of seals of sex $s$ and age $a$ that have survived spring hunting, and $\tilde{Q}^\text{sw}_t = Q^\text{sw}_t - H^\text{sw(1)}_t$ denotes the number of licenses remaining at the end of the spring hunting season.

The relative probability that a seal of sex $s$ and age $a$ becomes hunted in Sweden either during the spring or the fall can be approximated by normalizing Equations \eqref{rho_sw_spring_full} and \eqref{rho_sw_fall_full}, which results in:
\begin{equation}\label{eq:rel_prob_sw}
    \hat{\psi}^\text{sw(k)}_{s,a} = \frac{\hat{E}^\text{sw(k)}_{s,a} \phi_{s,a}^{\hat{\tau}_\text{sw(k)}}}
    {\sum_{i,j} \hat{E}^\text{sw(k)}_{i,j} \phi_{i,j}^{\hat{\tau}_\text{sw(k)}}}
\end{equation}
where $k \in \{1,2\}$ indicates either the spring or the fall season, and any effect of spring hunting on the relative probabilities of becoming hunted in the fall is assumed to be negligible. Similarly for Finland,
\begin{equation}\label{eq:rel_prob_fi}
    \hat{\psi}^\text{fi}_{s,a} = \frac{\hat{E}^\text{fi}_{s,a} \phi_{s,a}^{\hat{\tau}_\text{fi}}\hat{\omega}_{s,a}}
    {\sum_{i,j} \hat{E}^\text{fi}_{i,j} \phi_{i,j}^{\hat{\tau}_\text{fi}} \hat{\omega}_{s,a}} 
\end{equation}
where the ice extent is assumed to be large enough that relative haul-out probabilities are not affected.

Even though we did not explicitly model bycatch, an expression analogous to Equations \eqref{eq:rel_prob_sw} and \eqref{eq:rel_prob_fi} can be obtained to approximate the relative probabilities of becoming bycaught. Denoting by $B_{s,a}$ the actual number of bycaught seals with sex $s$ and age $a$, and ignoring the effects of demographic stochasticity, $B_{s,a}$ can be written as
\begin{equation}
    B_{s,a} = P(B|s,a)n_{s,a} = P(B|D,s,a)P(D|s,a)n_{s,a}
\end{equation}
where $n_{s,a}$ is the number of seals with sex $s$ and age $a$, $D$ denotes natural mortality, including bycatch, and $B$ denotes bycatch. The composition of the bycatch can then be written as
\begin{equation}\label{eq:expected_bycatch_comp}
    \frac{B_{s,a}}{\sum_{i,j} B_{i,j}} = 
    \frac{P(B|D,s,a) D_{s,a}}{\sum_{i,j} P_{i,j}(B|D,i,j) D_{i,j}}
\end{equation}
where $D_{s,a}=(1-\phi_{s,a})n_{s,a}$ is the number of seals of sex $s$ and age $a$ that died due to natural causes, including bycatch. A comparison of Equation \eqref{eq:expected_bycatch_comp} with the observation model for bycaught samples in Equation \eqref{eq:bycatch_obs_model} indicates that
\begin{equation}
    P(B|D,s,a) \propto \psi_{s,a}
\end{equation}
and therefore the relative probability that a seal of sex $s$ and age $a$ is bycaught is given by
\begin{equation}
    \hat{\psi}^\text{bc}_{s,a} = \frac{\psi_{s,a} (1-\phi_{s,a})}{\sum_{i,j} \psi_{i,j}(1-\phi_{i,j})} 
\end{equation}

\bibliographystyle{inter_research}
\bibliography{ringed_seal_citations}
\setcounter{page}{1}

\section{Density dependence}\label{app:density_dependence}

In resource-limited pinniped populations, increases in population density have shown associations with declines in body size and condition, which can lead to reduced fertility, postponed age at maturity, and heightened rates of juvenile mortality \citep{hammill2017}. In ringed seals, poor body condition can result in the abortion of the pregnancy before the embryo is implanted \citep{boyd1999}, and observations from other pinnipeds suggest that density dependence could lead to post-implantation abortions as well \citep{stenson2016}.

Given reports of significant late-term pregnancy losses in other pinniped populations and the increased energetic demands of late-stage gestation \citep{pitcher1998, mckenzie2005, stenson2016}, we assumed the rate of pregnancy losses increases linearly from zero at conception to a maximum at parturition. Moreover, we assumed that this rate, denoted by $\theta$, increases exponentially with population size:
\begin{equation}\label{eq:prob_pregnancy_failure}
    \theta(N, \tau) = \theta^\text{max}_0 e^{\theta_1 N}\tau.
\end{equation}
Here, $\theta^\text{max}_0$ is the "background" failure rate immediately after mating when population density is zero, $\theta_1$ is the rate at which the failure rate increases with population size and $\tau$ is the time since mating. The density dependent pregnancy rate at any point during the year can then be expressed as
\begin{equation}
    p(N, \tau) = p_0 e^{-\int_0^\tau \theta(N, \tau')d\tau'}
\end{equation}
\begin{equation}
    = p_0 e^{-\theta_0 e^{\theta_1 N}\tau^2}
    \label{eq:p_N_basic}
\end{equation}
where $\theta_0=\theta^\text{max}_0/2$ is the average background failure rate, and $p_0 = p(N, 0)$ is the probability that a female becomes pregnant during the breeding season. Assuming the pregnancy rate immediately before the birth of pups ($\tau \approx 1$) corresponds to the birth rate $b(N)$,
\begin{equation}
    p_0 = b(N) e^{\theta_0 e^{\theta_1 N}} = b(0) e^{\theta_0}
    \label{eq:p_0}
\end{equation}
where the second equality follows from the fact that $p_0$ is independent of $N$.
Substituting Equation \eqref{eq:p_0} into Equation \eqref{eq:p_N_basic},
\begin{equation}
    p_{t-1}(N_{t-1}, \tau) = b_t(N_{t-1}) e^{\theta_0 e^{\theta_1 N_{t-1}}(1-\tau^2)}
\end{equation}
or equivalently,
\begin{equation}
    p_{t-1}(N_{t-1}, \tau) = b_t(0) e^{\theta_0(1-e^{\theta_1 N_{t-1}} \tau^2)}
    \label{eq:p_N}
\end{equation}
where subscripts are included to make explicit the fact that both the population size and reproductive rates depend on the year, $t$. Setting $\tau=1$ in Equation \eqref{eq:p_N} yields an expression for the density dependent birth rate:
\begin{equation}
    b_t(N_{t-1}) = b_t(0) e^{\theta_0 (1- e^{\theta_1 N_{t-1}})}
    \label{eq:b_N}
\end{equation}

It is useful to parametrize Equation \eqref{eq:b_N} in terms of the carrying capacity $K$ instead of the parameter $\theta_1$ since information on historic population sizes of Baltic ringed seals is much more prevalent in the literature than information on the mechanisms of density dependent regulation. Evaluating Equation \eqref{eq:b_N} at the carrying capacity, and rearranging the terms to isolate $\theta_1$, gives
\begin{equation}
    \theta_1 = K^{-1}\log\left(1-\theta_0^{-1}\log\left(\frac{b(K)}{b_\text{max}}\right)\right).
    \label{eq:carrying_capacity}
\end{equation}
where we have assumed that $K$ is the carrying capacity for a ringed seal population in the absence of the effects of organochlorine contamination on reproduction. The value of $b(K)$ can be found by setting the net reproduction rate equal to one \citep{caswell2001, ersalman2024}:
\begin{equation}
    R_0 = \frac{b(K)}{2}(1-\phi_{f,5+})^{-1} \prod_{a=0}^4 \phi_{f,a} = 1
    \label{R0}
\end{equation}

Since there is little evidence of late-term pregnancy losses in Baltic ringed seals, and \citet{boyd1991} has postulated that the influence of nutritional factors on pinniped reproduction is greatest at the point of implantation, we also considered a model where the rate of pregnancy losses was assumed to decrease linearly:
\begin{equation}\label{eq:prob_pregnancy_failure_2}
    \theta(N, \tau) = \theta^\text{max}_0 e^{\theta_1 N}(1-\tau).
\end{equation}
With this assumption, the density dependent pregnancy rate is given by
\begin{equation}
    p_{t-1}(N_{t-1}, \tau) = b_t(0) e^{\theta_0(1-e^{\theta_1 N_{t-1}} (2-\tau) \tau)}.
    \label{eq:p_N_2}
\end{equation}

\bibliographystyle{inter_research}
\bibliography{ringed_seal_citations}
\setcounter{page}{1}

\section{Observation models for sampling bias in Swedish hunting during the spring}\label{app:sampling_bias_sw}

Since 2016, Swedish Museum of Natural History (SMNH) has preferentially requested samples of certain seals from hunters between May 1 and July 31. The specific sampling rules have changed over the years, and are summarized in Table~\ref{table:sampling_rules}.

\begin{table}[H]
\begin{center}
    \caption{Samples requested during the spring}
    \begin{tabular}{ |l||l|  }
         \hline
         Year & Requested samples \\
         \hline
         2015   & Any seal \\
         2016-2019 & Seals $>$ 100cm  \\
         2020-2021 & Females $>$ 100cm  \\
         \hline
    \end{tabular}
    \label{table:sampling_rules}
\end{center}
\end{table}

Since no rules have been specified in 2015, we assumed that seals were sampled randomly from the harvests (i.e. $v_{s,a}=1, \ \forall s,a$).

Since 2019, records of seals that have been hunted in Sweden include the hunters assessment of the seals sex and body length, along with an indication of whether or not the seal has been received by SMNH as a sample. This information makes it possible to model the sampling probability of each seal based on whether or not it meets the sampling criteria. 

We assumed that any sampling bias would be identical between 2016 and 2019. Using hunting records from 2019, we modelled whether or not seal $i$ is sampled from the harvests using the Bernoulli distribution
\begin{equation}
    o_i \sim \text{Bernoulli}(l_i x_1 + (1-l_i)x_0)
\end{equation}
where $o_i \in \{0,1\}$ indicates whether or not seal $i$ was sampled, $l_i$ is an indicator that equals one if seal $i$ is over $100$cm and zero otherwise, and $x_0$ and $x_1$ are the sampling probabilities for seals that are under and over $100$cm, respectively.

To estimate the age and sex dependent probability that a seals body length exceeds $100$cm, we utilized hunted seal samples obtained from Finland between May and July, when the sampling rules shown in Table~\ref{table:sampling_rules} are applicable in Sweden. We relied only on Finnish samples due to the known size bias in the Swedish samples, which could lead to overestimation of the probability that seals of a given sex and age exceed $100$cm. We then used the Bernoulli observation model
\begin{equation}
    l_i|s_i,a_i \sim \text{Bernoulli}(q_{s,a})
\end{equation}
where $q_{s,a}$ is the probability that a seal with sex $s$ and age $a$ is over $100$cm. 

The probability $v_{s,a}$ that a seal with sex $s$ and age $a$ is sampled from the Swedish harvest during the spring was then calculated as
\begin{equation}
    v_{s,a} = q_{s,a} x_1 + (1-q_{s,a}) x_0
\end{equation}

Similarly to the period 2016-2019, we assumed that sampling bias in the spring was identical during 2020-2021. During these years, virtually all of the spring samples were females (a single male individual was deemed an exception and excluded). 
We therefore assumed that sampling probability was zero for males (i.e. $v_{m,a}=0 \ \forall a$) during 2020-2021. The sampling probability for female seals was modelled identically to the period 2016-2019, except only data on female seals were used.

\begin{figure}[h!]
\begin{center}
 \includegraphics[width=0.5\linewidth]{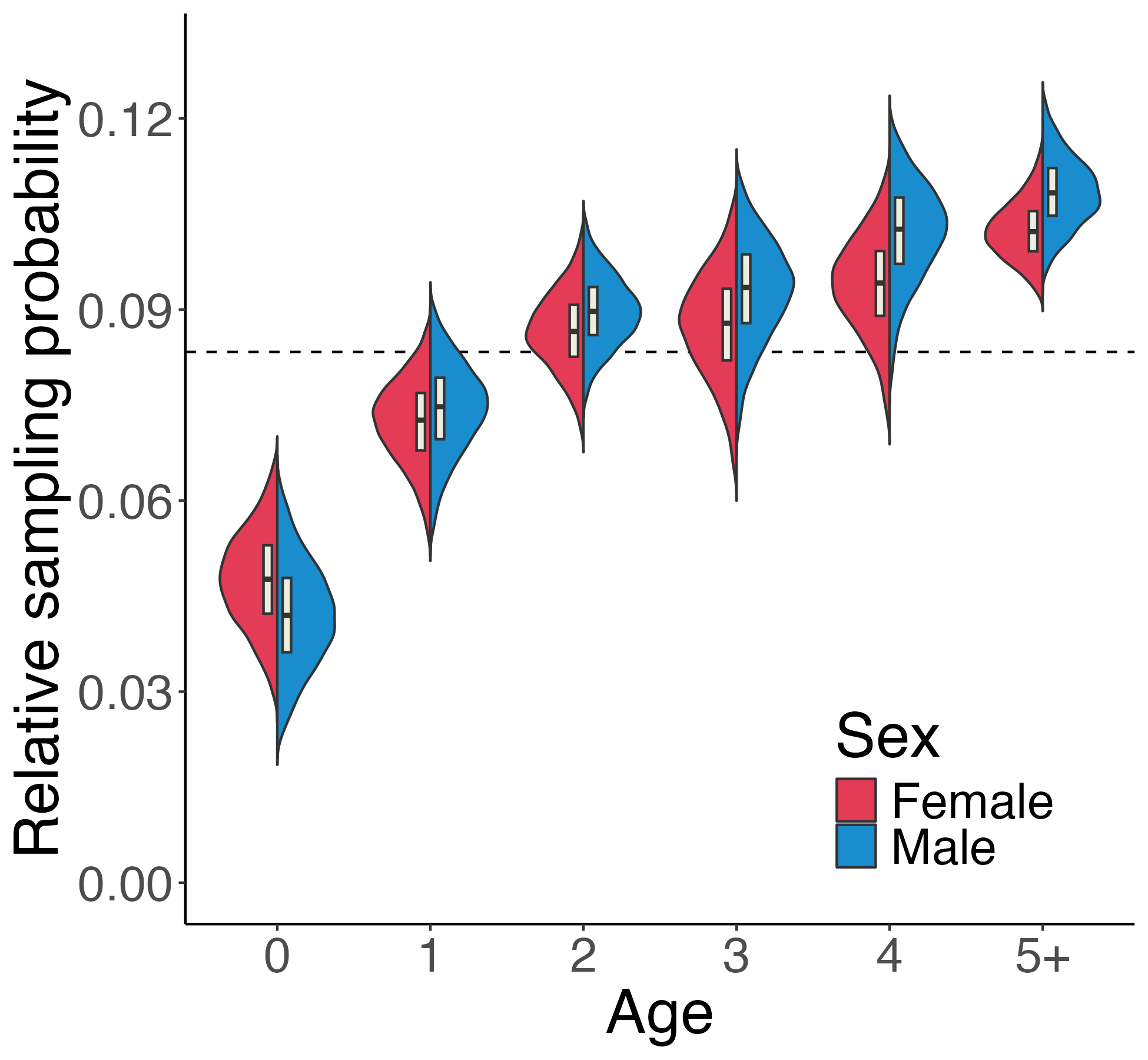}
 \caption{\setstretch{1.0} Estimated age and sex dependent sampling bias for spring hunting in Sweden during 2016-2019. The dashed line indicates random sampling.}
 \label{fig:sampling_bias}
\end{center}
\end{figure}

The estimated sampling bias $\mathbf{v}$ for 2016-2019 is shown in Figure~\ref{fig:sampling_bias}. No age-dependent differences in sampling probability was found among female seals during the spring in 2020-2021.

\setcounter{page}{1}

\section{Observation models for incomplete data}\label{app:obs_model_incomplete_info}

Hunted or bycaught seal samples occasionally had missing age or sex information. We assumed that the age and sex information  was missing at random from those samples. We therefore modelled samples with missing sex or age information using a multinomial distribution similar to those presented in Section~\ref{sec:Hunting_bag_and_hunting_samples}, but with probability vectors given respectively by the age and sex composition of the true harvests and bycatch.

Samples obtained prior to 2007 were not evaluated for placental scars, and we excluded all placental scar evaluations for samples obtained after April in each year from our analysis (Section~\ref{sec:demographic_data}). Our multinomial observation model for reproductive assessments of the remaining samples required evaluations of both a placental scar and a CA (Section~\ref{sec:reproduction_obs}). However, among these samples 16 were evaluated only for a CA and two were evaluated only for a placental scar. To model the samples which were evaluated only for a placental scar, we simply replaced the observation model in Equation \eqref{reproductive_category_obs_model} with the Bernoulli observation model
\begin{equation}
    z_{k,t}^{1\bullet}|
    \boldsymbol{\gamma}_t \sim \text{Bernoulli}\big(\gamma_t^{11}+\gamma_t^{10}\big)
\end{equation}
This approach was not appropriate for samples evaluated only for a CA, since the proportion of seals with a CA differed between samples with and without evaluations of a placental scar, suggesting that the probability of a missing placental scar evaluation was not independent of the outcome of the CA evaluation. To handle samples with missing placental scar evaluations, we included two additional categories, $z_{k,t}^{\bullet 1}$ and $z_{k,t}^{\bullet 0}$, in Equation \eqref{reproductive_category_obs_model}. The probability vector for the extended categorical model, $\tilde{\boldsymbol{\gamma}}_t$, was defined as:
\begin{equation}
    \tilde{\gamma}_t^{ij} = \gamma_t^{ij} 
    (1-P(\text{scar}=\text{NA}|\text{CA}=j)), \quad i,j \in \{0,1\}
\end{equation}
\begin{equation}
    \tilde{\gamma}_t^{\bullet 0} = 
    (\gamma_t^{00} + \gamma_t^{10}) 
    P(\text{scar}=\text{NA}|\text{CA}=0)
\end{equation}
\begin{equation}
    \tilde{\gamma}_t^{\bullet 1} = 
    (\gamma_t^{01} + \gamma_t^{11}) 
    P(\text{scar}=\text{NA}|\text{CA}=1)
\end{equation}
Here, $P(\text{scar}=\text{NA}|\text{CA}=0)$ and $P(\text{scar}=\text{NA}|\text{CA}=1)$ are model parameters denoting the probability of a missing placental scar evaluation conditional on the outcome of the CA evaluation, both of which were given Uniform(0,1) priors.

\setcounter{page}{1}

\section{Implementation in Stan}\label{app:continuous_approximations}

For computational efficiency, we generated samples from the posterior distributions of the parameters using the probabilistic programming language Stan \citep{stan2018}. Because Stan uses gradient-based methods, it does not permit discrete valued parameters. All discrete probability distributions in our model were thus approximated using continuous distributions. In particular, we approximated the binomial distribution with the Gaussian distribution, and the multinomial distribution with the logit-Gaussian distribution. 

We use the Gaussian approximation also to the binomial distribution in our model of pup production. Instead of modelling pup production directly as a multinomial distribution as stated, we formulated the birth and sex allocation of pups as two successive binomial processes approximated with Gaussian distributions. 

The multinomial distribution appears in our model of the survival process. When the expected number of successes for any multinomial category is small, the corresponding multivariate Gaussian approximation would have a significant probability density at negative values, not only resulting in a poor approximation, but also causing major problems with sampling. 
Hence, instead of a multivariate Gaussian distribution the multinomial distribution is often approximated with a Dirichlet distribution where a multinomial distributed random variable $\mathbf{y}$ is  approximated as 
\begin{equation}
    \mathbf{y} \approx \mathbf{y'} = \mathbf{\tilde{p}}N
\end{equation}
where
\begin{equation}
    \mathbf{\tilde{p}} \sim \text{Dirichlet}(\eta \mathbf{p})
\end{equation}
Here, $N$ is the number of individuals, $\mathbf{p}$ is a vector of transition probabilities, $\eta$ is a scaling parameter that determines the variance of $\mathbf{\tilde{p}}$ and $\mathbf{y'}$ is the continuous approximation to $\mathbf{y}$. The parameter $\eta$ would then be chosen so that the resulting variance of $\mathbf{y'}$ is equal to the variance of $\mathbf{y}$.
However, when $N$ is large, the implied variance of $\mathbf{\tilde{p}}$ is very small. In our model, $N$ corresponds to the total population size which is never smaller than a few thousand individuals. Hence, the resulting small variance of $\mathbf{\tilde{p}}$ caused significant computational challenges with the Dirichlet approximation, stemming from high correlations between $\mathbf{\tilde{p}}$ and other parameters relating to mortality. Thus, in order to improve sampling performance, we used logit-Gaussian approximation instead. The computational advantage of the logit-Gaussian distribution is that it is constructed from Gaussian parameters, which can be de-correlated through a non-centered parametrization. 

The Kullback-Leibler divergence of a logit-Gaussian distributed random variable $\mathbf{\tilde{p}'} \in \mathcal{R}^D$ from a Dirichlet distributed random variable $\mathbf{\tilde{p}}$ with the same dimensions is minimized when the logit-Gaussian distribution has mean vector
\begin{equation}
    \mu_i = \psi(\eta p_i) - \psi(\eta p_D) \quad i = 1, \dots , D-1
\end{equation}
\begin{equation}
    \mu_D = 0
\end{equation}
and covariance matrix
\begin{equation}
    K_{ij} = \delta_{ij} \psi'(\eta p_i) + \psi'(\eta p_D) \quad i = 1, \dots , D-1
\end{equation}
\begin{equation}
    K_{iD} = K_{Dj} = 0 \quad \forall i,j
\end{equation}
where $\psi$ and $\psi'$ are the digamma and trigamma functions, respectively and $\delta_{ij}$ is the Kronecker delta \citep{atchison1980}. $\mathbf{\tilde{p}}$ can then be approximated as
\begin{equation}
    \mathbf{\tilde{p}} \approx \mathbf{\tilde{p}'} \sim \text{softmax}(N(\boldsymbol{\mu}, \boldsymbol{K}))
\end{equation}

Finally, the behavior of a Dirichlet distribution (and its corresponding logit-Gaussian approximation) can be undesirable for approximating a multinomial distribution when any element of $\eta \mathbf{p}$ is sufficiently small (typically when the expected number of hunted seals is less than one for a given demographic group). To ensure that all elements of $\eta \mathbf{p}$ are greater than one, we scale $\eta$ such that
\begin{equation}
    \hat{\eta} = \eta \bigg(1 + \frac{1}{\text{min}(\eta \mathbf{p})}\bigg)
\end{equation}
The cost of this scaling is that the variance of demographic stochasticity is underestimated when the size of the hunting bag is small.

\bibliographystyle{inter_research}
\bibliography{ringed_seal_citations}

\end{appendices}

\end{document}